\documentclass{emulateapj}
\usepackage{amsmath}
\usepackage{apjfonts}
\usepackage{xspace}
\usepackage{hyperref} 
\usepackage{color}
\hypersetup{%
    colorlinks=false, 
    urlcolor=magenta, 
    linkcolor=blue, 
    citecolor=blue,
    breaklinks=true, 
    bookmarksnumbered=true,
    bookmarksopen=false, 
    bookmarksopenlevel=0,
    hypertexnames=true, 
    unicode=true,          
    pdftoolbar=true,        
    pdfmenubar=true,        
    pdffitwindow=false,     
    pdfstartview={FitH}    
}

\usepackage{etoolbox}
\makeatletter

\patchcmd{\NAT@citex}
  {\@citea\NAT@hyper@{\NAT@nmfmt{\NAT@nm}\NAT@date}}
  {\@citea\NAT@nmfmt{\NAT@nm}\NAT@hyper@{\NAT@date}}
  {}
  {}

\patchcmd{\NAT@citex}
  {\@citea\NAT@hyper@{%
     \NAT@nmfmt{\NAT@nm}%
     \hyper@natlinkbreak{\NAT@aysep\NAT@spacechar}{\@citeb\@extra@b@citeb}%
     \NAT@date}}
  {\@citea\NAT@nmfmt{\NAT@nm}%
   \NAT@aysep\NAT@spacechar%
   \NAT@hyper@{\NAT@date}}
  {}
  {}

\patchcmd{\NAT@citex}
  {\@citea\NAT@hyper@{%
     \NAT@nmfmt{\NAT@nm}%
     \hyper@natlinkbreak{\NAT@spacechar\NAT@@open\if*#1*\else#1\NAT@spacechar\fi}%
       {\@citeb\@extra@b@citeb}%
     \NAT@date}}
  {\@citea\NAT@nmfmt{\NAT@nm}%
   \NAT@spacechar\NAT@@open\if*#1*\else#1\NAT@spacechar\fi%
   \NAT@hyper@{\NAT@date}}
  {}
  {}

\makeatother

\newcommand \dd[1]  { \,\textrm d{#1}}
\newcommand{\be}{\begin{displaymath}}
\newcommand{\ee}{\end{displaymath}}
\newcommand{\bea}{\begin{eqnarray*}}
\newcommand{\eea}{\end{eqnarray*}}

\DeclareMathOperator{\sech}{sech}

\begin{document}
\def\ltsima{$\; \buildrel < \over \sim \;$}
\def\lsim{\lower.5ex\hbox{\ltsima}}
\def\gtsima{$\; \buildrel > \over \sim \;$}
\def\gsim{\lower.5ex\hbox{\gtsima}}

\submitted{Submitted to the Astrophysical Journal, 2015 July 24; accepted 2015 September 4}
\slugcomment{Submitted to the Astrophysical Journal, 2015 July 24; accepted 2015 September 4} 

\title{Tests of the planetary hypothesis for PTFO\,8-8695b}

\author{
Liang Yu\altaffilmark{1,2},
Joshua N.\ Winn\altaffilmark{1},
Micha\"{e}l Gillon\altaffilmark{3}, 
Simon Albrecht\altaffilmark{4},
Saul Rappaport\altaffilmark{1},\\
Allyson Bieryla\altaffilmark{5},
Fei Dai\altaffilmark{1}, 
Laetitia Delrez\altaffilmark{3},  
Lynne Hillenbrand\altaffilmark{6},
Matthew J.\ Holman\altaffilmark{5},
Andrew W.\ Howard\altaffilmark{7}, 
Chelsea X.\ Huang\altaffilmark{8},
Howard Isaacson\altaffilmark{9},
Emmanuel Jehin\altaffilmark{3},  
Monika Lendl\altaffilmark{3,10},
Benjamin T.\ Montet\altaffilmark{5,6}, 
Philip Muirhead\altaffilmark{11},
Roberto Sanchis-Ojeda\altaffilmark{9,12},
Amaury H.M.J.\ Triaud\altaffilmark{1,13,14}
}

\altaffiltext{1}{Department of Physics, and Kavli Institute for
  Astrophysics and Space Research, Massachusetts Institute of
  Technology, Cambridge, MA 02139, USA}

\altaffiltext{2}{\href{mailto:yuliang@mit.edu}{yuliang@mit.edu}}

\altaffiltext{3}{Institut d'Astrophysique et de G\'eophysique,
  Universit\'e de Li\`ege, all\'ee du 6 Ao\^{u}t 17, B-4000 Li\`ege,
  Belgium}

\altaffiltext{4}{Stellar Astrophysics Centre, Department of Physics
  and Astronomy, Aarhus University, Ny Munkegade 120, DK-8000 Aarhus
  C, Denmark}
  
\altaffiltext{5}{Harvard-Smithsonian Center for Astrophysics,
  Cambridge, MA 02138, USA}
  
\altaffiltext{6}{Cahill Center for Astronomy and Astrophysics,
  California Institute of Technology, Pasadena, CA, 91125, USA}

\altaffiltext{7}{Institute for Astronomy, University of Hawaii, 2680
  Woodlawn Drive, Honolulu, HI 96822, USA}

\altaffiltext{8}{Department of Astrophysical Sciences, Princeton
  University, Princeton, NJ 08544, USA}

\altaffiltext{9}{Department of Astronomy, University of California,
  Berkeley, CA 94720, USA}
  
\altaffiltext{10}{Observatoire Astronomique de l'Universit\'e de
  Gen\`eve, Chemin des Maillettes 51, CH-1290 Sauverny, Switzerland}
  
\altaffiltext{11}{Department of Astronomy, Boston University, 725
  Commonwealth Ave., Boston, MA 02215, USA}

\altaffiltext{12}{NASA Sagan Fellow}  

\altaffiltext{13}{Centre for Planetary Sciences, University of Toronto
  at Scarborough, 1265 Military Trail, Toronto, ON M1C 1A4, Canada}
  
\altaffiltext{14}{Department of Astronomy \& Astrophysics, University
  of Toronto, Toronto, ON M5S 3H4, Canada}

\keywords{planetary systems -- stars: individual (PTFO\,8-8695) --
  stars: pre-main sequence}

\begin{abstract}

  The T~Tauri star PTFO\,8-8695 exhibits periodic fading events that
  have been interpreted as the transits of a giant planet on a
  precessing orbit. Here we present three tests of the planet
  hypothesis. First, we sought evidence for the secular changes in
  light-curve morphology that are predicted to be a consequence of
  orbital precession. We observed 28 fading events spread over several
  years, and did not see the expected changes. Instead we found that
  the fading events are not strictly periodic. Second, we attempted to
  detect the planet's radiation, based on infrared observations
  spanning the predicted times of occultations. We ruled out a signal
  of the expected amplitude. Third, we attempted to detect the
  Rossiter-McLaughlin effect by performing high-resolution
  spectroscopy throughout a fading event. No effect was seen at the
  expected level, ruling out most (but not all) possible orientations
  for the hypothetical planetary orbit. Our spectroscopy also revealed
  strong, time-variable, high-velocity H$\alpha$ and Ca~H~\&~K
  emission features.  All these observations cast doubt on the
  planetary hypothesis, and suggest instead that the fading events
  represent starspots, eclipses by circumstellar dust, or occultations
  of an accretion hotspot.

\end{abstract}

\section{Introduction}
\label{introduction}

The discovery of close-in giant planets around very young stars ---
less than a few million years old --- would provide precious
information about the timing of planet formation, the structure of
newborn planets still cooling and contracting, and the mechanism for
shrinking planetary orbits and creating hot Jupiters.  Currently the
only candidate for such an object is PTFO\,8-8695b, found by
\citet{vaneyken12} (hereafter VE+12). PTFO\,8-8695 is a T~Tauri star
in the Orion-OB1a region, with a mass of $\approx$0.4~$M_\odot$, a
spectral type of M3, and an estimated age of 3~Myr
\citep{briceno05}. In addition to the quasi-sinusoidal variability
characteristic of T~Tauri stars, this star was found to exhibit
periodic fading events, during which the star dims by a few percent
for an interval of about 1.8~hours.  VE+12 reported on these and other
properties of the system, and advanced the hypothesis that the fading
events are transits of a close-in giant planet.
 
However, the planetary interpretation is not secure. The system has
some puzzling properties that seem incompatible with the planet
hypothesis, or at least demand that the system has somewhat exotic
properties. In the first place, the ``transit'' light curves do not
have the customary morphology. They were seen to vary in depth and
duration over a timespan of a year, and in some cases to lack the
expected symmetry around the time of minimum light. \citet{barnes13}
proposed that these changes are caused by a large misalignment between
the planet's orbit and the star's equatorial plane. This misalignment,
when combined with an asymmetric intensity profile on the stellar disk
due to gravity darkening, can produce asymmetric transit light
curves. Furthermore, the misalignment leads to nodal precession of the
orbit, which could explain the secular changes in morphology.

By itself this would not be too unusual. Hot Jupiters with spin-orbit
misalignments are now commonplace \citep[see, e.g.][]{albrecht12}, and
nodal precession has been observed in at least one other misaligned
system \citep{szabo11, szabo12}. \citet{barnes13} constructed a model
that quantitatively fits the two light curves measured by VE+12 in
2009 and 2010.  However, in the case of PTFO\,8-8695 the ``transit''
period is equal to the stellar rotation period (as estimated from the
quasi-sinusoidal variability): both are consistent with 0.448~days or
10.8~hours. It seems strange that the system would have reached
spin-orbit synchronization without also achieving spin-orbit
alignment.\footnote{\citet{kamiaka15} explored models in which the
  orbital and rotation periods are not necessarily synchronized, under
  the premise that the stellar rotation period could have any value up
  to 16~hours (an upper limit set by the measured $v\sin
  i_\star$). However, the quasi-sinusoidal flux variations outside the
  fading events are likely due to rotation, and have a period that
  agrees with that of the ``transit'' events to within a percent. Thus
  it seems unnecessary to consider non-synchronized models.} The
coincidence between the ``transit'' and rotation periods raises the
possibility that the fading events are actually due to starspots, or
eclipses by a corotating structure within a circumstellar disk or
accretion flow.

Another striking property of PTFO\,8-8695 is that the planetary radius
inferred by VE+12 was 1.9~$R_{\rm Jup}$, making it essentially tied
with WASP-17b \citep{triaud+10} for the largest known planetary
radius. Perhaps this should be expected, for a planet that is still
contracting from an initially distended state. Somewhat more worrying
is that the orbital period of 10.8~hours is within or at least near
the Roche limit for a gas giant \citep{rappaport+13}. This suggests
that the planet would be actively losing mass through Roche lobe
overflow.

The conventional way to confirm the existence of a transiting planet
is to detect the expected radial-velocity variation of the host
star. VE+12 attempted to detect such a signal but were foiled by the
spurious radial-velocity variations caused by stellar activity, which
are larger than the amplitude of the expected orbital velocity.  Even
if an apparently sinusoidal radial-velocity signal were detected, it
would be difficult to ascertain whether the signal is planetary in
origin or arises from stellar activity, due to the coincidence between
the ``transit'' and rotation periods.

Given the high scientific stakes, we attempted three less conventional
tests of the planetary hypothesis:
\begin{enumerate}

\item According to the gravity-darkening model of \citet{barnes13},
  continued nodal precession should produce variations in the
  asymmetry, duration, and depth of the fading events, with a period
  of a few years. \citet{barnes13} also predicted that there should be
  intervals of several months during which the fading events cease,
  because the planet's trajectory does not cross the face of the
  star. Therefore, we undertook time-series photometry of as many
  fading events as possible over a timespan of several years, to
  detect the expected changes in morphology.

\item Close-in giant planets emit relatively strongly at infrared
  wavelengths, due to a combination of reflected starlight and the
  planet's own thermal radiation.  Therefore, we attempted to detect
  the loss of light when the planet is hidden by the star, by
  performing time-series infrared photometry spanning the expected
  times of occultations (halfway between transits).

\item A key premise of the planet hypothesis is that the orbit is
  misaligned with the stellar equator. In contrast, starspots move in
  a prograde direction, aligned with stellar rotation. The angle
  between the trajectory of a transiting feature and the
  (sky-projected) stellar equator can be measured by observing the
  Rossiter-McLaughlin effect
  \citep{rossiter24,mclaughlin24}. Therefore, we undertook
  high-resolution optical spectroscopy throughout a fading event to
  measure the spin-orbit angle of whatever is apparently blocking the
  starlight. We also used the spectra to check for time variations in
  the sky-projected rotation rate ($v \sin i_{\star}$), which would be
  expected if the star is precessing.

\end{enumerate}

This paper is organized as follows. Section~\ref{sec:phot} presents
time-series photometry of the candidate transits and occultations,
using several ground-based telescopes and an archival observation with
the {\it Spitzer} Space Telescope.  Section~\ref{sec:spec} presents
our time-series spectroscopy and our attempt to detect the
Rossiter-McLaughlin effect. Section~\ref{sec:discussion} analyzes the
preceding results and their implications for the planetary hypothesis
as well as other possible explanations for the fading events.


\section{Time-series photometry}
\label{sec:phot}

\subsection{Overview}

\begin{figure}[t]
\epsscale{1}
\includegraphics[width=0.5\textwidth]{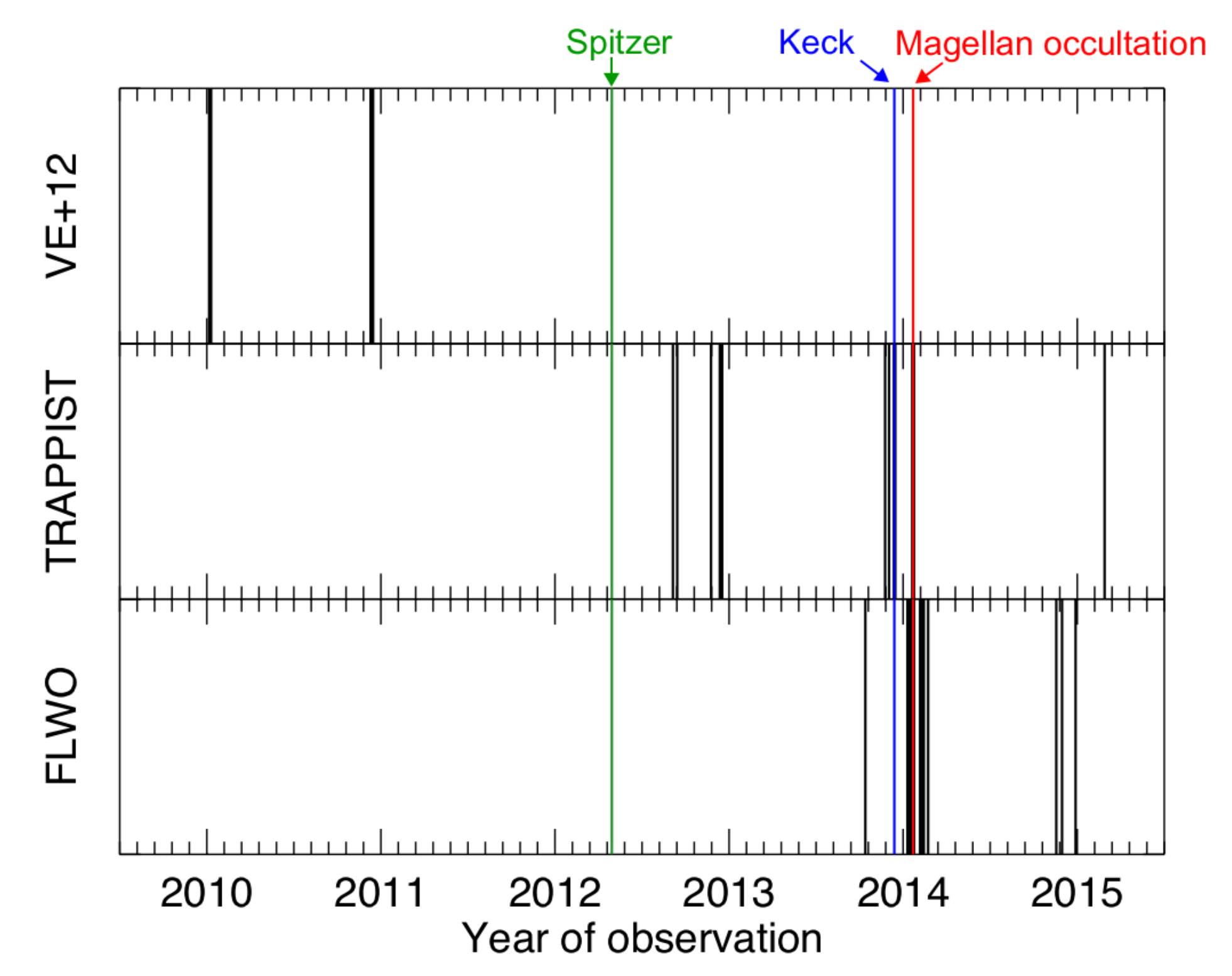}
\caption{Dates of observations of fading events with the FLWO~1.2m and
  TRAPPIST 0.6m telescopes, along with previous observations by VE+12.
  Also indicated are the dates of the observations with {\it Spitzer},
  Magellan, and Keck.}
    \label{fig:sampling} 
    \vspace{2mm}
\end{figure}

\begin{figure*}[t]
\epsscale{1}
\includegraphics[width=\textwidth]{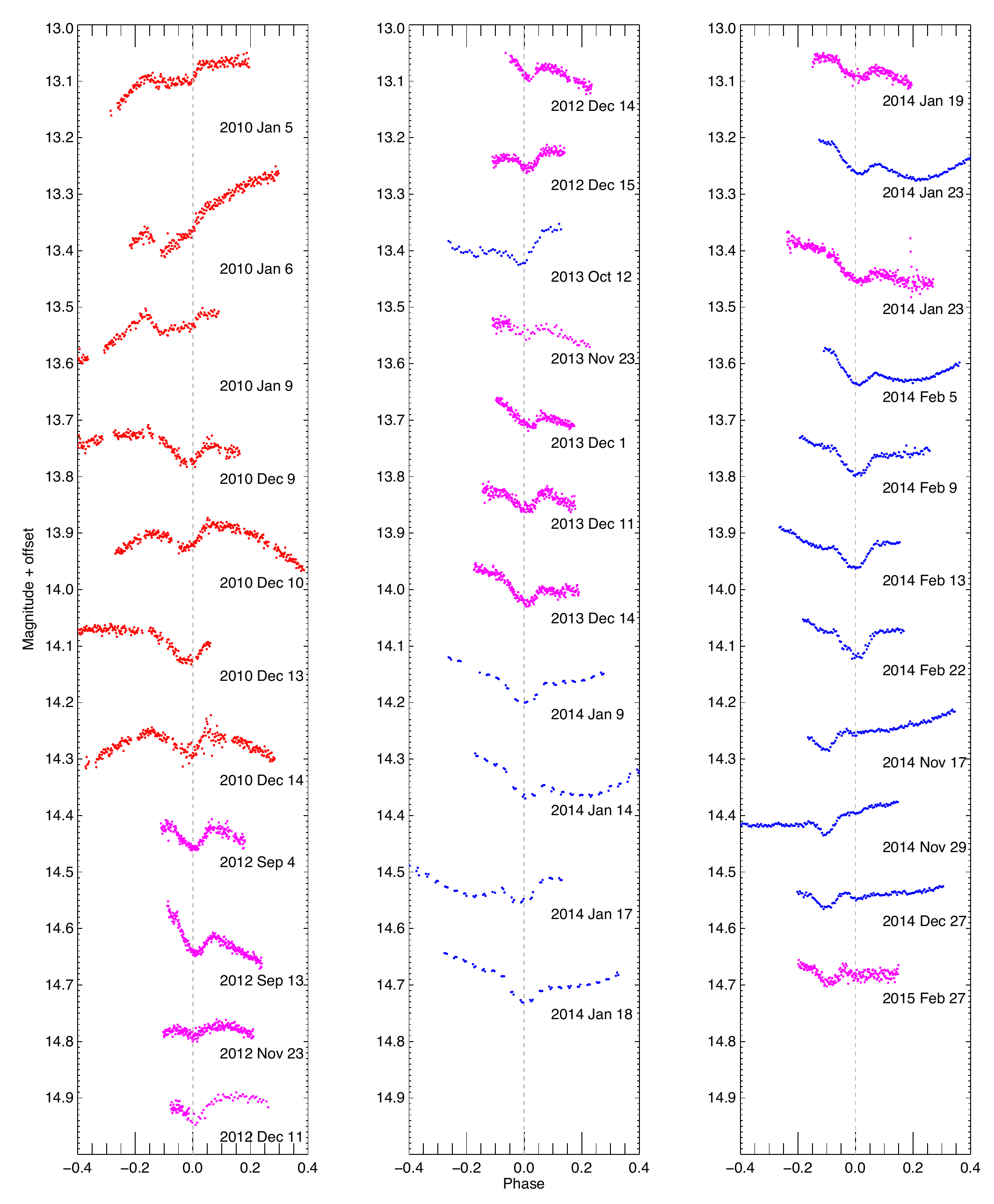}
\caption{Time-series photometry of fading events of PTFO\,8-8695.  The
  vertical scale is the same for all events; vertical offsets have
  been applied to separate the different time series.  Included in
  this plot are the 7 highest-quality light curves from VE+12 (red), as well
  as 26 new light curves from the FLWO~1.2m telescope (blue) and the
  TRAPPIST~0.6m telescope (magenta). The FLWO observations between
  2014~Jan~9-18 were conducted in both $i'$ and $g'$ band.}
\label{fig:unrectified} 
\end{figure*}

We conducted time-series photometric observations of fading events
between 2012 and 2015. Below, in \S~\ref{sec:ground-tra}, we present
ground-based observations of 26 different events.
Figure~\ref{fig:sampling} shows their distribution in time, and
Figure~\ref{fig:unrectified} shows the light curves (including the 7
highest-quality light curves presented previously by VE+12, for
reference). In a few cases we observed the event through multiple
broadband filters. The resulting multi-band light curves are shown in
Figure~\ref{fig:multiband}. We also observed a candidate occultation
at infrared wavelengths with one of the Magellan 6.5m telescopes;
those data are described in \S~\ref{sec:magellan} and plotted in
Figure~\ref{fig:magellan-occ}. Finally, we analyzed the available {\it
  Spitzer} data, spanning a fading event as well as an expected
occultation. Those data are described in \S~\ref{sec:spitzer} and
shown in Figure~\ref{fig:spitzer_all}. The dates of the {\it Spitzer}
and Magellan observations are also indicated on
Figure~\ref{fig:sampling}, along with the Keck spectroscopic
observations described in Section~\ref{sec:spec}.

\begin{deluxetable*}{lcccccc}
\tabletypesize{\scriptsize}
\setlength{\tabcolsep}{-2in} 
\tablewidth{0pt}
\tablecaption{Best-fitting parameters of fading events (see
  Eq.~\ref{eq:model}). \label{tbl:avefit}}
\tablecolumns{6}
\tablehead{
\colhead{UT date}&
\colhead{Filter}&
\colhead{$\chi^2_{\rm min}$}&
\colhead{No.\ of}&
\colhead{Time of minimum light, $t_0$} &
\colhead{Fractional loss} &
\colhead{Duration, $w$}  \\
\colhead{ }&
\colhead{ }&
\colhead{ }&
\colhead{data points}&
\colhead{[HJD$-$2,455,200]} &
\colhead{of light, $\delta$} &
\colhead{[days]}
}
\startdata
 2010 Jan 5 & $R$ & 193.4 & 228 & $ 1.8053 \pm 0.0014$ & $0.0442 \pm 0.0052$ & $ 0.0282 \pm 0.0033$ \vspace{0.5mm} \\
 2010 Jan 6 & $R$ & 216.5 & 244 & $ 2.6961 \pm 0.0017$ & $0.0789 \pm 0.0208$ & $ 0.0349 \pm 0.0059$ \vspace{0.5mm} \\
 2010 Jan 9 & $R$ & 217.0 & 191 & $ 5.8431 \pm 0.0014$ & $0.0979 \pm 0.0142$ & $ 0.0370 \pm 0.0042$ \vspace{0.5mm} \\
 2010 Dec 9 & $R$ & 169.9 & 270 & $ 339.9069 \pm 0.0006$ & $0.0498 \pm 0.0023$ & $ 0.0190 \pm 0.0013$ \vspace{0.5mm} \\
 2010 Dec 10 & $R$ & 248.4 & 298 & $ 340.8020 \pm 0.0006$ & $0.0425 \pm 0.0011$ & $ 0.0167 \pm 0.0007$ \vspace{0.5mm} \\
 2010 Dec 13 & $R$ & 191.1 & 271 & $ 343.9467 \pm 0.0010$ & $0.0558 \pm 0.0038$ & $ 0.0229 \pm 0.0017$ \vspace{0.5mm} \\
 2010 Dec 14 & $R$ & 473.2 & 288 & $ 344.8401 \pm 0.0009$ & $0.0412 \pm 0.0018$ & $ 0.0158 \pm 0.0012$ \vspace{0.5mm} \\
 2012 Sep 4 & $I+z$ & 194.0 & 209 & $ 974.8427 \pm 0.0005$ & $0.0510 \pm 0.0027$ & $ 0.0160 \pm 0.0011$ \vspace{0.5mm} \\
2012 Sep 13 & $I+z$ & 388.8 & 265 & $ 983.8121 \pm 0.0004$ & $0.0521 \pm 0.0014$ & $ 0.0138 \pm 0.0006$ \vspace{0.5mm} \\
2012 Nov 23 & $I+z$ & 240.1 & 243 & $ 1054.6614 \pm 0.0009$ & $0.0189 \pm 0.0013$ & $ 0.0128 \pm 0.0014$ \vspace{0.5mm} \\
2012 Dec 11 & $I+z$ & 94.2 & 94 & $1072.5994 \pm 0.0009$ & $0.0366 \pm 0.0024$ & $ 0.0132 \pm 0.0015$ \vspace{0.5mm} \\
2012 Dec 14 & $I+z$ & 154.0 & 176 & $1075.7379 \pm 0.0006$ & $0.0298 \pm 0.0014$ & $ 0.0098 \pm 0.0008$ \vspace{0.5mm} \\
2012 Dec 15 & $I+z$ & 152.8 & 182 & $1076.6357 \pm 0.0005$ & $0.0338 \pm 0.0029$ & $ 0.0128 \pm 0.0013$ \vspace{0.5mm} \\
2013 Oct 12 & $i'$ & 9.7 & 54 & $1377.9469 \pm 0.0009$ & $0.0324 \pm 0.0024$ & $ 0.0099 \pm 0.0012$ \vspace{0.5mm} \\
2013 Nov 23 & $I+z$ & 102.7 & 93 & $1419.6524 \pm 0.0022$ & $0.0177 \pm 0.0029$ & $ 0.0114 \pm 0.0030$ \vspace{0.5mm} \\
 2013 Dec 1 & $I+z$ & 245.9 & 195 & $1427.7243 \pm 0.0007$ & $0.0316 \pm 0.0026$ & $ 0.0163 \pm 0.0016$ \vspace{0.5mm} \\
2013 Dec 11 & $I+z$ & 190.3 & 233 & $1437.5864 \pm 0.0007$ & $0.0385 \pm 0.0030$ & $ 0.0153 \pm 0.0016$ \vspace{0.5mm} \\
2013 Dec 14 & $I+z$ & 255.1 & 262 & $1440.7266 \pm 0.0005$ & $0.0344 \pm 0.0016$ & $ 0.0139 \pm 0.0011$ \vspace{0.5mm} \\
2014 Jan 9 & $i'$ & 127.6 & 46 & $1466.7304 \pm 0.0007$ & $0.0385 \pm 0.0016$ & $ 0.0156 \pm 0.0012$ \vspace{0.5mm} \\
2014 Jan 14 & $i'$ & 183.0 & 57 & $1471.6634 \pm 0.0011$ & $0.0337 \pm 0.0025$ & $ 0.0128 \pm 0.0015$ \vspace{0.5mm} \\
2014 Jan 17 & $i'$ & 245.0 & 59 & $1474.8039 \pm 0.0010$ & $0.0377 \pm 0.0042$ & $ 0.0160 \pm 0.0023$ \vspace{0.5mm} \\
2014 Jan 18 & $i'$ & 177.0 & 57 & $1475.7000 \pm 0.0007$ & $0.0375 \pm 0.0014$ & $ 0.0144 \pm 0.0009$ \vspace{0.5mm} \\
2014 Jan 19 & $I+z$ & 253.6 & 240 & $1476.5974 \pm 0.0007$ & $0.0326 \pm 0.0023$ & $ 0.0167 \pm 0.0016$ \vspace{0.5mm} \\
2014 Jan 23 & $i'$ & 331.9 & 136 & $1480.6289 \pm 0.0004$ & $0.0628 \pm 0.0044$ & $ 0.0218 \pm 0.0013$ \vspace{0.5mm} \\
2014 Jan 23 & $I+z$ & 549.4 & 356 & $1480.6339 \pm 0.0010$ & $0.0295 \pm 0.0018$ & $ 0.0149 \pm 0.0015$ \vspace{0.5mm} \\
2014 Feb 5 & $i'$ & 404.4 & 111 & $1493.6353 \pm 0.0005$ & $0.0349 \pm 0.0013$ & $ 0.0140 \pm 0.0008$ \vspace{0.5mm} \\
2014 Feb 9 & $i'$ & 330.9 & 107 & $1497.6729 \pm 0.0005$ & $0.0402 \pm 0.0013$ & $ 0.0132 \pm 0.0008$ \vspace{0.5mm} \\
2014 Feb 13 & $i'$ & 246.4 & 99 & $1501.7075 \pm 0.0005$ & $0.0451 \pm 0.0022$ & $ 0.0147 \pm 0.0010$ \vspace{0.5mm} \\
2014 Feb 22 & $i'$ & 476.4 & 83 & $1510.6778 \pm 0.0007$ & $0.0463 \pm 0.0027$ & $ 0.0148 \pm 0.0013$ \vspace{0.5mm} \\
2014 Nov 17 & $i'$ & 208.1 & 119 & $1778.7654 \pm 0.0007$ & $0.0304 \pm 0.0020$ & $ 0.0116 \pm 0.0012$ \vspace{0.5mm} \\
2014 Nov 29 & $i'$ & 259.0 & 145 & $1790.8761 \pm 0.0009$ & $0.0185 \pm 0.0014$ & $ 0.0088 \pm 0.0011$ \vspace{0.5mm} \\
2014 Dec 27 & $i'$ & 199.7 & 118 & $1818.6748 \pm 0.0007$ & $0.0222 \pm 0.0012$ & $ 0.0094 \pm 0.0008$ \vspace{0.5mm} \\
2015 Feb 27 & $I+z$ & 200.0 & 206 & $1880.5567 \pm 0.0009$ & $0.0206 \pm 0.0017$ & $ 0.0087 \pm 0.0012$ \vspace{0.5mm} 

\enddata
\normalsize
\label{tab:dd}
\end{deluxetable*}

\begin{deluxetable*}{lccccc}
\tabletypesize{\scriptsize}
\setlength{\tabcolsep}{-2in} 
\tablewidth{0pt}
\tablecaption{Best-fitting parameters of phase-folded light curves (see Eq.~\ref{eq:amodel})}
\tablecolumns{6}
\tablehead{
\colhead{Light curve}&
\colhead{Filter}&
\colhead{Time of minimum light, $t_0$} &
\colhead{Fractional loss} &
\colhead{Ingress duration, $w_1$} &
\colhead{Egress duration, $w_2$} \\
\colhead{ }&
\colhead{ }&
\colhead{[days]} &
\colhead{of light, $\delta$} &
\colhead{[days]}&
\colhead{[days]}
}
\startdata
 FLWO average & $i'$ & $ 0.0029 \pm 0.0028$ & $0.0360 \pm 0.0017$ & $0.0166 \pm 0.0023$ & $0.0124 \pm 0.0018$ \vspace{0.5mm} \\
 TRAPPIST average & $I+z$ & $ 0.0018 \pm 0.0018$ & $0.0336 \pm 0.0010$ & $0.0149 \pm 0.0014$ & $0.0128 \pm 0.0013$ \vspace{0.5mm} 
 \enddata
\normalsize
\label{tab:avefit}
\end{deluxetable*}

\begin{figure*}[t]
\epsscale{1}
\includegraphics[width=\textwidth]{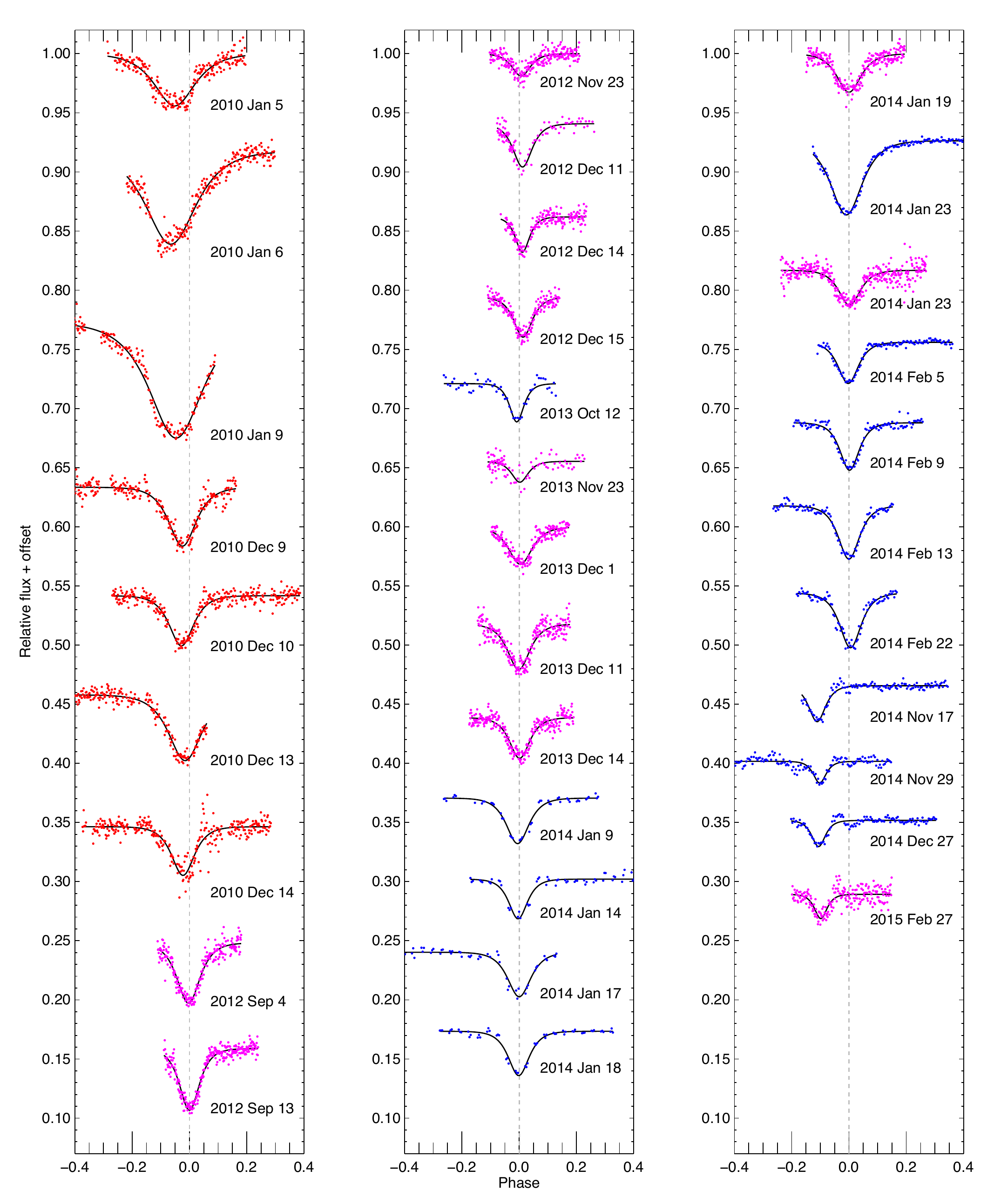}
\caption{Same as Figure~\ref{fig:unrectified}, but after
  ``flattening'' the out-of-transit flux variation by dividing by the
  best-fitting polynomial function of time.  Flux values are
  normalized to unity outside of the transit.  The black curves are
  the best-fitting models (see Eqn.~\ref{eq:model}), from which we
  derived the transit times, depths, and durations that are reported
  in Table~\ref{tbl:avefit}.}
\label{fig:rectified} 
\end{figure*}

\subsection{Ground-based observations of fading events}
\label{sec:ground-tra}

We observed 13 fading events with the 1.2m telescope at the Fred
Lawrence Whipple Observatory (FLWO) on Mt.\ Hopkins, Arizona.  The
instrument, Keplercam, has a single $4096 \times 4096$ CCD with a
$23\farcs 1$ field of view.  All the events were observed through an
$i'$ filter.  For the events between 2014 Jan 9--18, we interleaved
the $i'$-band exposures with $g'$-band exposures, although the
$g'$-band data were only useful in two cases. Calibration was
performed using standard IRAF\footnote{The Image Reduction and
  Analysis Facility (IRAF) is distributed by the National Optical
  Astronomy Observatory, which is operated by the Association of
  Universities for Research in Astronomy (AURA) under a cooperative
  agreement with the National Science Foundation.} procedures,
including bias and flat-field corrections. The time stamps were placed
on the $\mathrm{BJD_{TDB}}$ system using the code by
\citet{eastman10}. Circular aperture photometry was performed with the
Interactive Data Language (IDL).

Another 13 events were observed with the 0.6m TRAnsiting Planets and
PlanetesImals Small Telescope (TRAPPIST), located at ESO's La Silla
Observatory in Chile. This telescope is equipped with a
thermoelectrically-cooled $2048 \times 2048$ CCD with a $22'$ field of
view \citep{gillon11, jehin11}. The observations were conducted with a
custom ``$I + z$'' filter, which has transmittance $>$90\% between
750--1100~nm. We refer the reader to \citet{gillon13} for descriptions
of the procedures for observing and data reduction. The two events of
2012~Dec~14 and 15 were simultaneously observed with a Gunn $r'$
filter, using EulerCam on the 1.2m Euler-Swiss Telescope at the La
Silla site, Chile. EulerCam uses a $2048 \times 2048$ CCD with a field
of view of 14$\farcs$7 on a side. For details on the instrument and
data reduction procedures, please refer to \citet{lendl12}.

A single event on 2014~Jan~19 was observed with the 6.5m Magellan~I
(Baade) telescope at Las Campanas Observatory in Chile. The same event
was observed simultaneously by TRAPPIST in the $I+z$ band. With
Magellan, we observed in the $H$ band using FourStar, a $2048 \times
2048$ infrared array with a $10\farcm 9$ square field of view. The
data were reduced with IRAF and IDL procedures similar to those used
on the FLWO data.

In all cases the flux of PTFO\,8-8695 was divided by the summed flux
from several reference stars, leading to the light curves plotted in
Figure~\ref{fig:unrectified}. This figure also shows the 7 light
curves presented by VE+12 that cover the entire fading event; those
observations were performed with the 1.2m Palomar telescope and an $R$
filter.

Outside of the fading events, the star varies gradually by
$\sim$0.1~mag over several hours, in a manner consistent with its
young age and late spectral type. Superimposed on those relatively
gradual variations are periodic transit-like fading events lasting no
more than about 2~hours. The depth and duration of the fading seems to
vary from event to event. To derive the basic phenomenological
parameters of the dimming events --- depth, duration, and time of
minimum light --- we fitted a parameterized model describing both the
gradual out-of-transit variations as well as the transit-like loss of
light. We modeled the gradual variations as a polynomial function of
time (2nd or 3rd order, depending on the event). The additional
loss of light during the fading event, relative to the
polynomial-corrected
out-of-transit flux, was modeled as
\begin{equation}
\Delta f(t) = \delta \sech \left[\frac{t-t_0}{w}\right] = \frac{2\delta}{e^{(t-t_0)/w} + e^{-(t-t_0)/w}},
\label{eq:model}
\end{equation}
where $\delta$ is the maximum fractional loss of light (the ``transit
depth''), $w$ is the duration, and $t_0$ is the time of minimum
light. We chose this model instead of a more physically-motivated
transit model \citep[e.g., the model presented by][]{mandelagol02}
because the asymmetries and other odd features in the light curves do
not fit the standard models. Hence there is no advantage in fitting
the physical model when a much simpler model can provide estimates of
the basic transit parameters. One might be able to fit the data with a
model based on transits of an oblate, oblique, precessing,
gravity-darkened star \citep{barnes09}, but such a model is far more
demanding computationally. Our analytic model suffices to estimate the
basic parameters of each event.

Figure~\ref{fig:rectified} shows the light curves after dividing out
the best-fitting polynomial functions. This gives a clearer view of
the ``transits'' with most of the long-term trends
removed. Table~\ref{tab:dd} gives the model parameters, as well as the
value of $\chi^2_{\rm min}$ and the number of data points in each time
series.  In most cases, $\chi^2_{\rm min}$ is too large to be
statistically acceptable, i.e., the simplified model of
Eq.~\ref{eq:model} does not fit the data to within the photometric
uncertainties.  For this reason, Table~\ref{tab:dd} does not report
the formal parameter uncertainties defined by the usual criterion
$\Delta\chi^2=1$. Rather, the reported parameter uncertainties have
been enlarged by a factor of $\sqrt{\chi^2_{\rm min}/N_{\rm dof}}$
where $N_{\rm dof}$ is the number of degrees of freedom. These
enlarged uncertainties were also adopted for our subsequent
calculations.

\begin{figure}[h]
\epsscale{1}
\includegraphics[width=0.5\textwidth]{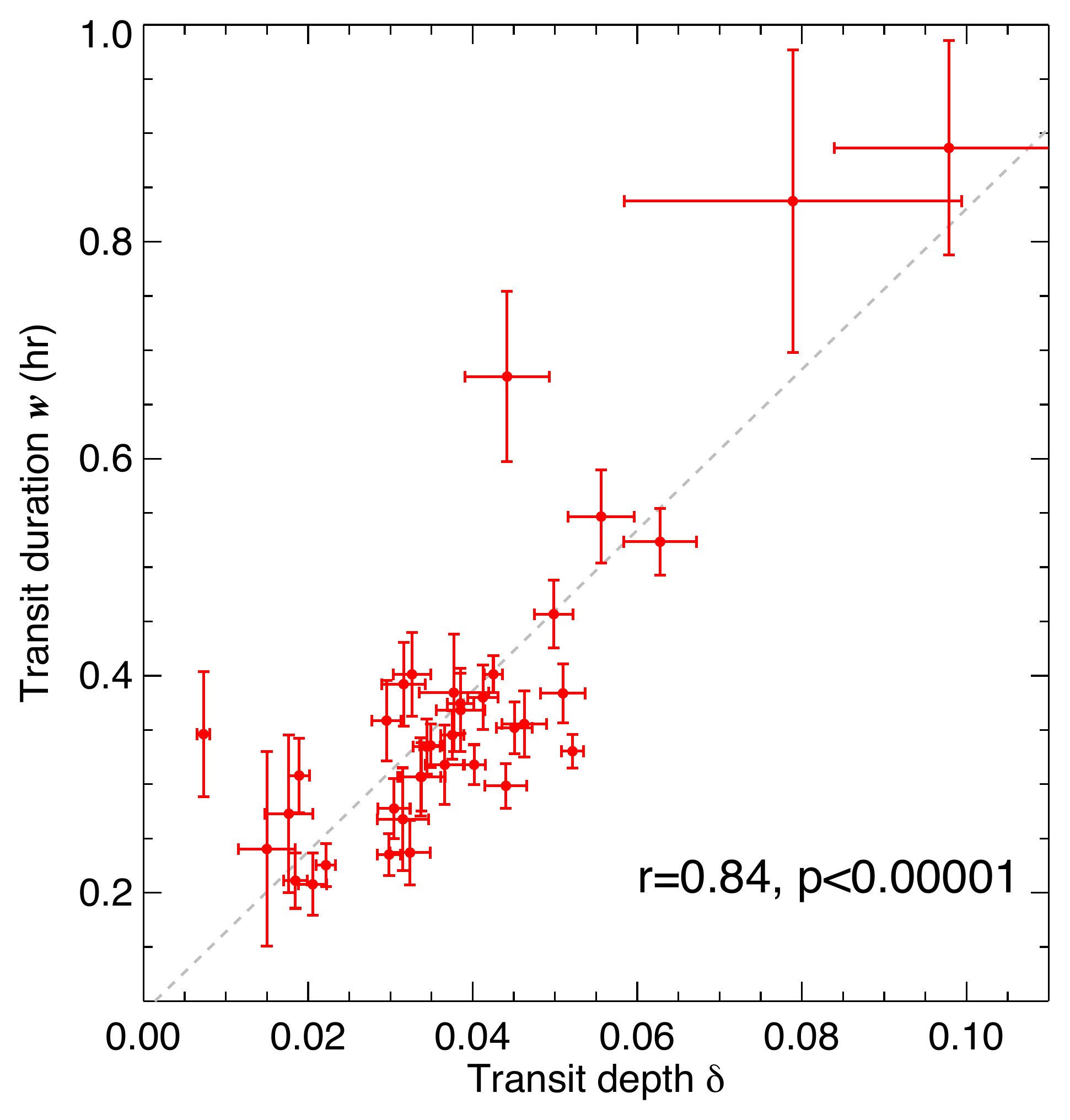}
\caption{Depths and durations of the fading events, estimated for each
  individual event by fitting a simple analytic model
  (Eq.~\ref{eq:amodel}). There is a positive correlation between
  depth and duration.  The best-fit straight line is shown as a dashed
  line. The Pearson correlation coefficient and its statistical
  significance are given in the bottom right corner.}
\label{fig:dd} 
\vspace{2mm}
\end{figure}

Figure~\ref{fig:dd} shows the measured depths and durations. When the
transits are deeper, they also tend to have longer durations; the
measured depth and duration have a Pearson correlation coefficient of
0.84 ($p<10^{-5}$).

\begin{figure}[t]
\minipage{0.5\textwidth}
  \includegraphics[width=\linewidth]{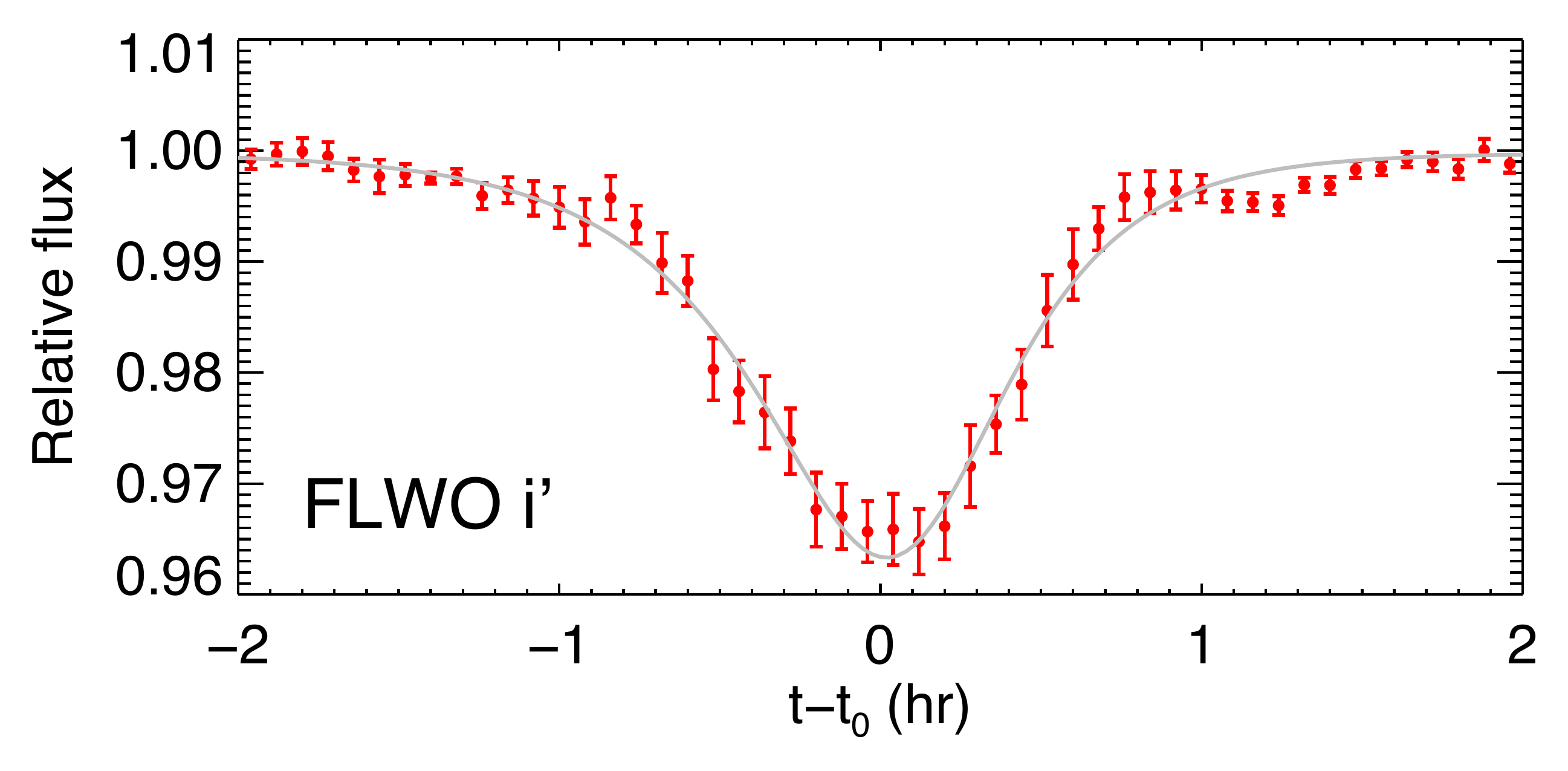}
\endminipage\hfill \\
\minipage{0.5\textwidth}
  \includegraphics[width=\linewidth]{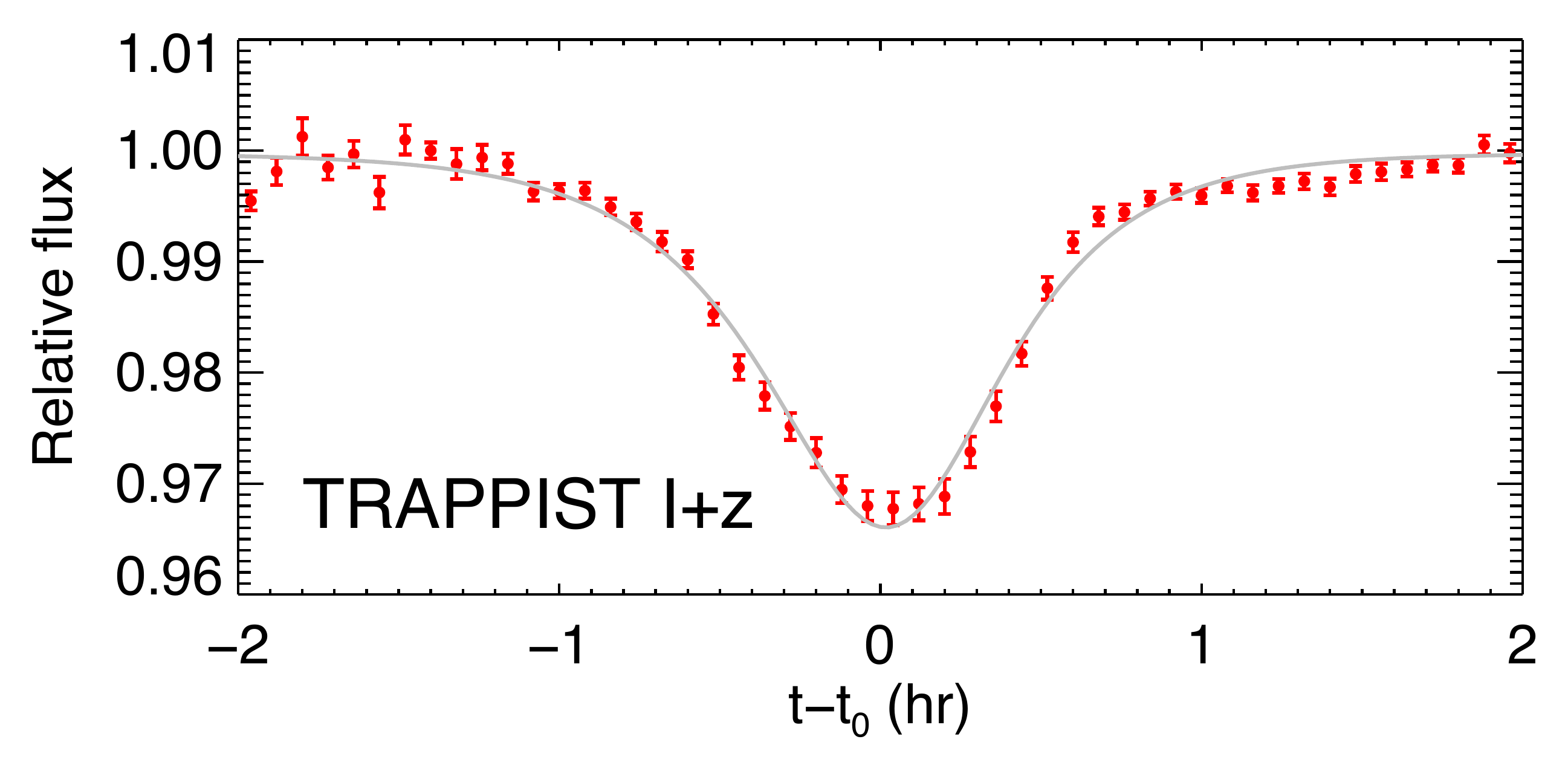}
\endminipage

\caption{{\it Top.}---Phase-folded light curve based on data from all
  the fading events observed with the FLWO~1.2m telescope in the $i'$
  band. {\it Bottom.}---Same, but for the $I+z$ data obtained with
  the 0.6m~TRAPPIST telescope.  In both cases the data were placed
  into 50 time bins spanning a 4-hour period bracketing the expected
  transit time. The grey line represents the best-fit asymmetric model
  for each light curve. The error bars represent the standard
  deviation of the mean in each time bin.}
\label{fig:phasefolded} 
\end{figure}

The VE+12 light curves showed strong asymmetries in at least 5 out of
the 7 complete light curves. None of our new light curves show strong
asymmetries, at least not as clearly as was seen by VE+12.  We tried
fitting a model in which the rates of brightness variation are not
symmetric about the time of minimum light, by using a non-standard
variant of the hyperbolic secant function \citep[see,
e.g.,][]{ruan00}:
\begin{equation}
\Delta f = \frac{2 \delta}{e^{(t-t_0)/w_1} + e^{-(t-t_0)/w_2}}.
\label{eq:amodel}
\end{equation}
The asymmetric model does not seem to improve the quality of the fit
to a significant degree. The number of cases for which the fitted
asymmetry obeyed $w_1>w_2$ (more prolonged ``ingress'') was nearly the
same as the number of cases with $w_2>w_1$, without any obvious
pattern.  Figure~\ref{fig:phasefolded} shows the FLWO and TRAPPIST
data as a function of $t-t_0$ (coverted to hours) after some averaging
in time to increase the signal-to-noise ratio. In these averaged light
curves there does seem to be a slight asymmetry, with a longer
``ingress'' than ``egress'' in both cases. The best-fitting asymmetric
model is shown in Figure~\ref{fig:phasefolded}, and the parameters are
given in Table~\ref{tab:avefit}.

\begin{figure*}[htb]
\minipage{\textwidth}
\includegraphics[width=0.5\linewidth]{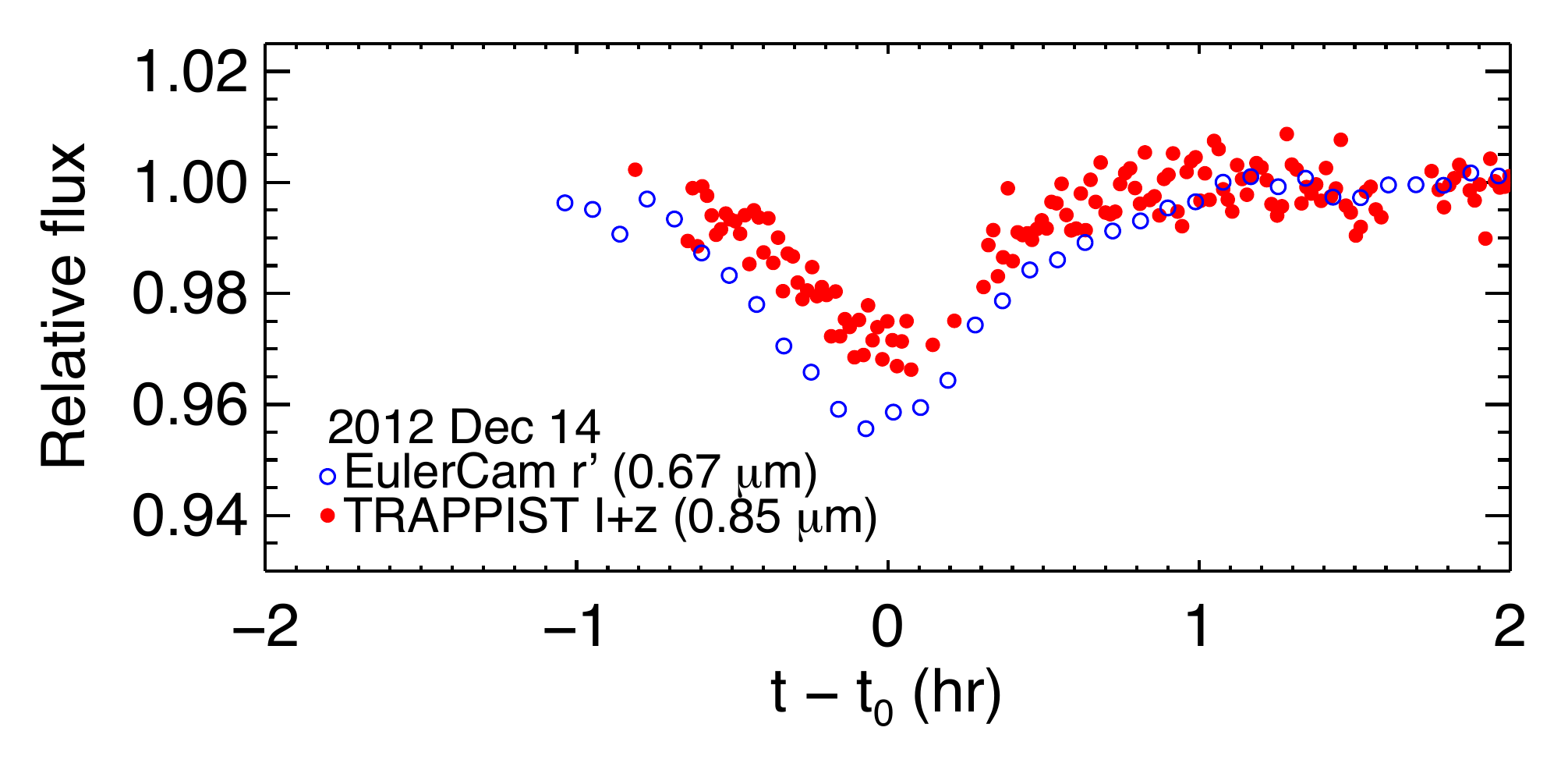}
  \includegraphics[width=0.5\linewidth]{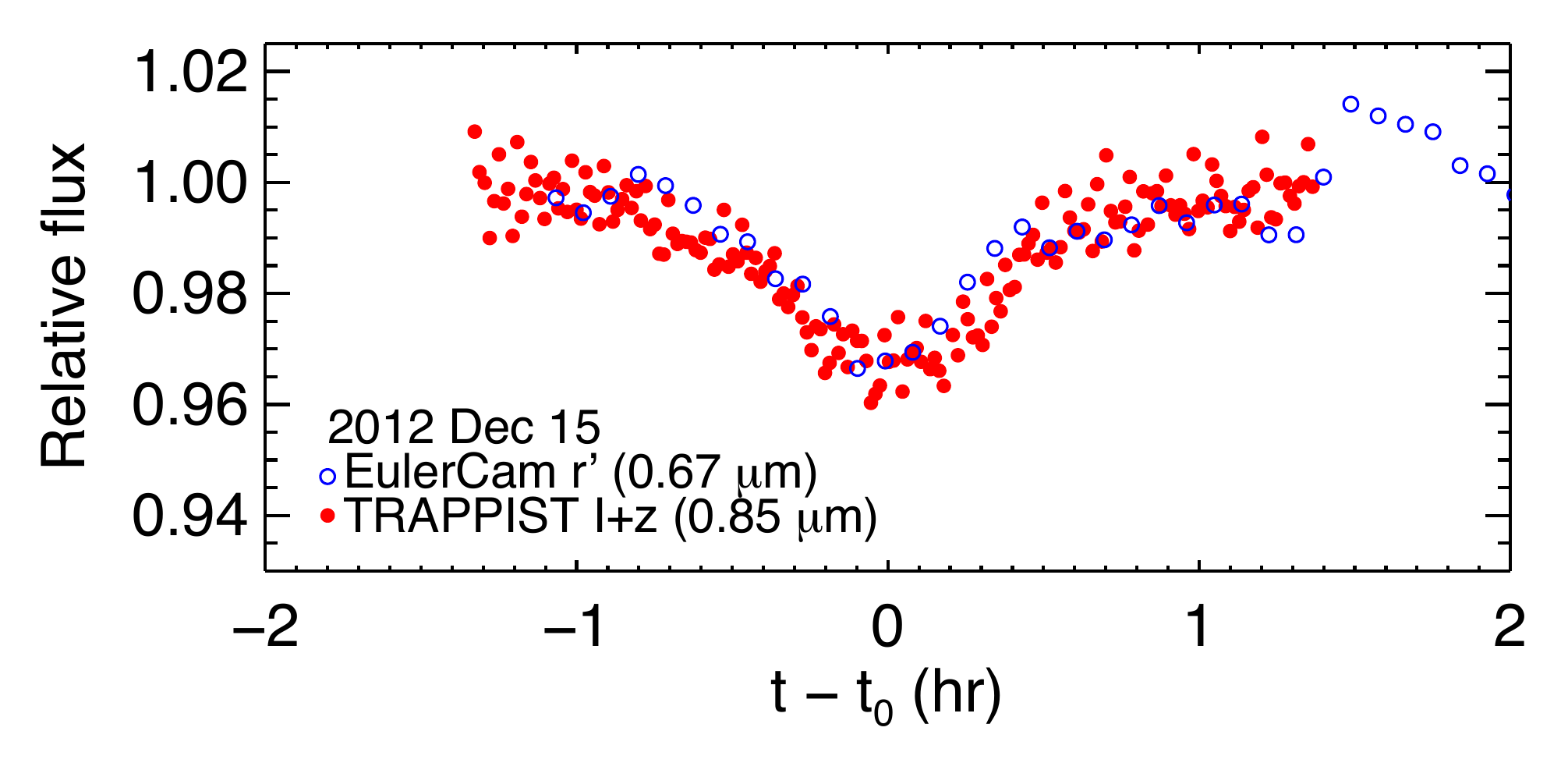}
\endminipage\hfill
\minipage{\textwidth}
\includegraphics[width=0.5\linewidth]{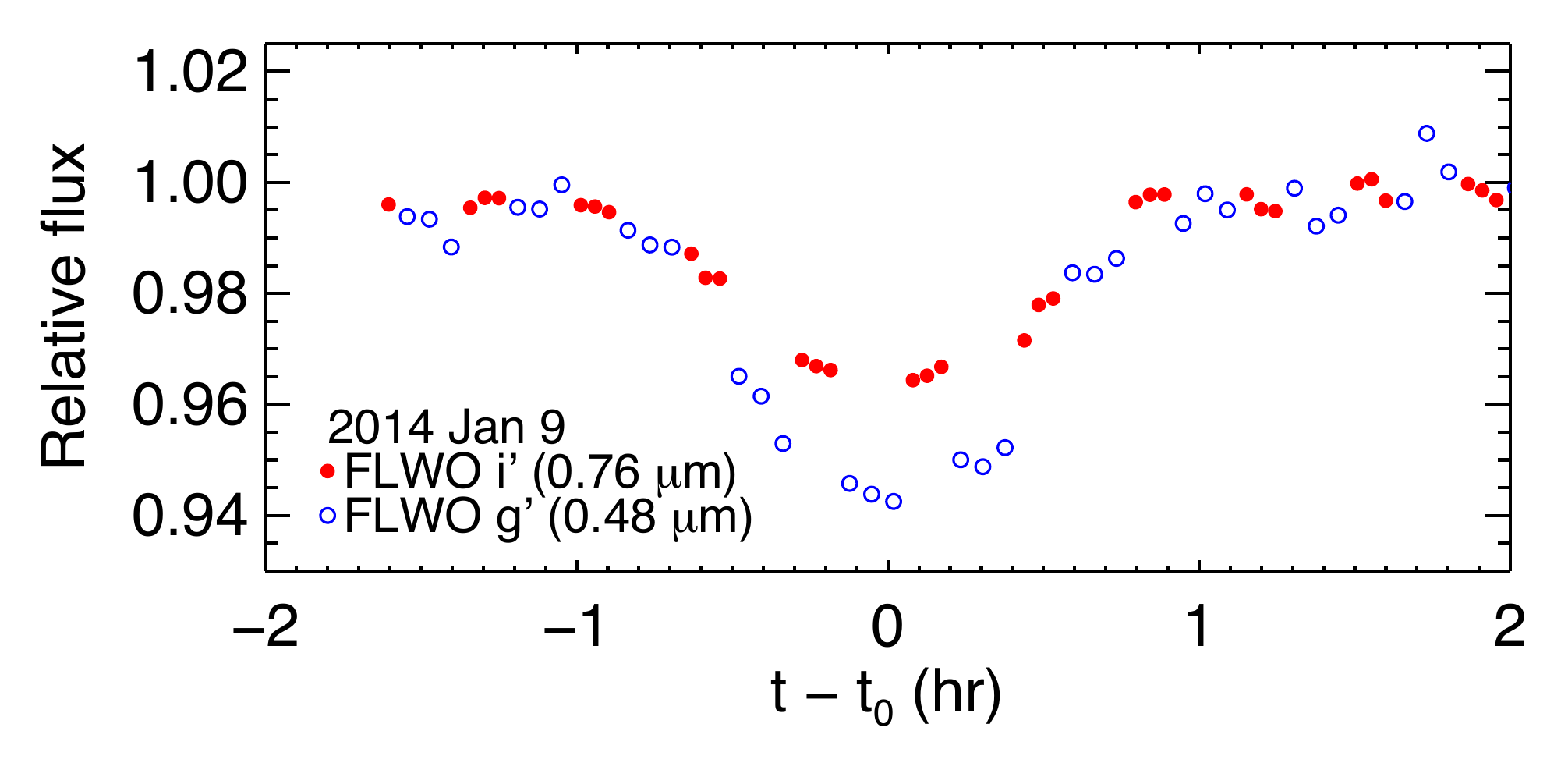}
  \includegraphics[width=0.5\linewidth]{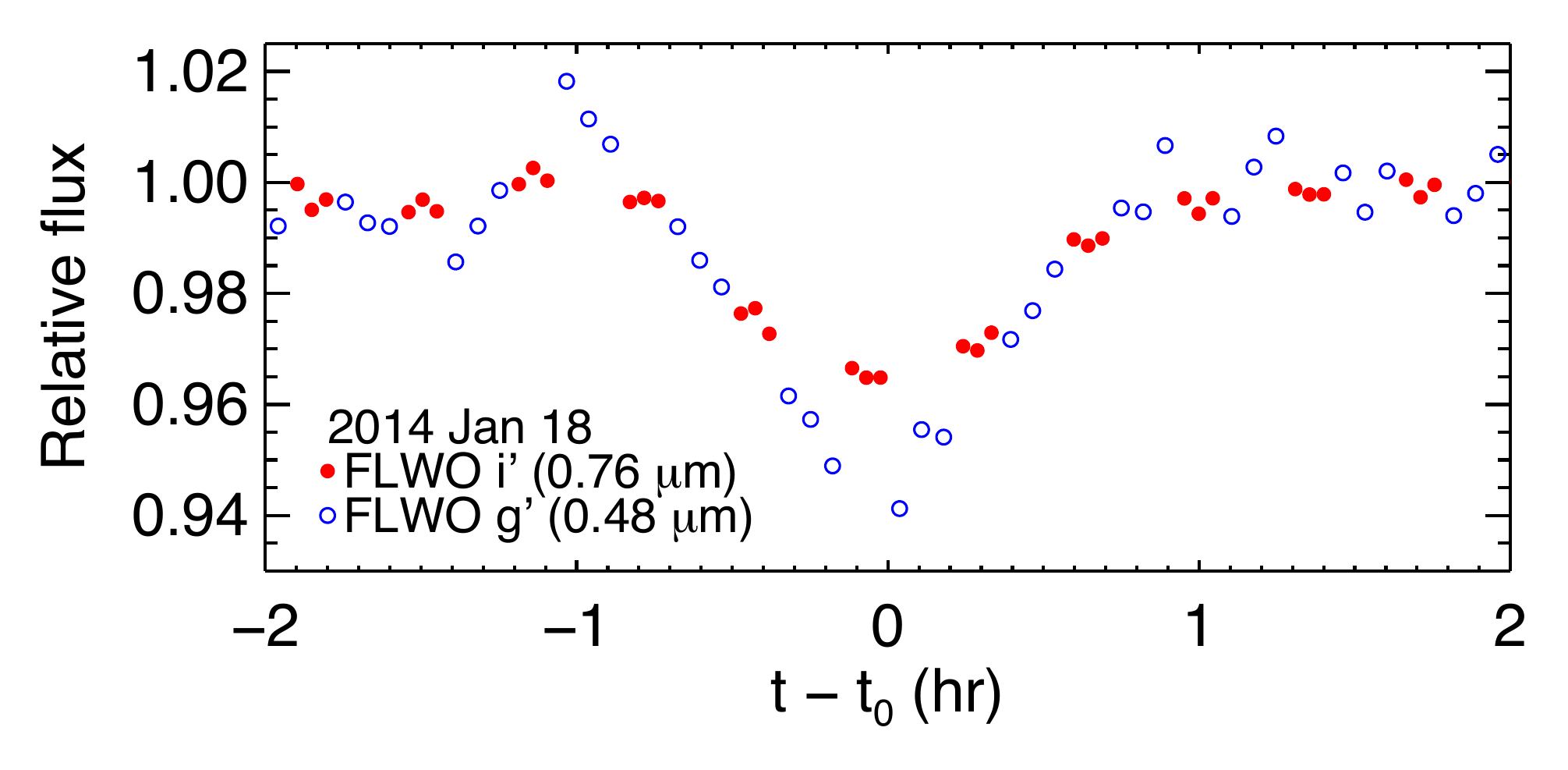}
\endminipage\hfill
\minipage{\textwidth}%
\centering
  \includegraphics[width=0.5\linewidth]{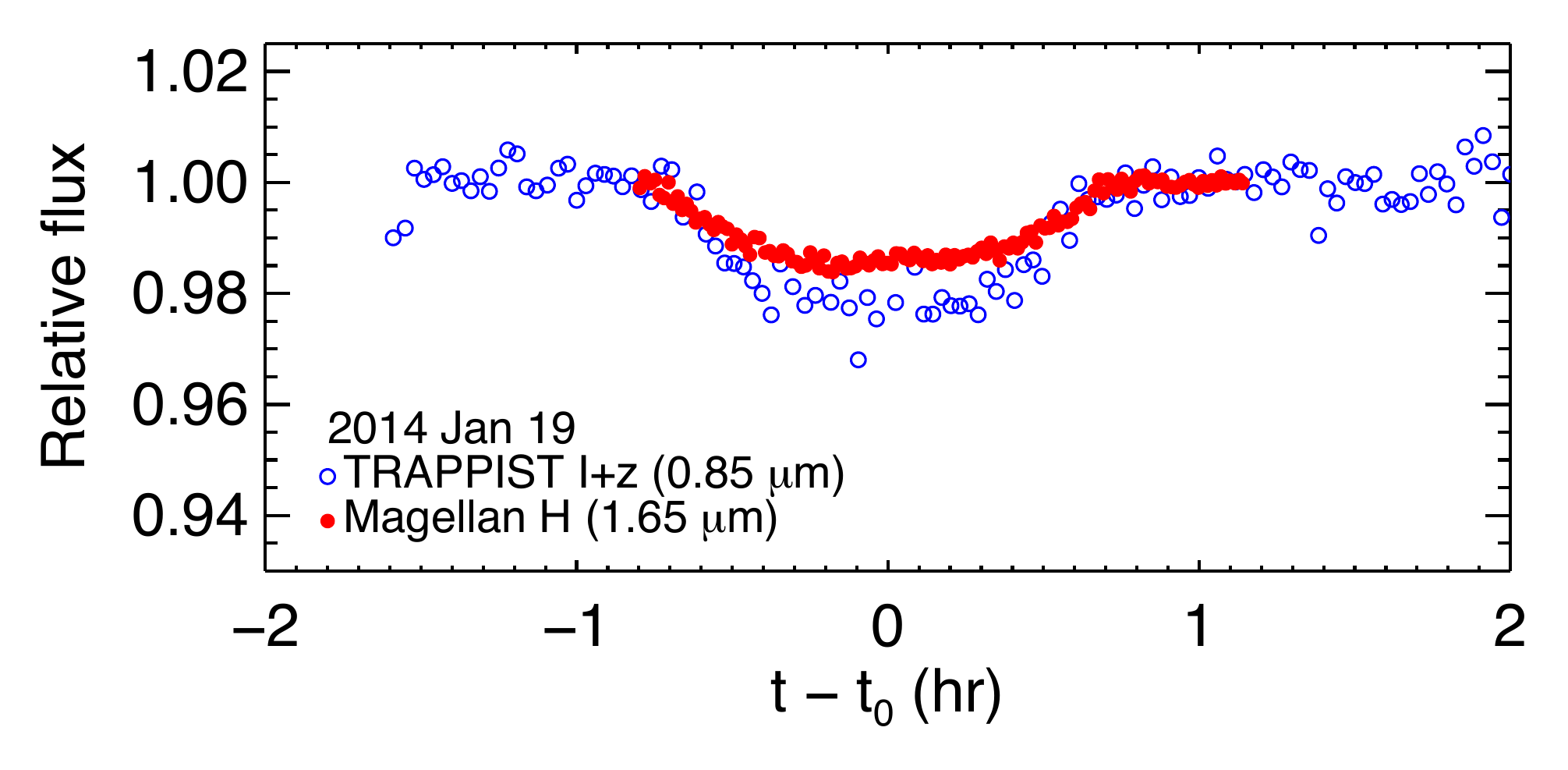}
\endminipage

\caption{Multiband observations of fading events.
  {\it Top row.}---Simultaneous observations in the $r'$ and $I+z$ bands
  (open blue and filled red circles, respectively).
  {\it Middle row.}---Interleaved observations in the $g'$ and $i'$ bands
  (open blue and filled red circles, respectively).
  {\it Bottom.}---Simultaneous observations in the $I+z$ and $H$ bands
  (open blue and filled red circles, respectively).
  In all but one case, the loss of light is greater in the bluer bandpass.}
\label{fig:multiband}
\end{figure*}

Another finding is that the loss of light is usually strongly
chromatic, as shown in Figure~\ref{fig:multiband}. For those two cases
in which we observed the same event in both the $i'$ and $g'$ bands,
we found the loss of light to be 30-50\% larger in the $g'$ band. For
the single case in which we observed in both $H$ and $I+z$, the loss
of light was also $\sim$40\% larger in the bluer band. We also
observed two events simultaneously in the Gunn $r'$ and $I+z$ bands. On the
first night the loss of light was 20-30\% larger in the bluer band. On
the second night, the loss of light in $r'$ was essentially the same
as in $I+z$.

\subsection{Ground-based observations of predicted occultations}
\label{sec:ground-occ}

The infrared emission from close-in giant planets is often strong
enough that it is possible to detect the decrement in total flux when
the planet is hidden by the star. The detection of such an occultation
would be strong evidence supporting the planet hypothesis.  We
searched for the predicted occultations in the Magellan and {\it
  Spitzer} time-series infrared photometry, as discussed below.

\subsubsection{Magellan observations}
\label{sec:magellan}

\begin{figure}[t]
\epsscale{1}
\includegraphics[width=0.5\textwidth]{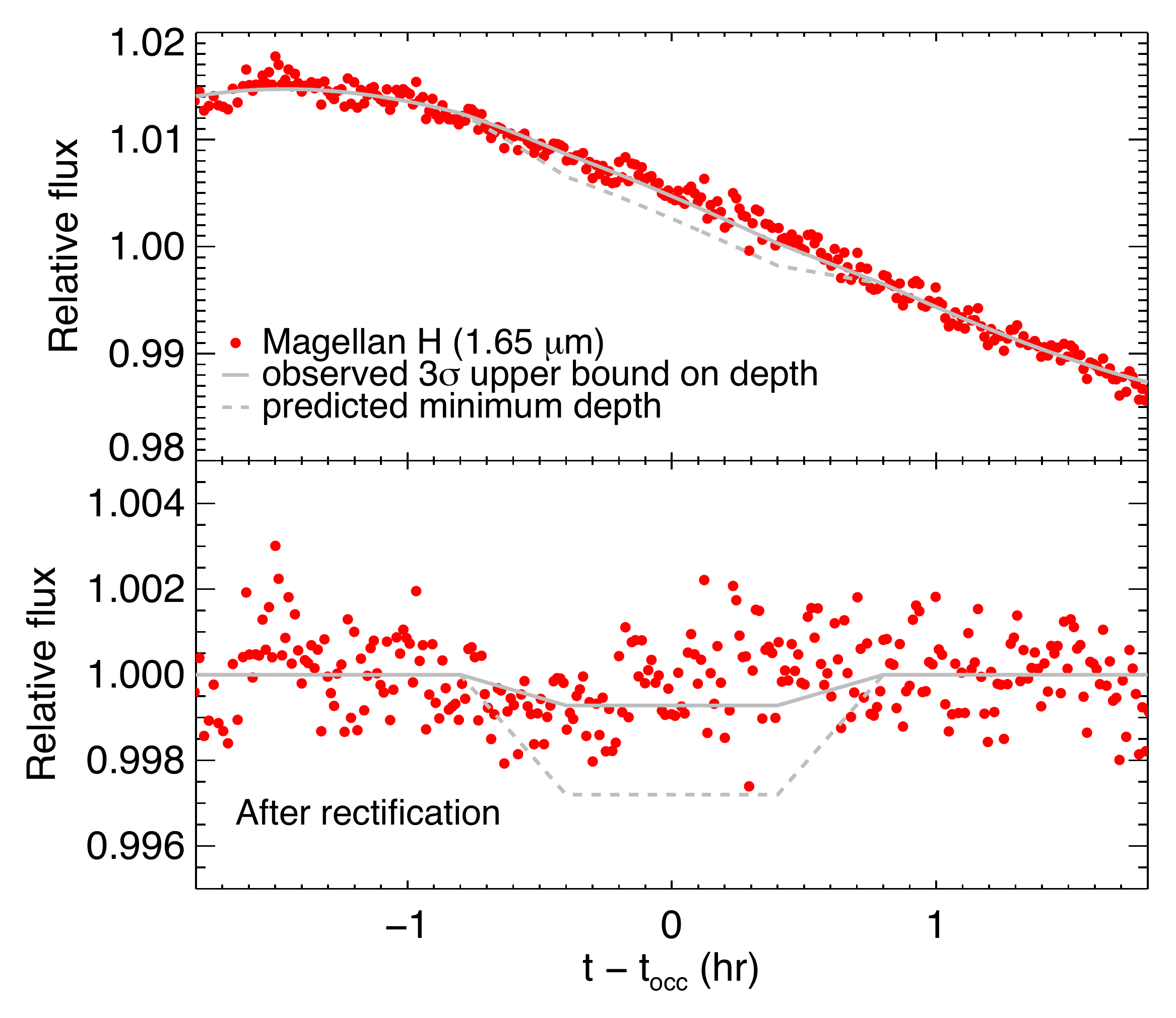}
\caption{{\it Top.}---Near-infrared photometry of PTFO\,8-8695
  spanning the predicted time of occultation. The data points are
  averaged in groups of 5 for clarity. The solid gray line shows the
  $3\sigma$ bound on the maximum depth of the occultation given by
  this light curve. The dashed grey line shows the minimum predicted
  occultation depth according to the model of \citet{barnes13}. {\it
    Bottom.}---Same, but after dividing through the light curve by the
  best-fit cubic function to the out-of-occultation region. }
\label{fig:magellan-occ} 
\end{figure}

The 6.5m Magellan~I (Baade) telescope and FourStar were also used to
monitor PTFO\,8-8695 for approximately 6~hours on 2014~Jan~21,
spanning the predicted time of a planetary occultation. The prediction
was based on the assumption that the planet's orbit is circular, i.e.,
that the occultations occur exactly halfway between transits. The
observations were conducted in the $H$ band and the data were
processed in the manner described in \S~\ref{sec:ground-tra}.
Figure~\ref{fig:magellan-occ} shows the resulting light curve. No
occultation signal is evident, although some gradual variability is
seen.

We determined an upper limit on the relative brightness of the
planetary dayside by fitting an occultation model to the data. The
model had the same total duration and ingress/egress durations as the
``transit'' light curve observed with Magellan two days earlier. As
with the transits, we fitted the gradual variations with a polynomial
function of time (in this case, a cubic function). We used a Markov
Chain Monte Carlo (MCMC) algorithm to calculate the posterior
probability distribution for the parameters describing the cubic
function as well as the loss of light during the occultation.  The
minimum $\chi^2$ value was 2082.5, with 1876 degrees of freedom,
indicating a statistically unacceptable fit; the cubic function is
evidently not a completely satisfactory description of the observed
flux variations. Rather than develop more elaborate models we simply
inflated the parameter uncertainties by the factor $\sqrt{\chi^2_{\rm
    min}/N_{\rm dof}}$.  The resulting occultation depth was
$\delta_{\rm occ} = 0.00024 \pm 0.00016$, corresponding to a 3$\sigma$
upper limit of $\delta_{\rm occ}< 0.00072$.

To decide if this upper limit rules out the planetary hypothesis, we
need to know the expected occultation signal.
Following the usual simplified model for transiting planets,
the fractional loss of light during an occultation is
\be
\delta_{\rm occ} = A_g \left( \frac{R_p}{a} \right)^2 +
\left(\frac{R_p}{R_\star}\right)^2 \frac{\int_{\lambda_1}^{\lambda_2}
  B_\lambda(T_p) d\lambda} {\int_{\lambda_1}^{\lambda_2}
  B_\lambda(T_\star) d\lambda}.
  \label{eq:deltaocc}
\ee
The first term is due to reflected starlight, in which $A_g$ is the
geometric albedo, and $R_p$ and $R_\star$ are the planetary and
stellar radii. The second term is due to the planet's thermal
emission, in which the observing
bandpass extends from $\lambda_1$ to $\lambda_2$ (1.48--1.76~$\mu$m,
for $H$ band), $B_\lambda(T)$ is the Planck function, $T_\star$ is the
star's effective temperature, and $T_p$ is the planet's dayside
effective temperature. The latter is calculated from the condition of
radiative equilibrium, giving
\begin{equation}
T_p = T_\star \left(\frac{R_\star}{a}\right)^{1/2} \left(
  \frac{1-A}{f} \right)^{1/4},
\end{equation}
where $A$ is the Bond albedo and $f$ is a dimensionless number
depending on the manner of radiation. If the entire surface radiates
isotropically as a blackbody, then $f=4$. If instead the dayside
radiates uniformly and the nightside radiation can be neglected, then
$f=2$. Furthermore, if the angular dependence of the planet's
radiation is assumed to follow Lambert's law, then $A_g = 2A/3$.

\begin{figure*}[ht]
  \plottwo{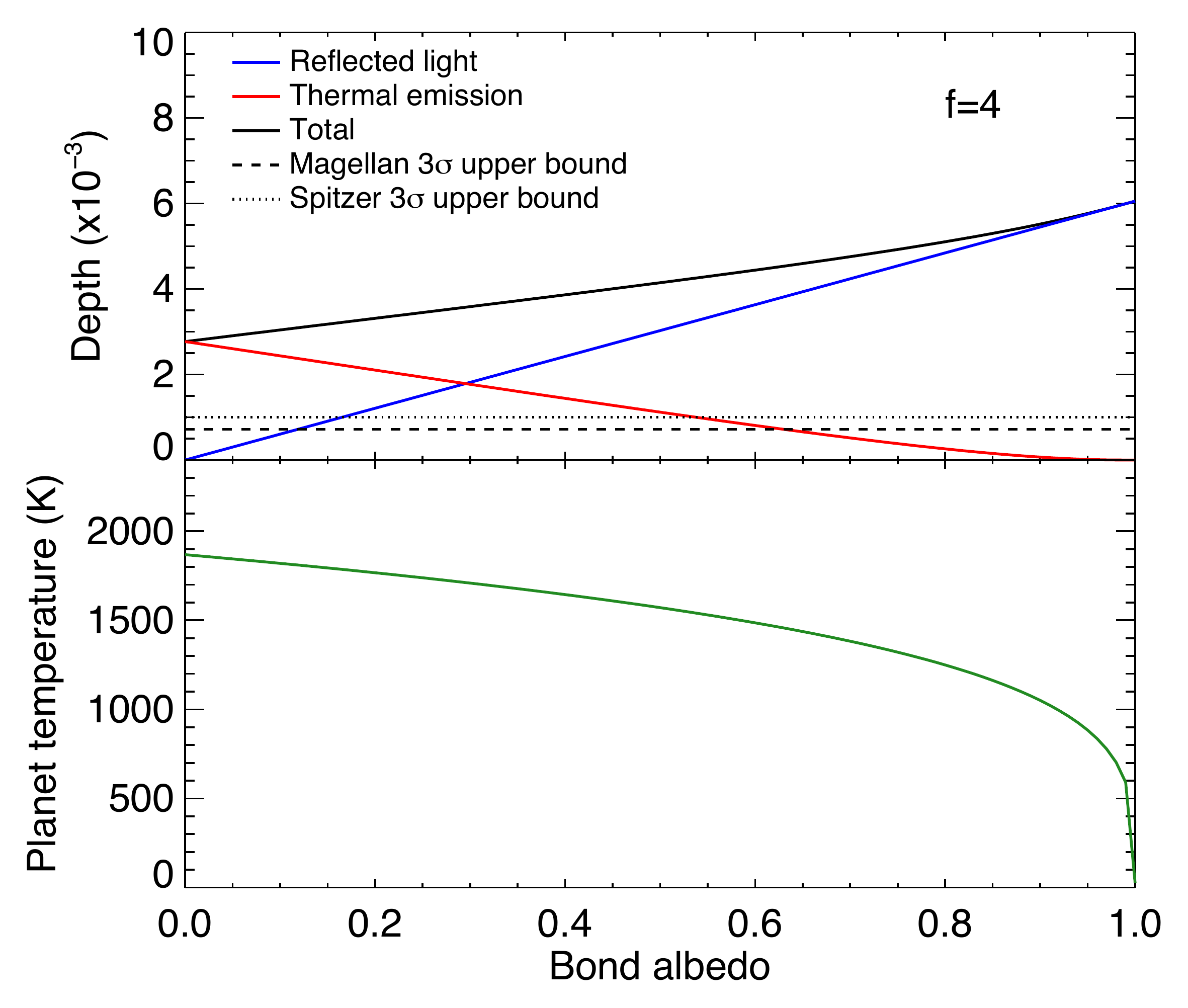}{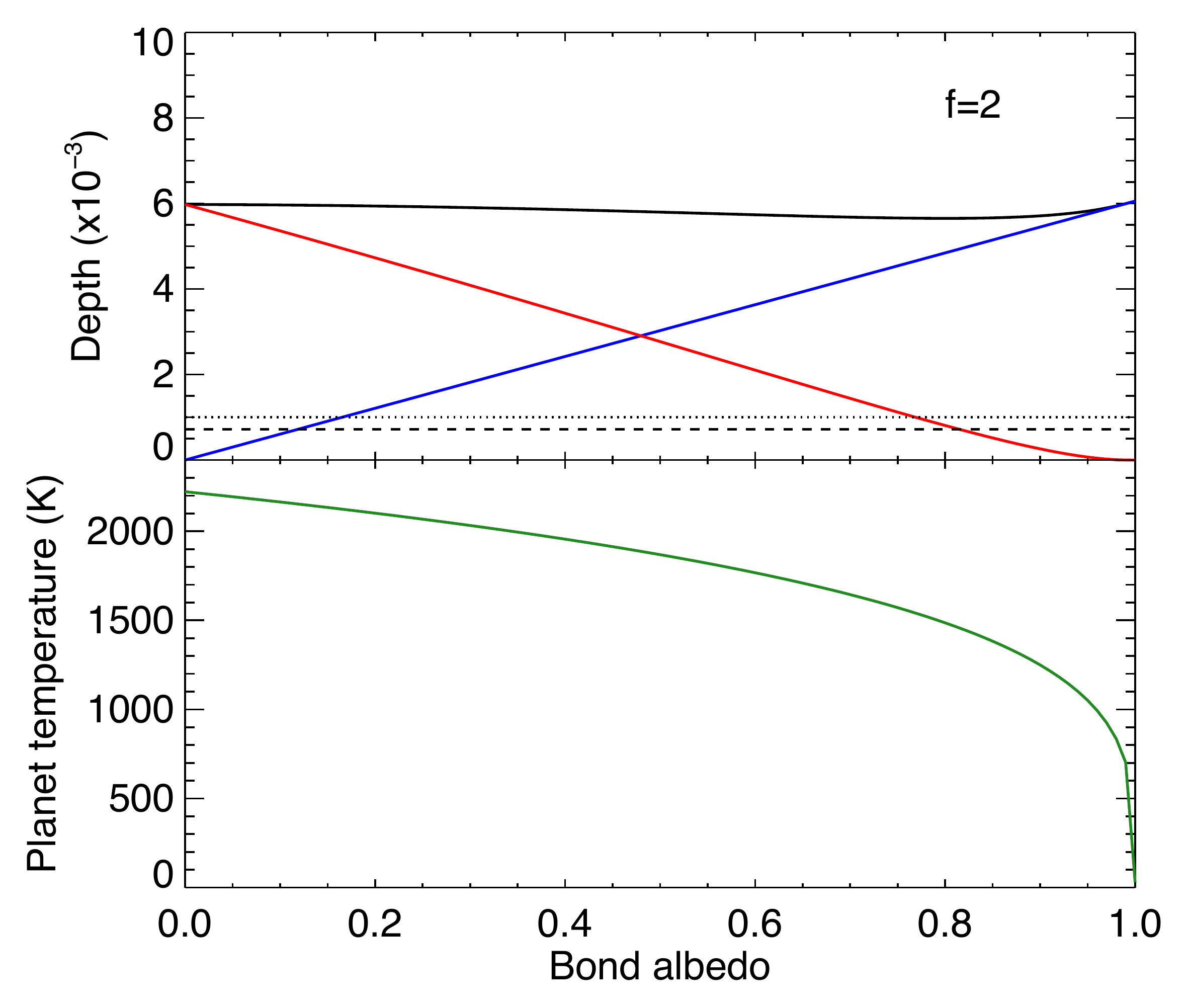}
  \caption{{\it Left.}---The top panel shows the expected occultation
    depth $\delta_{\rm occ}$ (solid black line) plotted as a function
    of the Bond albedo for $f=4$.  The red and blue lines represent
    the contributions to the total occultation depth from thermal
    emission of the planet and reflected light, respectively. For
    comparison, the $3\sigma$ upper bounds on $\delta_{\rm occ}$
    obtained from Magellan and {\it Spitzer} are shown as dashed and
    dotted lines. The bottom panel shows the corresponding dayside
    effective temperature of the planet. {\it Right.}---Same, but for
    $f=2$.}
\label{fig:albedo} 
\end{figure*}

In this case it is difficult to establish the key parameters
$(R_p/R_\star)^2$ and $R_\star/a$, because of the changing morphology
of the fading events with time and wavelength. \citet{barnes13} found
$(R_p/R_\star)^2 \approx 0.027$ and $R_\star/a \approx 0.58$ using a
model incorporating the effects of gravity darkening and orbital
precession. Using those parameters, we calculate the expected value of
$\delta_{\rm occ}$ and plot it as a function of the Bond albedo in
Figure~\ref{fig:albedo}, for both $f=4$ and $f=2$.  The expected
$\delta_{\rm occ}$ ranges from a minimum value of $0.0028$ for $f=4$
and $A=0$, to a maximum value of $0.0061$ for $f=2$ and $A=1$. Such
large occultation depths are ruled out by our Magellan observations.

\subsubsection{Spitzer observations}
\label{sec:spitzer}

The {\it Spitzer} Space Telescope monitored PTFO\,8-8695 on
2012~Apr~28 for about 12~hours, slightly longer than a full
photometric period.  The data consist of 1369 full-array images from
the IRAC detector operating at 4.5~$\mu$m, with an integration time of
26.8~seconds.  The data were calibrated by the {\it Spitzer} pipeline
version S19.1.0. These observations were carried out in non-cryogenic
mode under program no.\ 80257 (PI: Stauffer), and are publicly
available on the {\it Spitzer} Heritage Archive
database.\footnote{While this manuscript was in preparation,
  \citet{ciardi15} reported an independent analysis of these same
  data.}

We converted the data from the {\it Spitzer} units of specific
intensity (MJy~sr$^{-1}$) into photon counts, and then performed IRAF
aperture photometry on each subarray image. Best results were obtained
with an aperture radius of 2.5 pixels and a background annulus
extending from 11 to 15.5 pixels from the center of the point-spread
function (PSF). The center of the PSF was measured by fitting a
two-dimensional Gaussian function to each image. At this stage, 30
discrepant fluxes were discarded by applying a 5$\sigma$ median
clipping algorithm.

Next we needed to remove the apparent flux variations associated with
motion of the image on the detector, the main source of systematic
effects in time-series photometry with the IRAC InSb arrays
\citep{knutson+08}. This effect is caused by the combination of (i)
the coarse sampling of the PSF, (ii) the significant inhomogeneity of
the pixels, and (iii) fluctuations in the telescope pointing. To
mitigate this effect we chose the Bi-Linearly-Interpolated Sub-pixel
Sensitivity (BLISS) mapping method presented by \citet{stevenson+12}.
In this method, the flux data themselves are used to constrain a model
for the subpixel sensitivity variations. In our implementation of the
method, the detector area probed by the PSF center was divided into a
$13\times 13$ grid. With this degree of sampling, the PSF center
visited each grid point at least 10 times throughout the course of the
observations.

\begin{figure}[h]
\epsscale{1}
\includegraphics[width=0.5\textwidth]{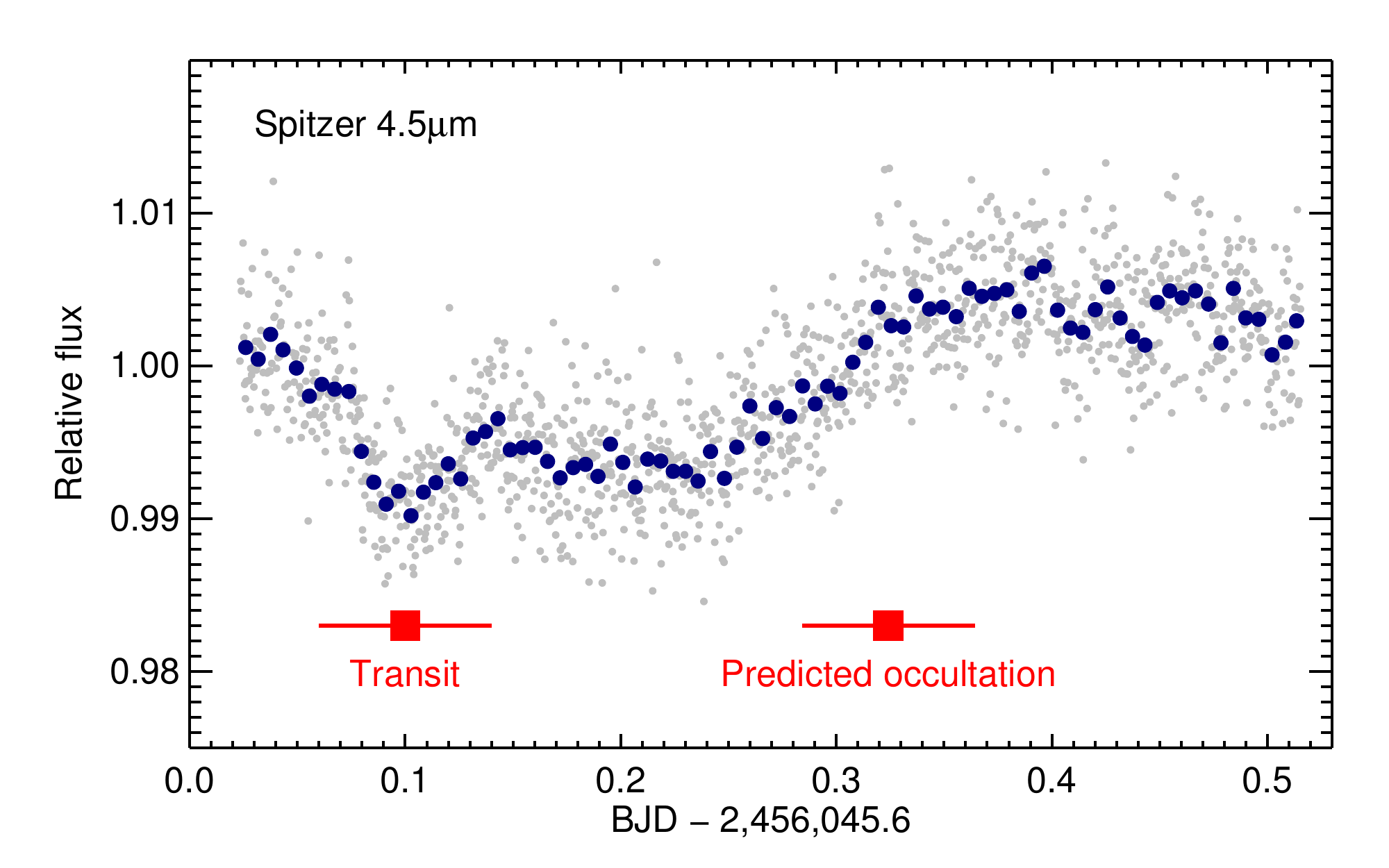}
\caption{{\it Spitzer} time-series 4.5~$\mu$m photometry of
  PTFO\,8-8695 from 2012~Apr~28. The small gray points represent
  individual measurements; the larger dark blue points are time
  averages. Red bars show the times of the fading event, and the
  predicted time of the planetary occultation.}
\label{fig:spitzer_all} 
\vspace{1mm}
\end{figure}

\begin{figure*}[ht]
\epsscale{1}
\includegraphics[width=1.0\textwidth]{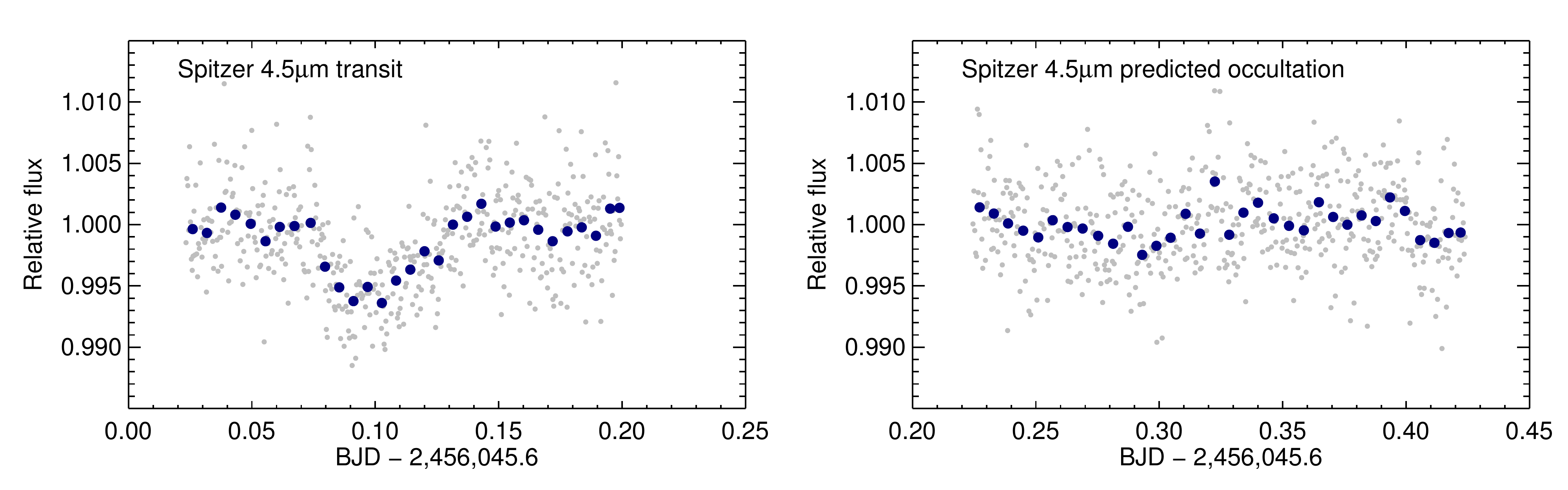}
\caption{Close-up of the ``transit'' and ``occultation'' data, after
  rectification to unit flux outside of the events.}
\label{fig:spitzer_tra_occ} 
\vspace{1mm}
\end{figure*}

Figure~\ref{fig:spitzer_all} shows the {\it Spitzer} light curve after
BLISS correction. The variability can be described as the combination
of quasi-sinusoidal variation with a period of $\approx$0.5~day, and
the transit-like dip in brightness at the expected time, with an
amplitude of 0.5\% and a duration of approximately 1.4~hours. No
occultation is seen at the expected time (0.224 days after the
transit). Figure~\ref{fig:spitzer_tra_occ} gives a better view of the
transit-like event, and the data surrounding the predicted time of
occultation. In this figure, only the data within 0.1~days of each
event are shown, and the data have been rectified by fitting a
quadratic function of time to data outside of the event and then
dividing by the best-fitting function.

Just as with the Magellan light curve, we determined an upper limit on
the occultation loss of light by fitting a parameterized model to the
data. The model included a quadratic function of time to describe the
out-of-occultation variations. The occultation model was required to
have the same durations between first, second, third and fourth
contacts as observed earlier with Magellan. The loss of light
$\delta_{\rm occ}$ was a free parameter.  We used an MCMC algorithm to
calculate the posterior probability distribution of $\delta_{\rm occ}$
and the parameters of the quadratic function. The minimum $\chi^2$
value was 565.9, with 535 degrees of freedom. The result for the
occultation depth was $\delta_{\rm occ} = -0.0008 \pm 0.0006$, i.e.,
the best-fitting value corresponds to a brightness increase rather
than a loss of light.  This corresponds to a 3$\sigma$ upper limit of
$\delta_{\rm occ}< 0.0010$. Again, as illustrated in
Figure~\ref{fig:albedo}, the upper bound on $\delta_{\rm occ}$ given
by {\it Spitzer} is smaller than the occultation depth implied by the
parameters of the \citet{barnes13} model.

Careful inspection of Figure~\ref{fig:spitzer_all} shows a candidate
flux dip of centered around a time coordinate of 0.425, with an
amplitude of $\approx$0.3\%. One might be tempted to attribute this
dip to the occultation of a planet on an eccentric orbit, for which
the occultation need not be halfway between transits. However, the
statistical significance of this dip is dubious, and the required
value of the eccentricity would be $e> 0.35$, using Eq.~33 of
\citet{winn_review}. Such a high eccentricity would be unprecedented
and unexpected for a short-period planet. In general, giant planets
with periods shorter than 3~days have nearly circular orbits, a fact
that is attributed to the gradual action of tidal dissipation. Given
the youth of the star, it is possible that there has not yet been
sufficient time for orbital circularization; however, a higher
eccentricity and a potentially smaller pericenter distance would also
put the planet in even more danger of violating the Roche limit.

Our non-detections of occultation signals at both 1.7~$\mu$m and
4.5~$\mu$m bands rule out the existence of a planet that radiates like
a blackbody in these two bands.  We have not pursued more realistic
models for the planetary emission spectrum, given that the atmospheric
composition is unconstrained, but seems unlikely that atmospheric
absorption features would suppress the planetary flux in both bands to
such a degree that it would be undetectable in our data.


\subsection{Departure from periodicity}
\label{sec:omc}

\begin{figure}[h]
\epsscale{1}
\includegraphics[width=0.5\textwidth]{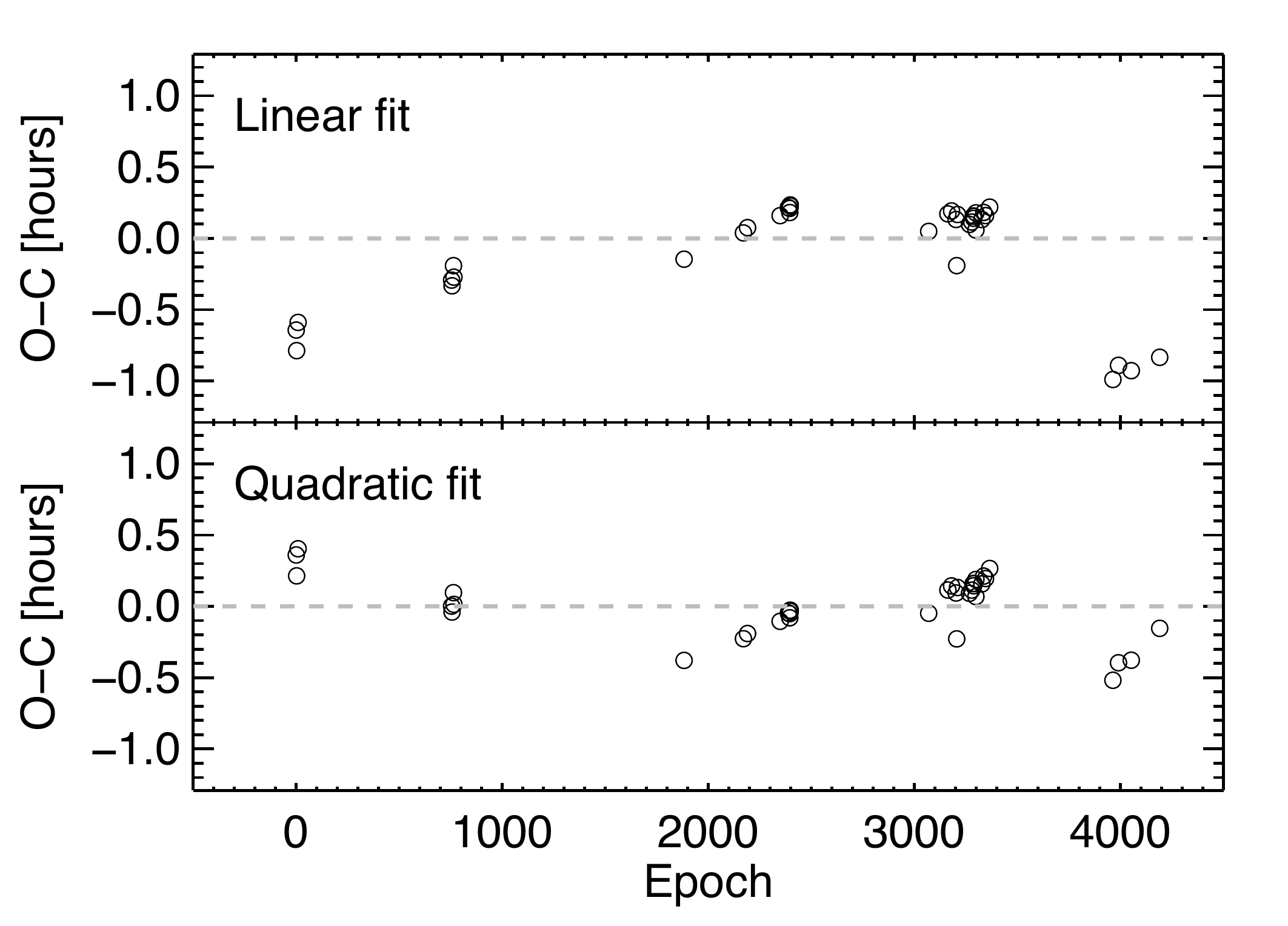}
\caption{{\it Top panel}: Timing residuals after subtracting the
  best-fitting linear function of epoch(constant period). {\it Bottom
    panel}: Timing residuals after subtracting the best-fitting
  quadratic function of epoch (steady decrease in period).}
\label{fig:omc} 
\end{figure}

Perhaps the most important finding of all the photometric observations
is that the fading events are not strictly periodic. The top panel of
Figure~\ref{fig:omc} shows the residuals after subtracting the
best-fitting linear function of epoch from the measured times of
minimum light,
\begin{equation}
t_n = t_0 + P n,
\label{eq:linear}
\end{equation}
for which $\chi^2_{\rm min} = 15573$ with 35 degrees of freedom. This
poor fit is the result of the large scatter ($\approx$15~min) of the
residuals within each season, and the even larger deviation
($\approx$1.3~hours) of the most recent season's residuals relative to
the earlier data.  The pattern of residuals suggests that the period
was nearly constant up until the 2014/5 observing season, when the
fading events began occurring earlier than expected.  This apparent
change in period or phase can be readily checked by gathering
additional data over the next few seasons.  The best-fitting
parameters of the linear ephemeris are
\bea
t_0 & = & 2455201.832\pm 0.007~{\rm days}, \\
P & = & 0.448391 \pm 0.000003~{\rm days}.
\eea
In these expressions the uncertainties have been scaled up 
by a factor of $\sqrt{\chi^2_{\rm min}/N_{\rm dof}}$ to account for
the statistically poor fit.

We also tried fitting a quadratic function of epoch,
\begin{equation}
t_n = t_0 + P_0 n + \frac{1}{2} \frac{\dd P}{\dd n} n^2,
\label{eq:quadratic}
\end{equation}
for which $\chi^2_{\rm min} = 4980$ with 34 degrees of freedom. 
After enlarging the parameter uncertainties as described above,
the best-fitting parameters are
\bea
t_0 & = &  2455201.790 \pm 0.006~{\rm days},\\
P_0 & = & 0.448438 \pm 0.000006~{\rm days},\\
\dd P / \dd n & = & (-2.09 \pm 0.25) \times 10^{-8}~{\rm days~epoch}^{-1}.
\eea
The bottom panel of Figure~\ref{fig:omc} shows the residuals between the
observed and calculated times. The implied fractional change in period
per epoch, calculated as $\frac{1}{P} \frac{\dd P}{\dd n}$, is equal
to $-4.66 \times 10^{-8}$. If this period change were to continue
steadily, the period would shrink to zero after $P_0/\dot{P} \sim
10^4$~years.

\section{Spectroscopic observations}
\label{sec:spec}

\begin{figure}[h]
\epsscale{1}
\includegraphics[width=0.5\textwidth]{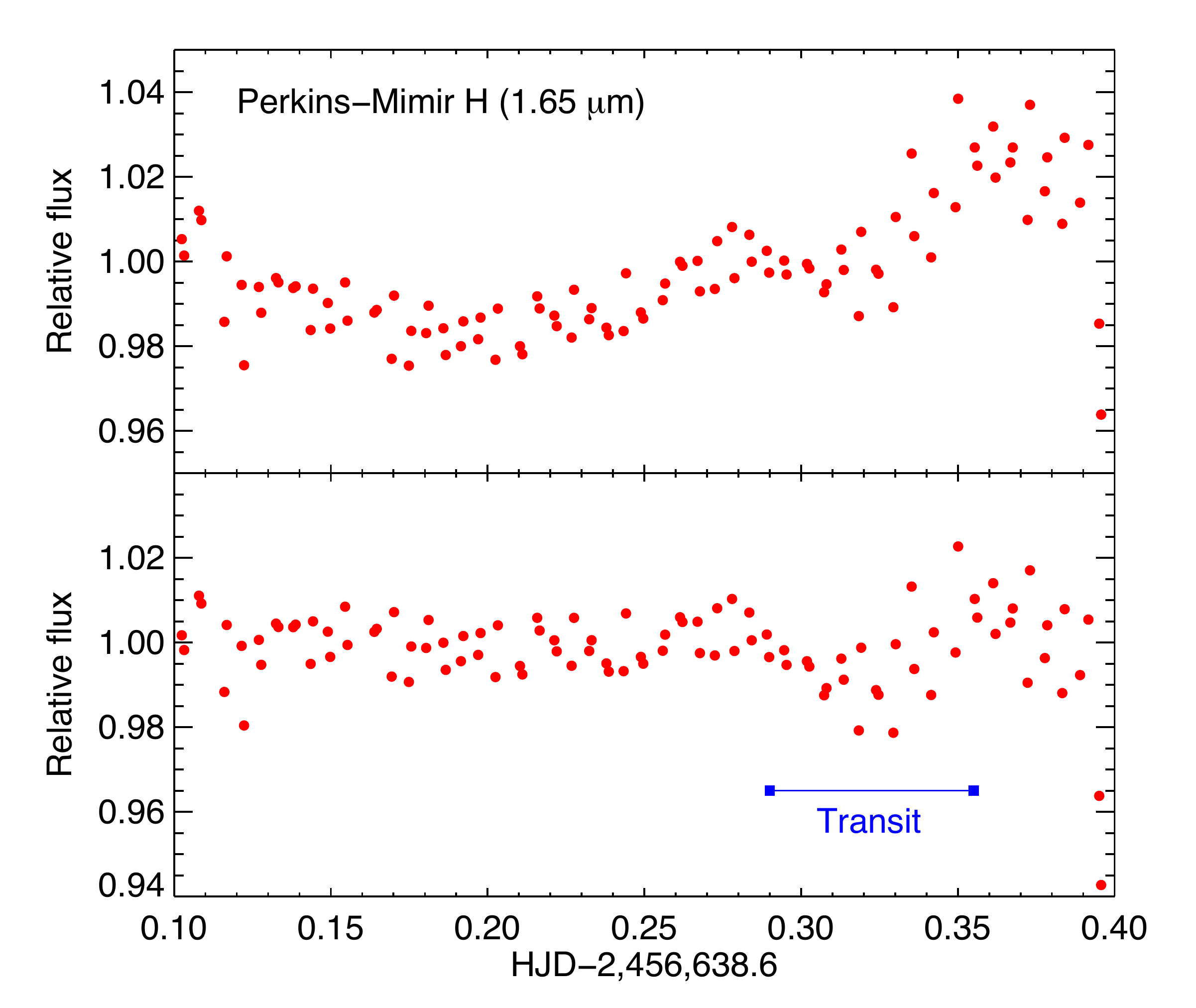}
\caption{{\it Top panel}: Perkins-Mimir 1.65~$\mu$m photometry of
  PTFO\,8-8695 from 2013~Dec~12. Data points are averaged in groups of
  5 for clarity. {\it Bottom panel}: Same, but after rectification to
  unit flux outside of the transit. A transit signal can be seen in
  the region marked by the blue bar.}
\label{fig:perkins} 
\end{figure}

We monitored the optical spectrum of PTFO\,8-8695 on the night of
2013~Dec~12 UT, employing the Keck\,I telescope and its High Resolution
Spectrograph \citep[HIRES;][]{vogt1994}.  A total of 22 observations
with integration times of 14~min were obtained. This sequence covered
2~hr spanning the event, and 1~hr after the event. HIRES was used in
its standard setting, but without an iodine cell in the light path.

To confirm that a fading event was indeed taking place during the
spectroscopic observations, we attempted to gather simultaneous
photometry with several small telescopes, but in only one case was the
weather at least somewhat cooperative. We obtained data in the $H$
band with Mimir, a cryogenic, facility-class near-infrared instrument
on the 1.83~m Perkins telescope outside Flagstaff, Arizona
\citep{clemens07}. Figure~\ref{fig:perkins} shows the light curve. A
transit-like dip of $\sim$2\% was seen at the expected time,
confirming that a fading event did occur, although the data are too
noisy to extract much further information.  We also note that fading
events were seen by TRAPPIST on 2013~Dec~11 and 14, bracketing our
Keck observation.

\subsection{Search for the Rossiter-McLaughlin effect}

The primary purpose of the spectroscopic observations was to seek
evidence for the Rossiter-McLaughlin (RM) effect, the spectroscopic
anomaly that is seen during a planetary transit due to stellar
rotation. During a transit, a planet blocks different portions of the
rotating stellar photosphere, leaving a particular imprint on the
rotationally-broadened stellar absorption lines. The exact shape and
time development of the spectral deformations depend on the transit
parameters, and in particular on the angle between the stellar
rotation axis and the orbital axis as projected on the sky plane. For
PTFO\,8-8695, the planet hypothesis requires a large misalignment
between these angles. We attempted to detect the RM effect and test
that prediction.

After the initial data reduction we corrected for the blaze function
by using calibration lamp exposures to estimate the blaze function for
each order, and then fitting a linear function of wavelength to remove
the residual variations and normalize the continuum to
unity. Following the barycentric correction, all of the out-of-transit
exposures were co-added to create a single spectrum with a higher
signal-to-noise ratio. This was used for a final differential
normalization, wherein the summed spectrum was subtracted from each
observed spectrum, and a 4th-order polynomial was fitted to the
residuals in each order. These polynomials were subsequently
subtracted from the corresponding spectrum. This was done to minimize
the potential influence of any time variations in the blaze function
throughout the night. We verified that the details of this
normalization process did not have a significant influence on the
following analysis.

For each spectrum, we calculated the cross-correlation function
(CCF) with reference to a synthetic spectrum. The synthetic spectrum
was obtained from the PHOENIX database \citep{husser2013}, for a star
with $T_{\rm eff}=3500$~K, $\log g=3.5$ and solar metallicity. We
selected the appropriate wavelength ranges for creating the CCF via a
visual inspection. We needed to locate areas for which the
normalization seemed reliable, and where there were at least a few
well-defined absorption lines. As had been reported by VE+12, there
are only a few regions between 5000--7000 \AA \ suitable for this work.

\begin{figure}[t]
\epsscale{1}
\includegraphics[width=0.5\textwidth]{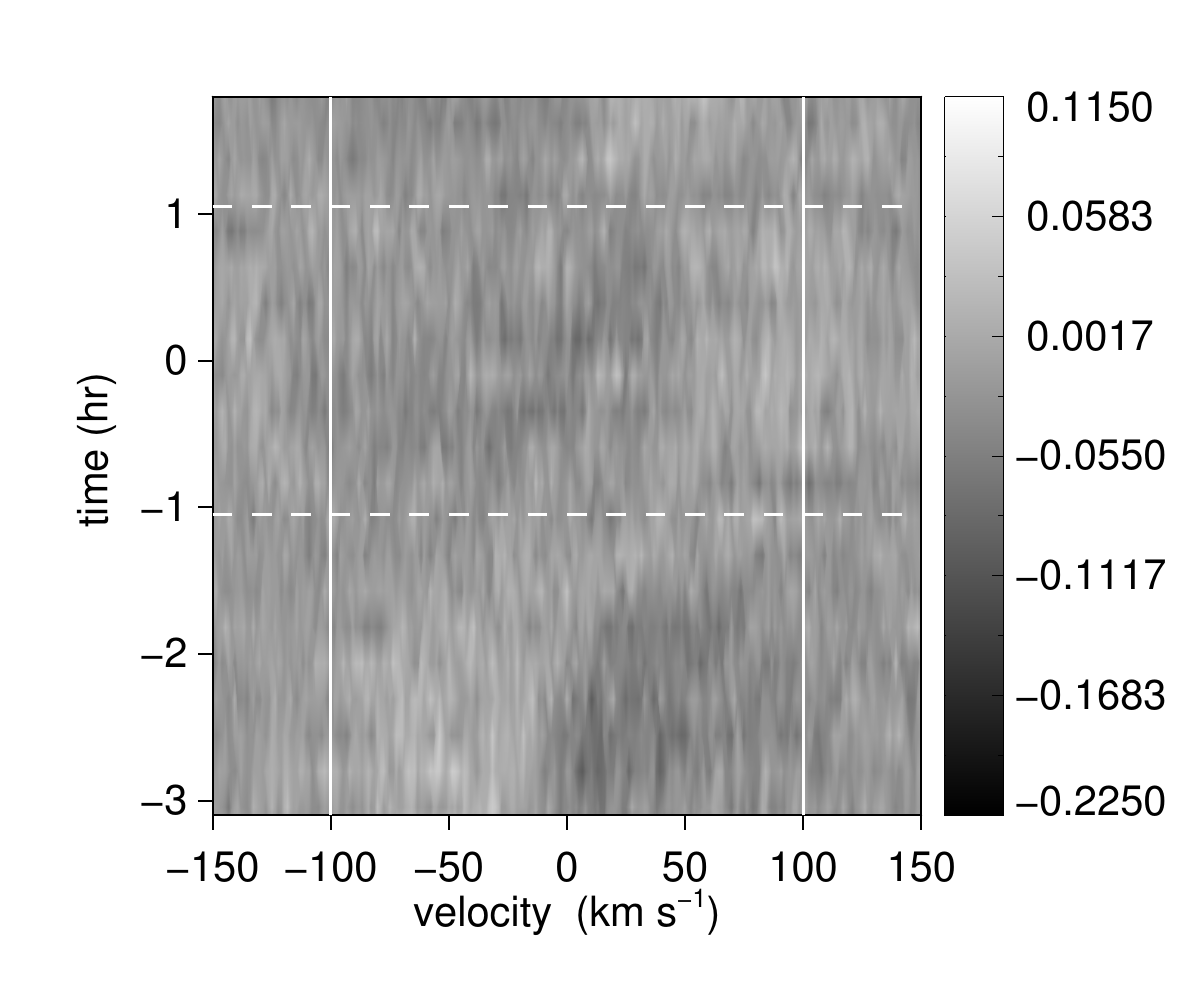}
\caption{Time variations of the CCF during the night of 2013
  Dec.~11/12. Dark areas indicate a deficit relative to the mean CCF
  on that night. The CCFs have been
  normalized to have a peak value of unity. The dashed lines
  indicate the predicted times of first and last contact.  Vertical
  lines indicate our estimate for $v \sin i_\star$. One can see
  variations of the CCFs before, during, and after the transit. No
  clear sign of a planetary transit is visible.}
\label{fig:RM} 
\end{figure}

We calculated the mean CCF based on all of the out-of-transit
observations.  Then we subtracted the mean CCF from each individual
CCF. When ordered in time, the resulting ``differential CCFs'' should
display the shadow of the transiting object in velocity space.  The
deformation due to a transiting object would be seen as a dark line.
The slope of this line in the velocity-time plane would depend on
the projected obliquity. For example, in the case of good spin-orbit
alignment, there would be a deficit of blue light (negative radial
velocities) in the first half of the transit, followed by a deficit of
redshifted light during the second half. As can be seen in
Figure~\ref{fig:RM}, no such signal --- neither aligned nor
misaligned --- can be discerned.

Could we have detected the spectroscopic transit of the hypothetical
planet, given the quality of our data? To answer this question, we
simulated the RM signal of a transiting planet with $(R_p/R_\star)^2 =
0.026$, the approximate transit depth measured by TRAPPIST during the
events of 2013~Dec~11 and 14, closely bracketing the event observed
with Keck. We assumed $v \sin i_\star=100$~km\,s$^{-1}$, a value
consistent with the line broadening seen in our Keck spectra, and
adopted a macroturbulent velocity of $15$~km\,s$^{-1}$. Then we
injected RM signals into the data, for various choices of $\lambda$,
the sky-projected stellar obliquity.

\begin{figure*}[t]
\epsscale{1}
\includegraphics[width=0.32\textwidth]{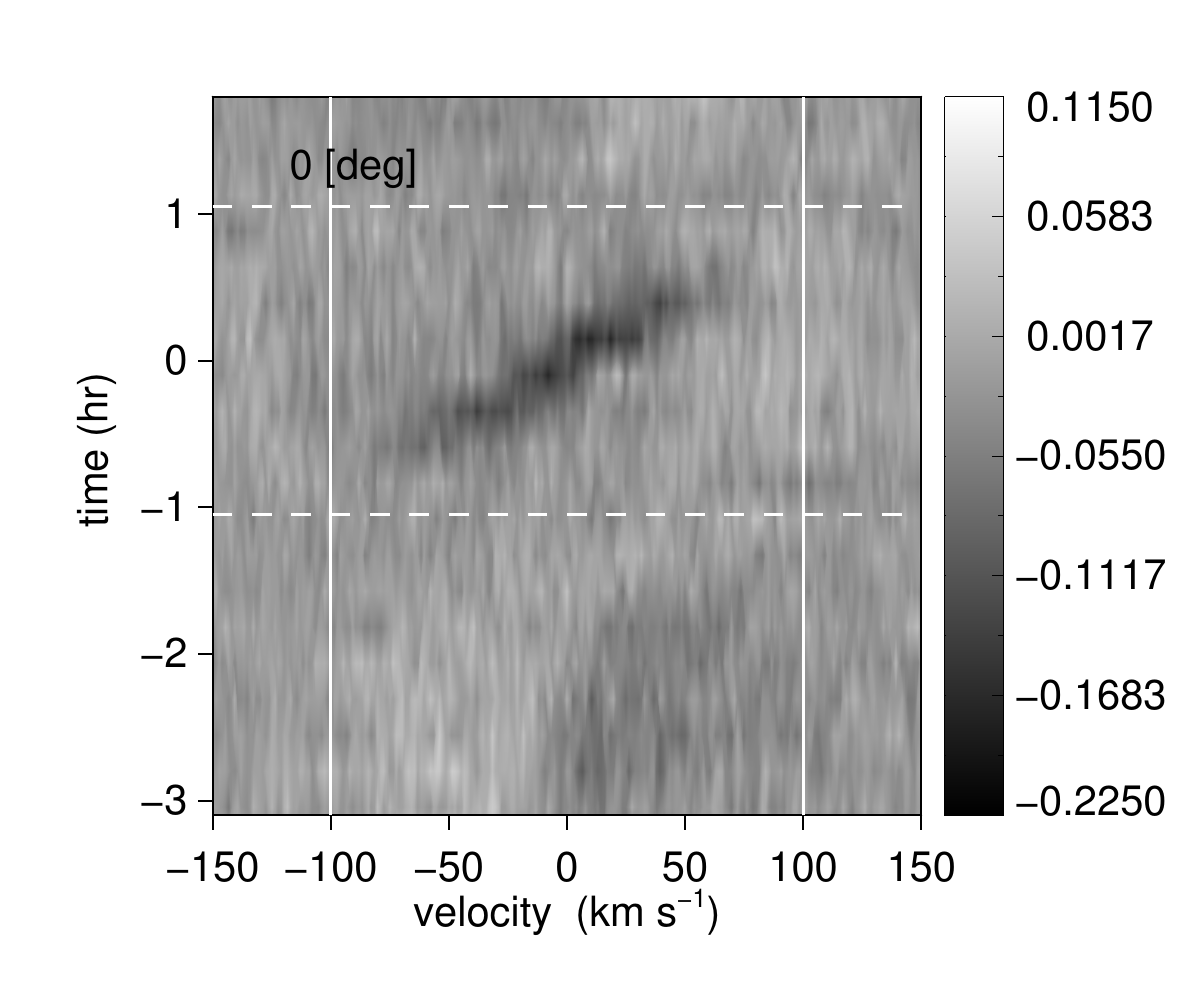}
\includegraphics[width=0.32\textwidth]{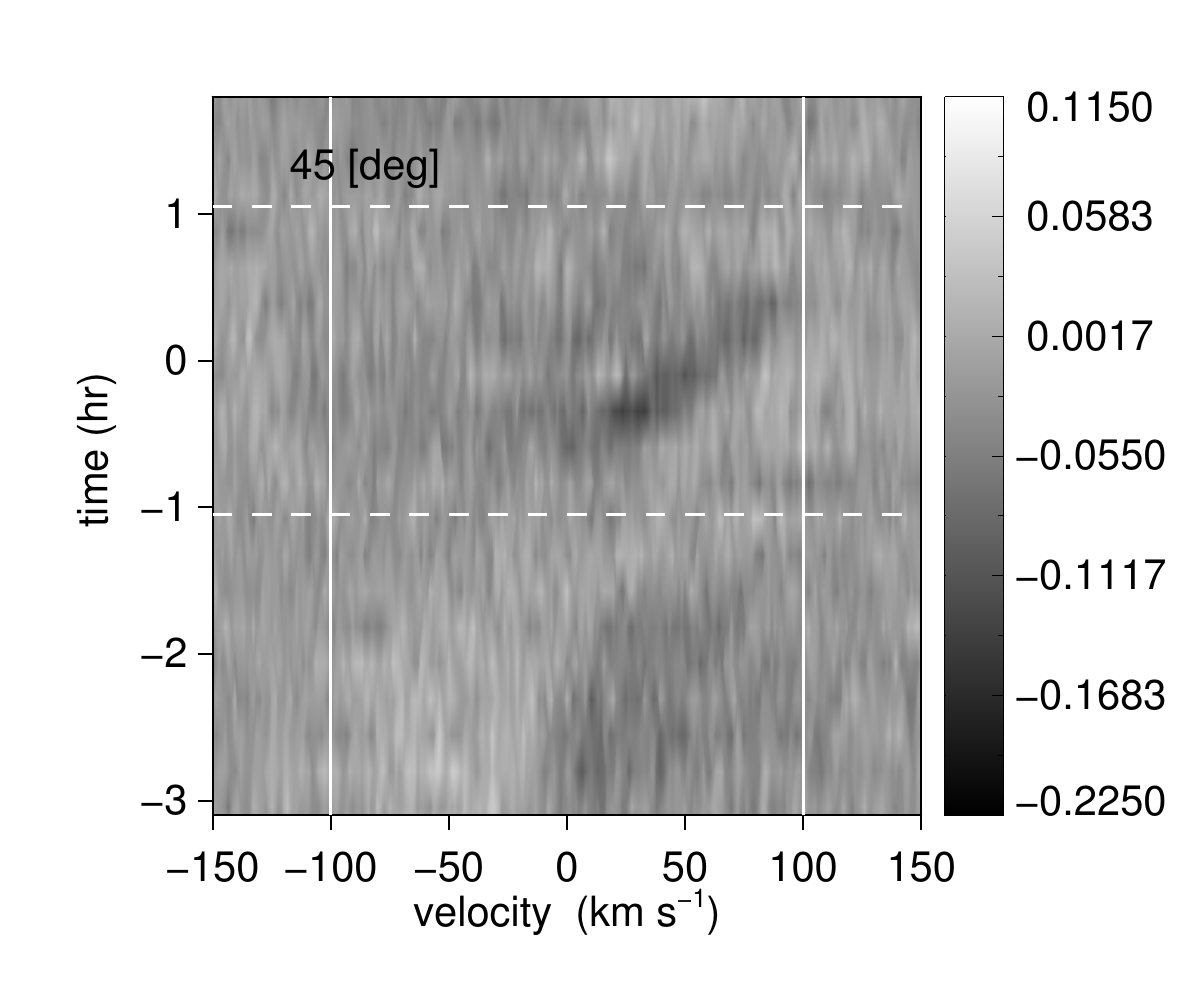}
\includegraphics[width=0.32\textwidth]{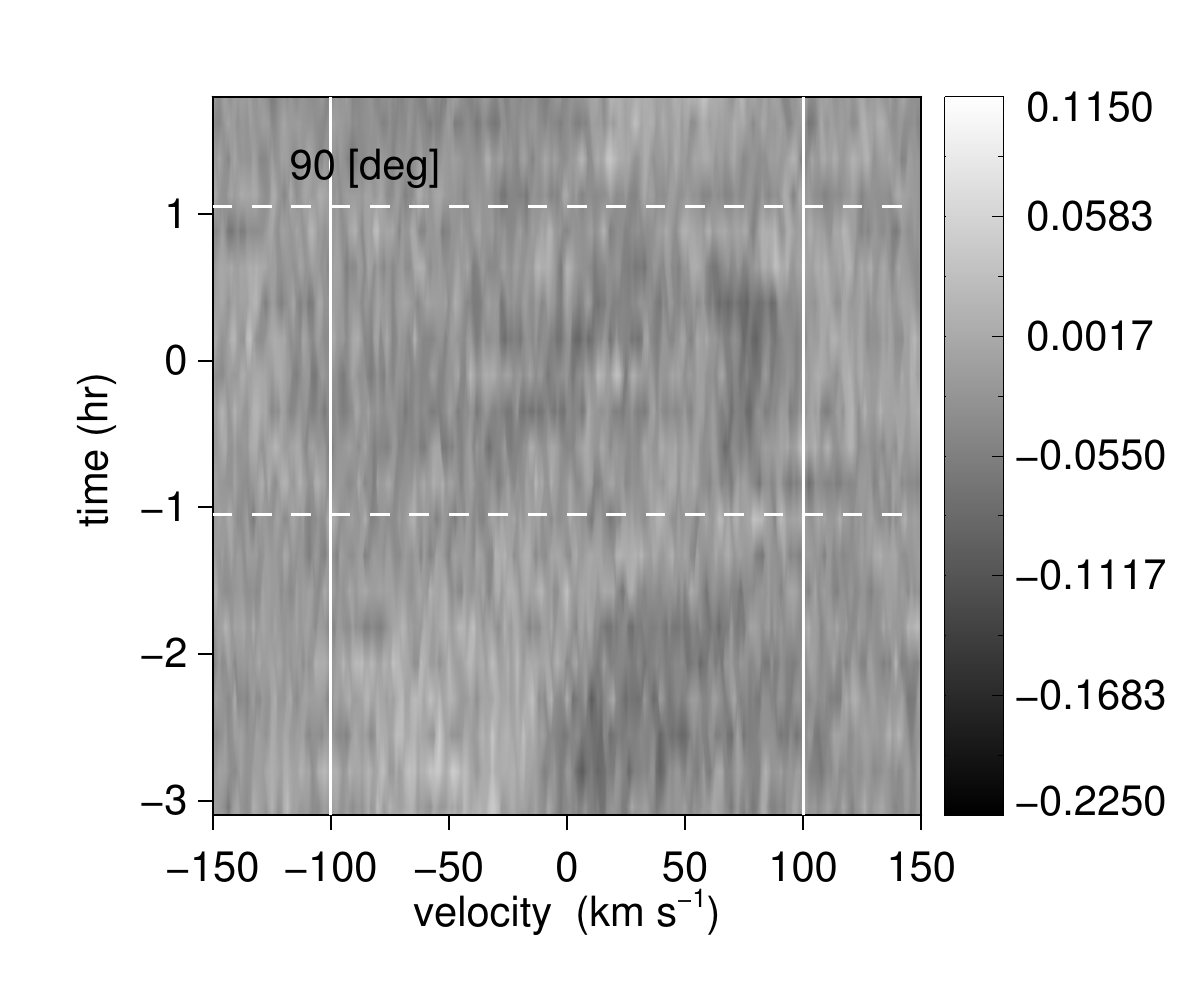}
\caption{Same as Figure~\ref{fig:RM}, but in this case an artificial
  signal of a transiting planet has been injected.
  The left panel shows the case $\lambda=0^\circ$ (spin-orbit alignment),
  the middle panel shows $\lambda=45^\circ$ and the right panel shows
  $\lambda=90^\circ$.}
\label{fig:RM_injected} 
\end{figure*}

Some representative examples of the simulated RM effect are shown in
Figure~\ref{fig:RM_injected}. The left panel shows the simulated RM
signal of a well-aligned planet ($\lambda=0^\circ$). Such a signal
would easily have been detectable with the data at hand. The same is
true for $\lambda=45^\circ$, shown in the middle panel. However, for
$\lambda = 90^\circ$, the signal would have been more difficult to
detect. This is because in this case the RM signal is nearly
stationary in velocity, as shown in the right panel, making it more
difficult to separate from the noise in the velocity--time plane.
Based on visual inspection of figures similar to
Figure~\ref{fig:RM_injected} we conclude that we can rule out any
trajectory except for those within about 15$^\circ$ of
perpendicularity ($\lambda=90^\circ$ or 270$^\circ$). Given the
non-Gaussian and correlated nature of the noise, it is difficult to
make a firmer statistical statement.

\subsection{Projected rotation rate}

A secondary goal was to seek changes in the sky-projected rotation
rate ($v\sin i_\star$) that would be expected if the star's rotation
axis is precessing around the total angular momentum of the system.
If PTFO\,8-8695 does consist of a star and a planet whose rotation
axes and orbits precess around the common angular momentum, then $v
\sin i_\star$ should change with time. We searched for such a change
between the two epochs for which Keck/HIRES data have been
obtained. Five spectra were obtained in April 2011 and presented by
VE+12. Another epoch is represented by our December 2013 data.

\begin{figure}[t]
\epsscale{1}
\includegraphics[width=0.5\textwidth]{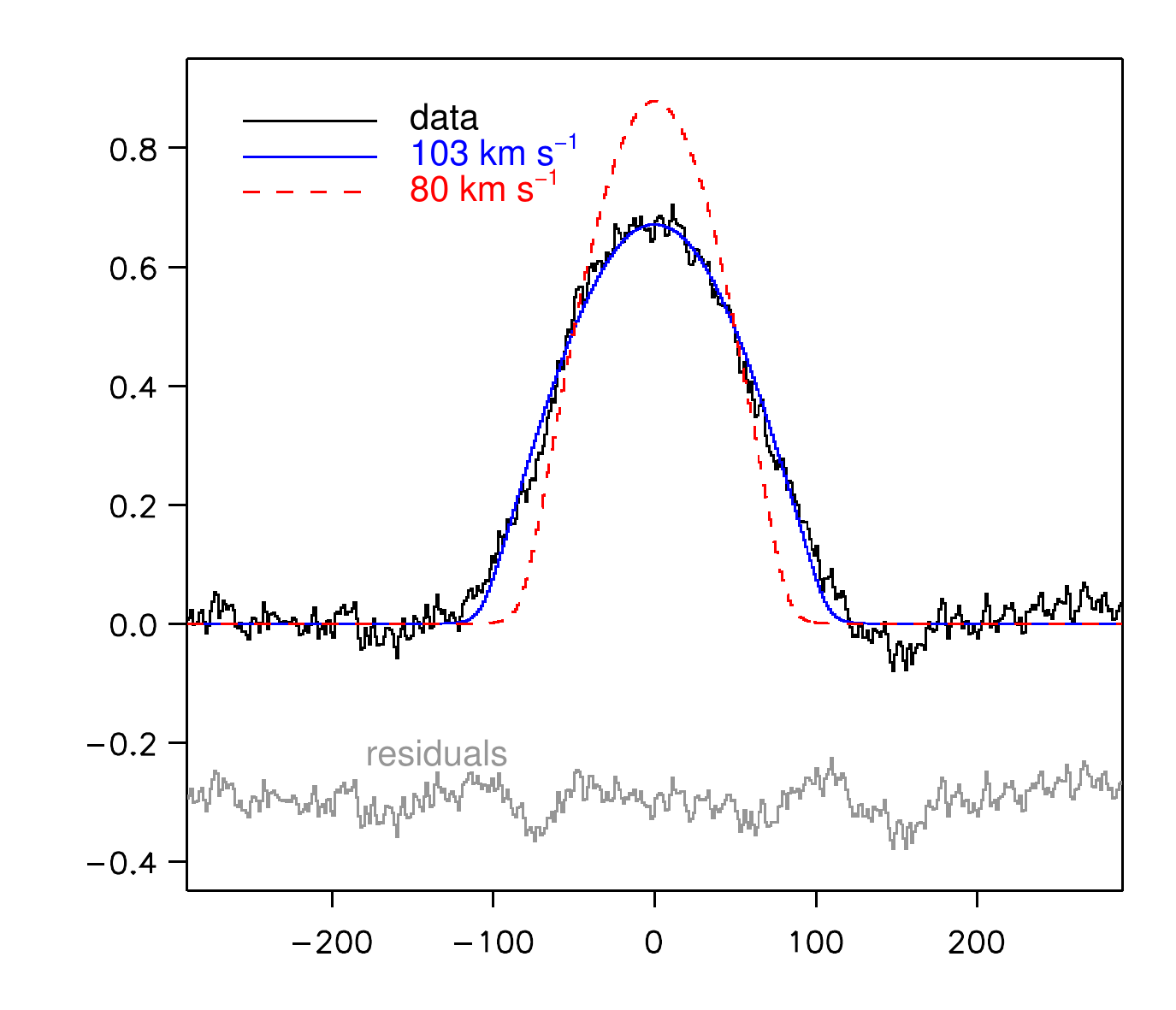}
\caption{CCF for the pre-transit observation from Dec.~2013 (black),
  along with the best-fitting model with $v\sin
  i_\star=103$~km\,s$^{-1}$ (blue). The red dashed line represents a
  model with $v\sin i_\star=80$~km\,s$^{-1}$, the value reported by
  VE+12. The gray line shows the differences between the data and the
  best-fitting (blue) model, vertically offset by 0.3 units for
  clarity.}
\label{fig:vsini} 
\end{figure}

We derived CCFs from the 2011 data in exactly the same way as for the
2013 data. We compared the CCFs with theoretical absorption lines
taking into account uniform rotation, limb darkening, macroturbulence,
and gravity darkening. We adopted the quadratic limb darkening
parameters from the tables of \citet{claret2012} for $\log g=3.5$ and
$T_{\rm eff}=3500$~K, solar metallicity, and the Johnson $V$ band. We
adopted a macroturbulent velocity of $15$~km\,s$^{-1}$.  To model the
gravity darkening we assumed the same effective temperature, a
rotation period of 0.448 days, a stellar mass of $0.4$~$M_\odot$, a
stellar radius of $1.4$~$R_\odot$, a stellar inclination angle of
$90^\circ$, and a gravity-darkening exponent of $\beta=0.25$.  We
neglected any oblateness of the stellar photosphere.  As an example,
Figure~\ref{fig:vsini} shows the CCF for one of the pre-transit
observations from Dec~2013, along with the best-fitting model.

The results for $v\sin i_\star$ are $103.6\pm0.3$~km\,s$^{-1}$ in
2011, and $104.1\pm0.7$~km\,s$^{-1}$ in 2013. The quoted uncertainties
are based on the scatter between the different observations for each
epoch, and do not include any additional systematic uncertainties due
to the limitations of the model (such as uncertainties in the
treatment of limb darkening and gravity darkening, or the neglect of
differential rotation and oblateness).  Therefore the relative
variation in $v \sin i_\star$ is bounded to less than a percent,
although the absolute value is probably uncertain by at least 10\%.

To obtain a better idea about the absolute value of $v \sin i_\star$
and its uncertainty, we tried fitting individual absorption lines
rather than the CCF.  Specifically, we fitted seven apparently
isolated lines between $5300$ and $7700$~\AA{}. The standard deviation
in the $v \sin i_\star$ measurement form these seven absorption lines
varies between $7-15$~km\,s$^{-1}$ for the different observations,
which suggests that an uncertainty of $10$~km\,s$^{-1}$ in $v \sin
i_\star$ should be a reasonable estimate. This leads to our
final estimate of $103\pm10$~km\,s$^{-1}$.

Our result for $v\sin i_\star$ is higher than the value of
$80.6\pm8.1$~km\,s$^{-1}$ reported by VE+12, but we do not think that
this necessarily (or even likely) implies that the projected rotation
rate is varying in time. This is because it is difficult to compare
the results directly, given that VE+12 used a completely different
instrument and analysis technique. Our internal comparison is much
more sensitive, since it is between two Keck/HIRES spectra obtained at
different times and analyzed in exactly the same way.

\subsection{Time variations in emission-line profiles}
\label{sec:emission}

\begin{figure}[t]
\epsscale{1}
\includegraphics[width=0.5\textwidth]{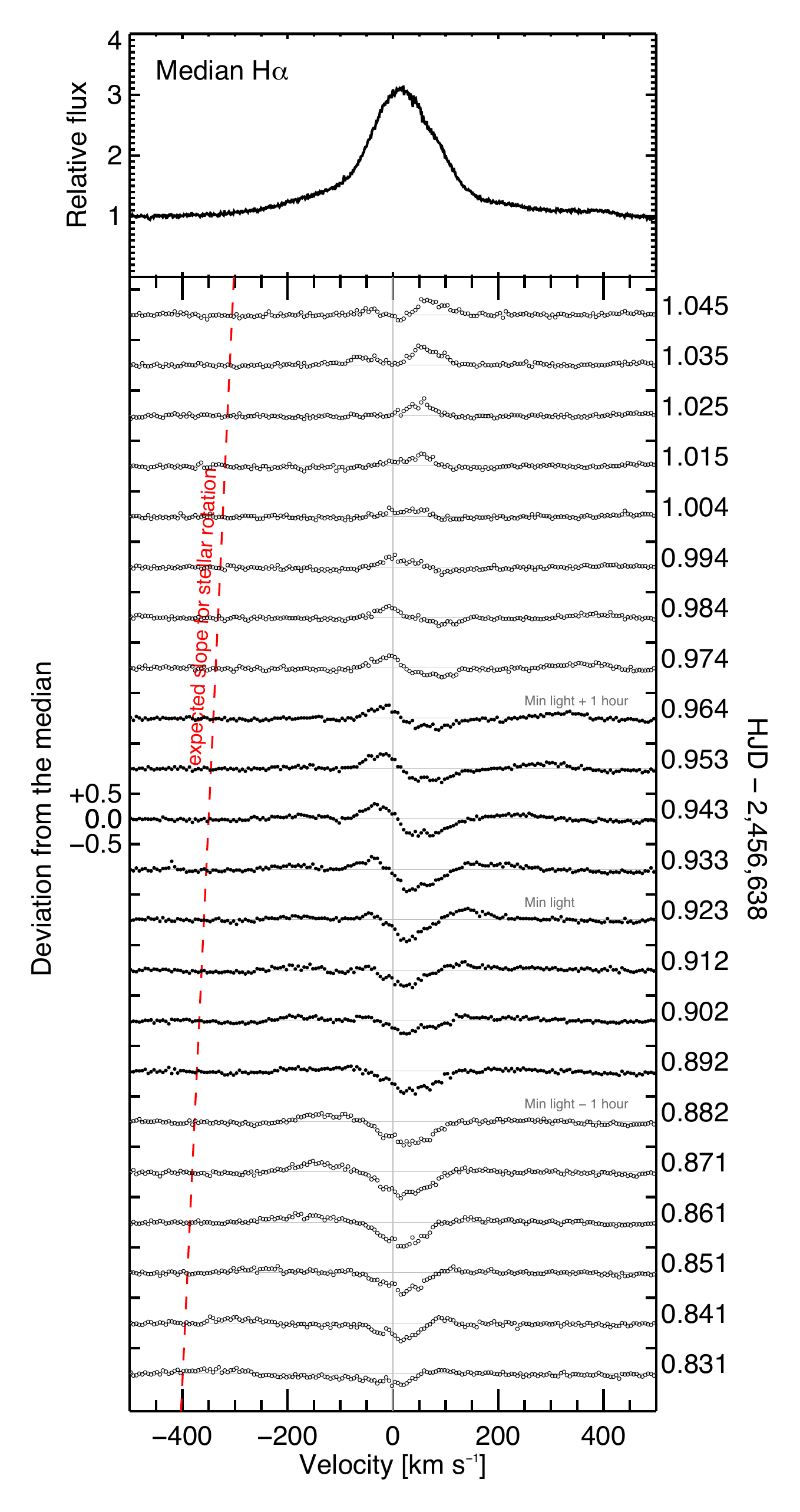}
\caption{The H$\alpha$ line profile of PTFO\,8-8695 on
  2013~Dec~12. {\it Top.}---Median line profile, based on all spectra
  more than one hour away from minimum light. {\it Bottom.}---Time
  series of residuals between each observed spectrum and the median.
  Open symbols are the ``out-of-transit'' spectra used to create the
  median spectrum. Filled symbols are the spectra within one hour of
  minimum light.  On the left axis, each vertical tick mark represents
  one unit of relative flux, on the same scale as the top panel. The
  right axis gives the time of each spectrum.}
\label{fig:halpha} 
\end{figure}

Our optical spectra also reveal strong emission lines from the
hydrogen Balmer series as well as the Ca~II~H~\&~K transitions. The
top panel of Figure~\ref{fig:halpha} shows the median H$\alpha$ line
profile, based on all the ``out-of-transit'' spectra observed on
2013~Dec~12 (at least an hour before or after the time of minimum
light). The line is very broad. Most of the emission is confined to
velocities $\lesssim$100~km~s$^{-1}$, consistent with the star's
rotation rate, but the velocity profile extends to at least
300~km~s$^{-1}$, particularly on the blue side. This is suggestive of
at least a low level of ongoing accretion. Material that falls onto
the star from large distances could attain the free-fall velocity
$\sqrt{2GM_\star/R_\star} \approx 330$~km~s$^{-1}$, given the nominal
parameters $M_\star=0.4~M_\odot$ and $R_\star=1.4~R_\odot$. The
equivalent width of the H$\alpha$ line is 8.7~\AA, placing the star
near the traditional borderline between the categories of
``weak-lined'' and ``classical'' T~Tauri stars. For simplicity it is
often said that the classical stars are actively accreting while the
weak-lined stars are not accreting, although in reality there seems to
be no sharp distinction between these categories, and PTFO\,8-8695
presents an intermediate case.

Figure~\ref{fig:halpha} also shows the time sequence of observed
changes in the H$\alpha$ line profile. Specifically, for each of the
22 spectra, we plotted the residuals between the observed line profile
and the median ``out-of-transit'' line profile. Also indicated are the
times of minimum light, as well as ``ingress'' (one hour prior) and
``egress'' (one hour afterward). The red dashed line indicates a slope
of 37~km~s$^{-1}$~hr$^{-1}$. This is the expected radial acceleration
of any feature attached to the stellar photosphere, which would move
from $-100$~km~s$^{-1}$ to $+100$~km~s$^{-1}$ over the course of
$P_{\rm rot}/2 = 0.224$~days.

Evidently the line profile varied in a complex pattern on a timescale
of minutes. Several excess-emission features do seem to be rotating
along with the star; for example, a pattern of positive residuals
appears in the fourth-to-final spectrum (time coordinate $t=1.015$) at
velocity $+50$~km~s$^{-1}$ and shifted redward at the expected rate
throughout the final three observations.  A similar pattern -- perhaps
originating from the same feature -- is seen starting at minimum light
at velocity $-50$~km~s$^{-1}$ and proceeding redward until about an
hour after minimum light.  These particular components of the emission
line seems likely to be caused by active regions on the stellar
surface.

During the fading event, the residuals show relative absorption at a
redshifted velocity of 25--100~km~s$^{-1}$. The absorption seemed to
disappear at around the same time as the end of the fading event. The
onset of the absorption feature was at least 2 hours before minimum
light, which is at least an hour before what seems to be ``ingress''
of the fading event. Thus, the transient redshifted absorption does
not seem to be exactly coincident with the fading event, although it
does at least suggest that the fading event was associated with hot
infalling material in front of the star.

\begin{figure}[t]
\epsscale{1}
\includegraphics[width=0.5\textwidth]{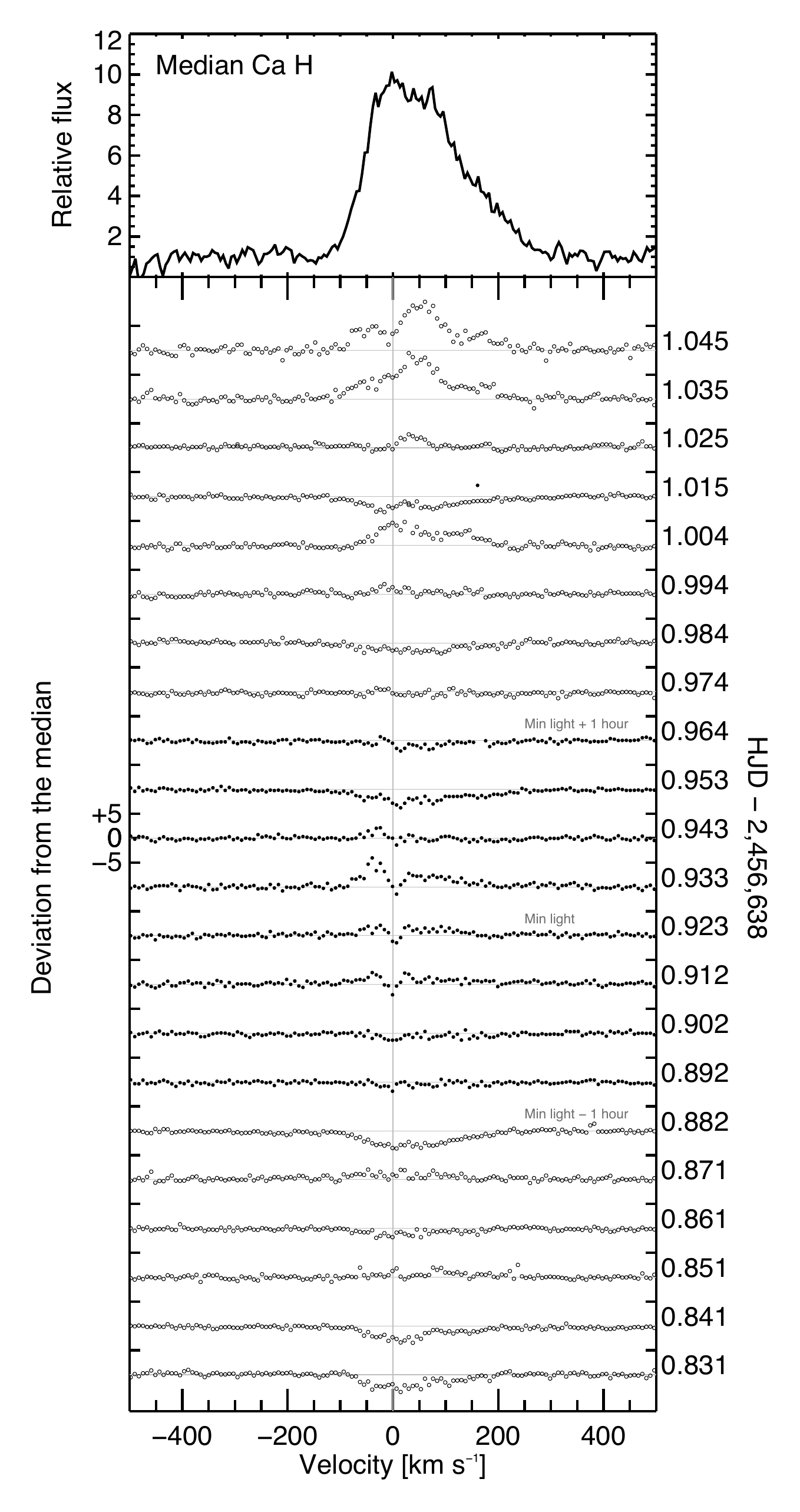}
\caption{The Ca~II~H line profile of PTFO\,8-8695 on
  2013~Dec~12. {\it Top.}---Median line profile, based on all spectra
  more than one hour away from minimum light. {\it Bottom.}---Time
  series of residuals between each observed spectrum and the median.
  Open symbols are the ``out-of-transit'' spectra used to create the
  median spectrum. Filled symbols are the spectra within one hour of
  minimum light.
  On the left axis, each vertical tick mark represents 5 units of relative flux, on the
  same scale as the top panel. The right axis gives the time of each
  spectrum.}
\label{fig:cah} 
\end{figure}

The H$\gamma$ line profiles (not shown here) tell a similar story but
with a lower signal-to-noise ratio. The H$\beta$ line was not
observed, given the spectral format. Figure~\ref{fig:cah} shows the
median Ca~II~H line profile, along with the time series of deviations
from the median. In this case the fractional variations were even
stronger and seemingly faster; there is no straightforward narrative
to the sequence of residuals.

\begin{figure}[t]
\epsscale{1}
\includegraphics[width=0.5\textwidth]{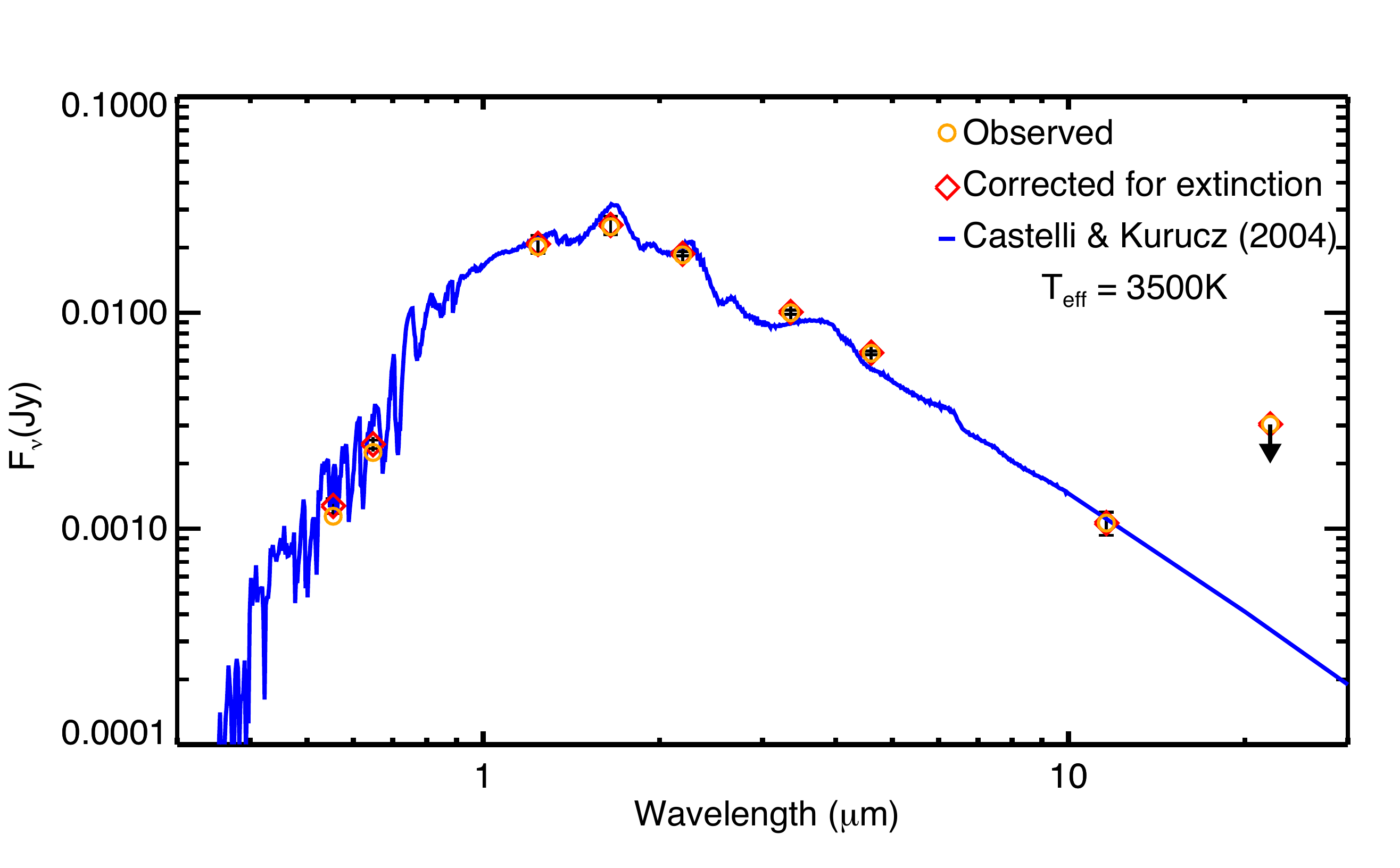}
\caption{Spectral energy distribution of PTFO\,8-8695 based on
  publicly available broadband photometry (orange diamonds), along
  with the best-fitting stellar atmosphere model (blue line). The data
  were corrected for extinction (red diamonds) prior to fitting.
  There is no evidence for any infrared excess that could be
  attributed to a circumstellar disk.}
\label{fig:sed} 
\end{figure}

To seek independent evidence for ongoing accretion, we used the
available broadband photometry to construct the spectral energy
distribution of PTFO\,8-8695 and search for any infrared excess.
Figure~\ref{fig:sed} shows the result, based on a query of the VizieR
website\footnote{http://vizier.u-strasbg.fr}, which gave measurements
in the $VRJHK$ bands as well as the {\it WISE} $W1$-$W4$ bands. (The
$W4$ observation gave an upper limit.) We corrected for dust
extinction with the dust map from the NASA/IPAC
website\footnote{http://irsa.ipac.caltech.edu}, and fitted the results
to a grid of zero-metallicity stellar atmosphere models from the
library of \citet{castelli04}. The best-fitting stellar parameters
were $T_{\rm eff}=3500\pm 120$~K and $\log g = 4.0\pm 0.9$. The
effective temperature is in agreement with the previously reported
value of 3470\,K \citep{briceno05}. The apparent lack of an infrared
excess out to 10~$\mu$m is characteristic of weak-lined T~Tauri stars. This lack of evidence for the existence of an accretion accretion disk within 1 AU does not necessarily rule out accretion, but does suggest that any accretion is relatively weak.

\section{Discussion}
\label{sec:discussion}

We now summarize the results of the three tests that we undertook to
test the hypothesis that the fading events of PTFO\,8-8695 are the
transits of a close-in giant planet:
\begin{enumerate}

\item The new light curves show variations in depth and duration from
  event to event. However we did not find strong evidence for
  asymmetries or other secular changes in morphology indicative of the
  changing trajectory of a transiting planet. Furthermore, in all
  cases, a fading event was observed at the appointed time, even
  though the cessation of transits was predicted to be likely by
  \citet{barnes13}.

\item Infrared photometry spanning the predicted times of planetary
  occultations has ruled out signals of the expected amplitude.

\item The Rossiter-McLaughlin effect could not be detected, ruling out
  a transiting planet with the expected parameters, unless the
  planet's trajectory is nearly perpendicular to the projected stellar
  equator. Nor did we detect any changes in $v \sin i_{\star}$ between
  2011 and 2013, which would have been produced by precession of the
  stellar rotation axis.

\end{enumerate}

Any of these tests could have resulted in a confirmation of the planet
hypothesis. In all cases, though, the planet hypothesis was
disfavored. In addition we found that the fading events are not
strictly periodic. In the most recent observing season the events
occurred more than one hour earlier than expected. This finding is
incompatible with the strict periodicity that one expects for a
planetary orbit.

While this paper was in preparation we learned of the work by
\citet{ciardi15} and \citet{koen15}, who also pursued follow-up
observations of PTFO\,8-8695. Among the observations of
\citet{ciardi15} was a nondetection of any transit-like event on
2012~Dec~21, based on observations in the $r'$ band. They gave an
upper limit of 0.7\% on the transit depth. Their simultaneous
spectroscopy also revealed no evidence of the Rossiter-McLaughlin
effect. They interpreted these nondetections as evidence for the
predicted cessation of the transits. However, our observations reveal
that the fading events did indeed take place on 2012~Dec~11, 14, and
15, with a depth of approximately 3\% in all cases. It seems unlikely
that orbital precession could have abruptly reduced the transit depth
from 3\% to below 0.7\% in less than one week. Therefore our results
cast doubt on this aspect of the interpretation of
\citet{ciardi15}. Likewise, \citet{koen15} reported non-detections of
predicted fading events on 2015~Jan~3-4. However, the predicted times
were based on the assumption of strict periodicity, which our
observations have shown to be false. Judging from Fig.~2 of
\citet{koen15}, it seems possible that the fading events were recorded
in the SAAO observations a few hours earlier than expected.  In any
case we detected a clear 2\% fading event on 2014 Dec 27, only one
week earlier than the SAAO observations.

\begin{figure*}[t]
\epsscale{1}
\minipage{\textwidth}
\includegraphics[width=0.33\textwidth]{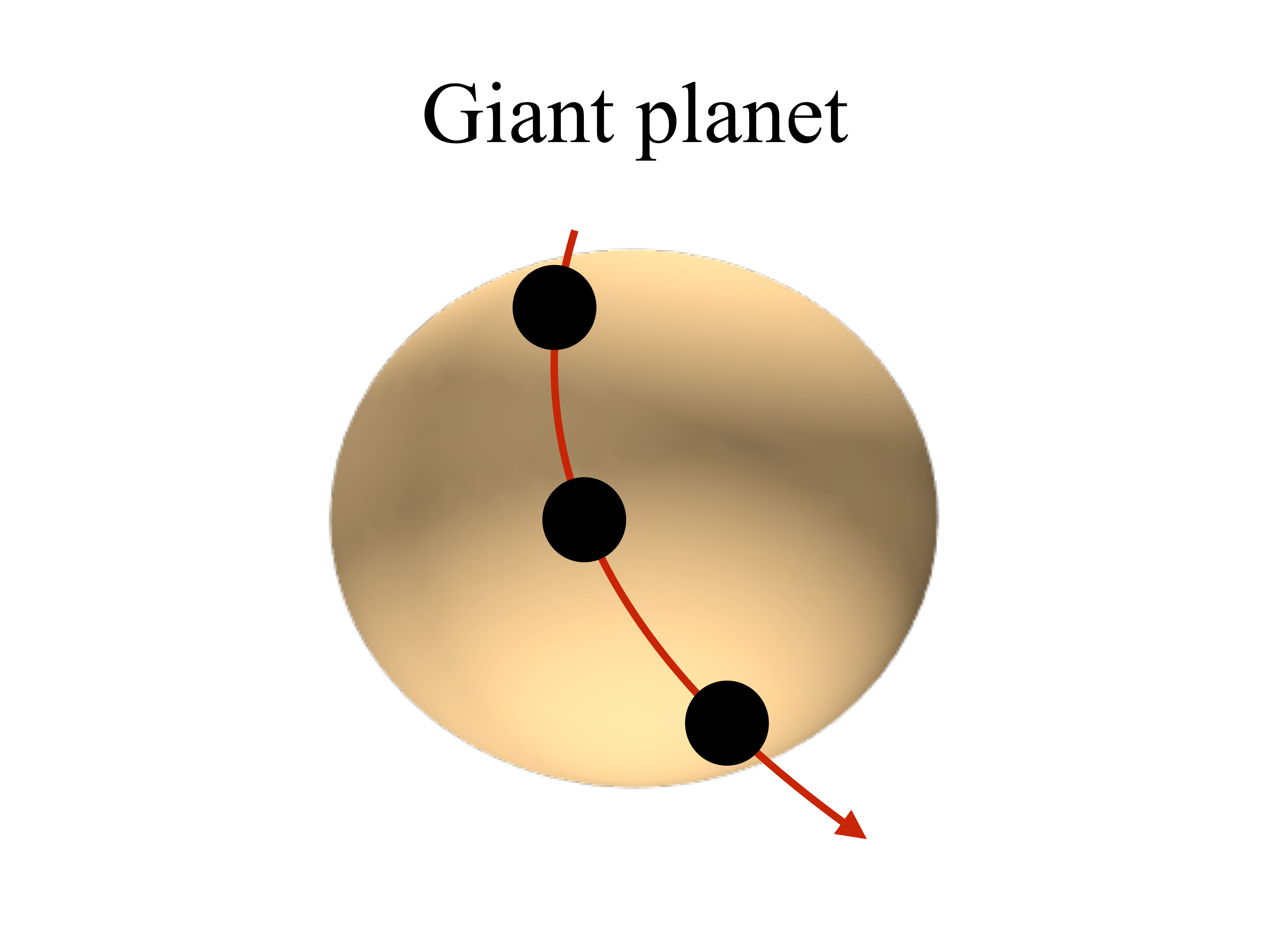}
\includegraphics[width=0.33\textwidth]{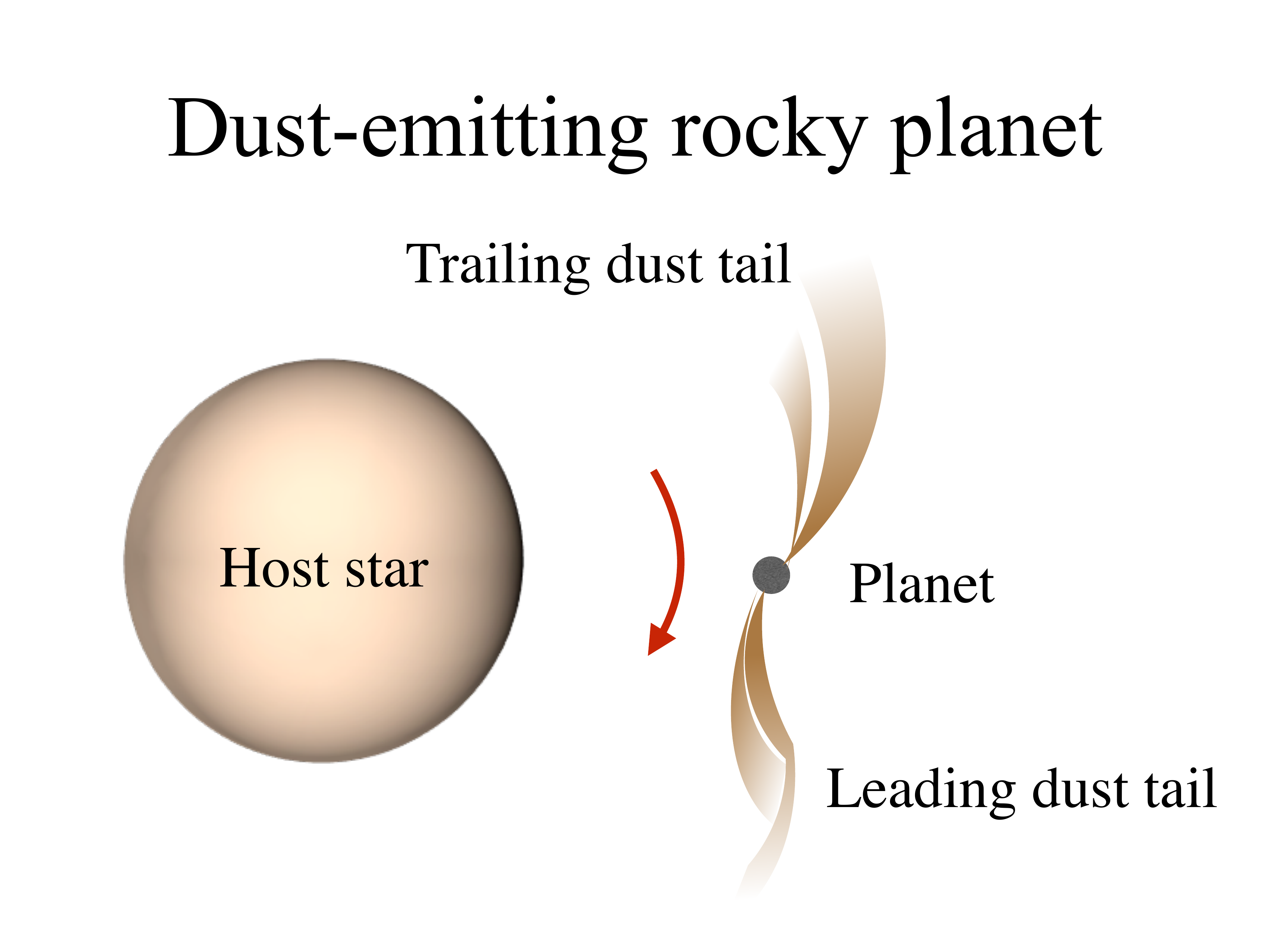}
\includegraphics[width=0.33\textwidth]{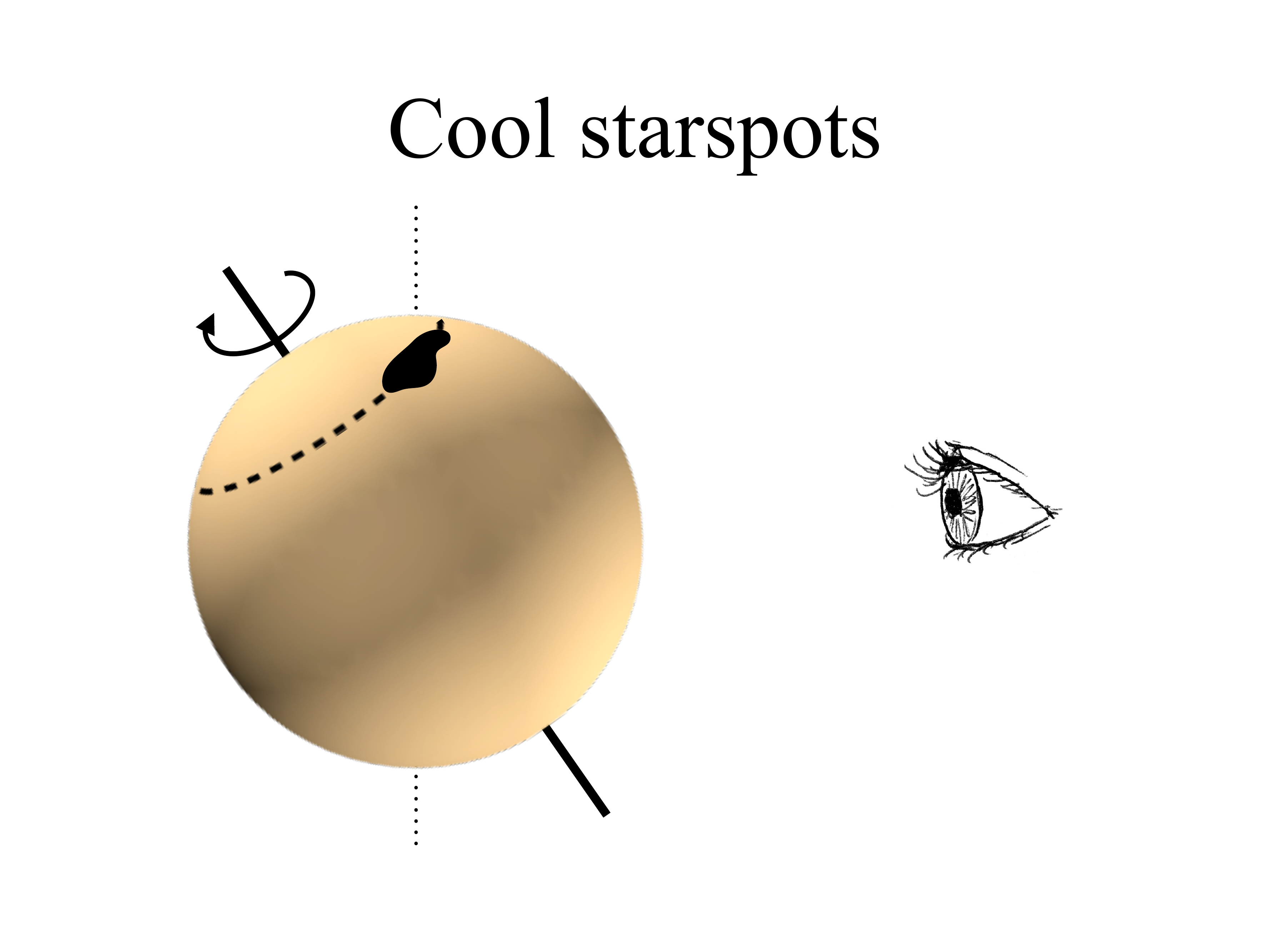}
\endminipage\hfill
\minipage{\textwidth}
\centering
\includegraphics[width=0.33\textwidth]{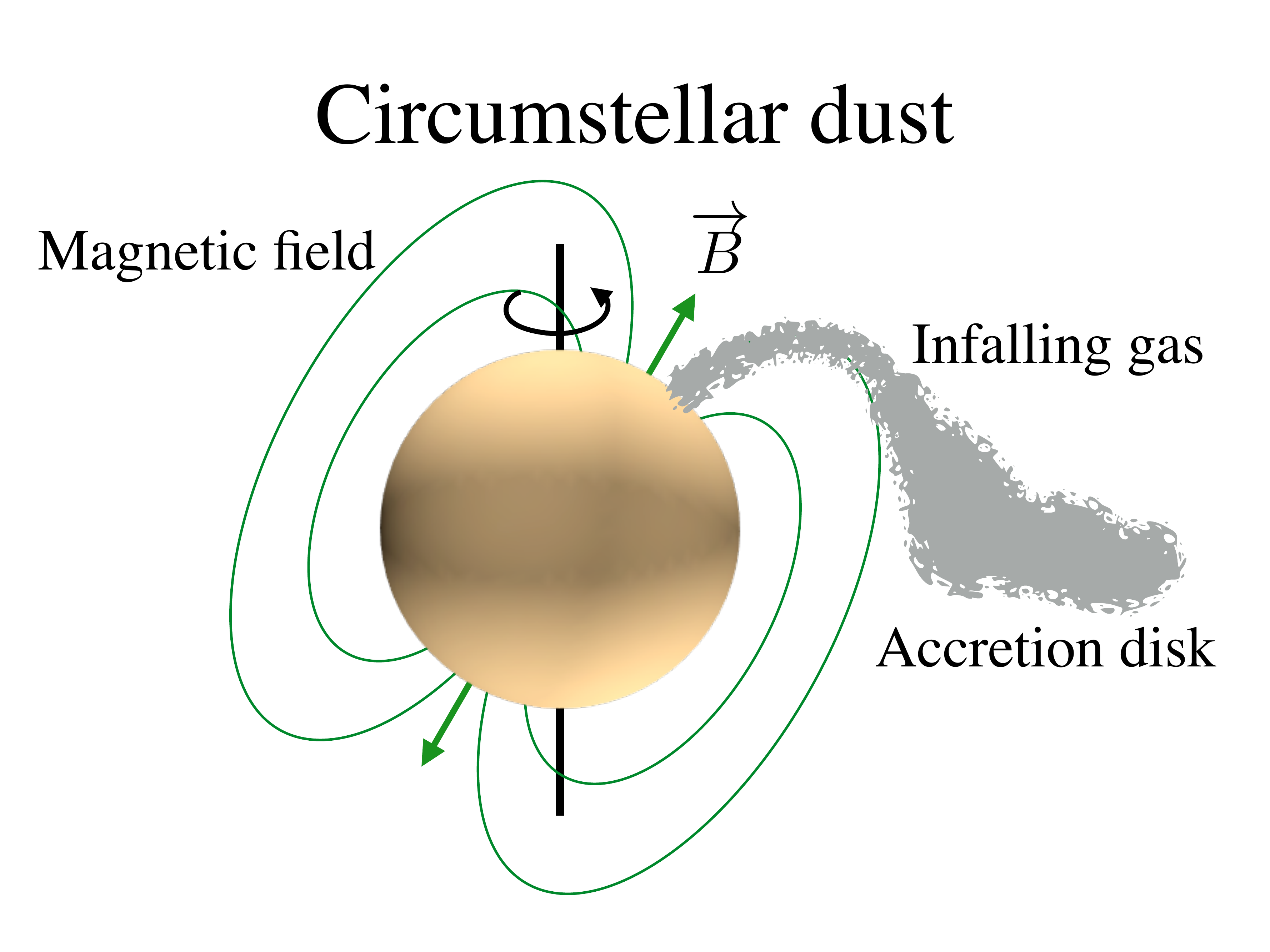}
\includegraphics[width=0.33\textwidth]{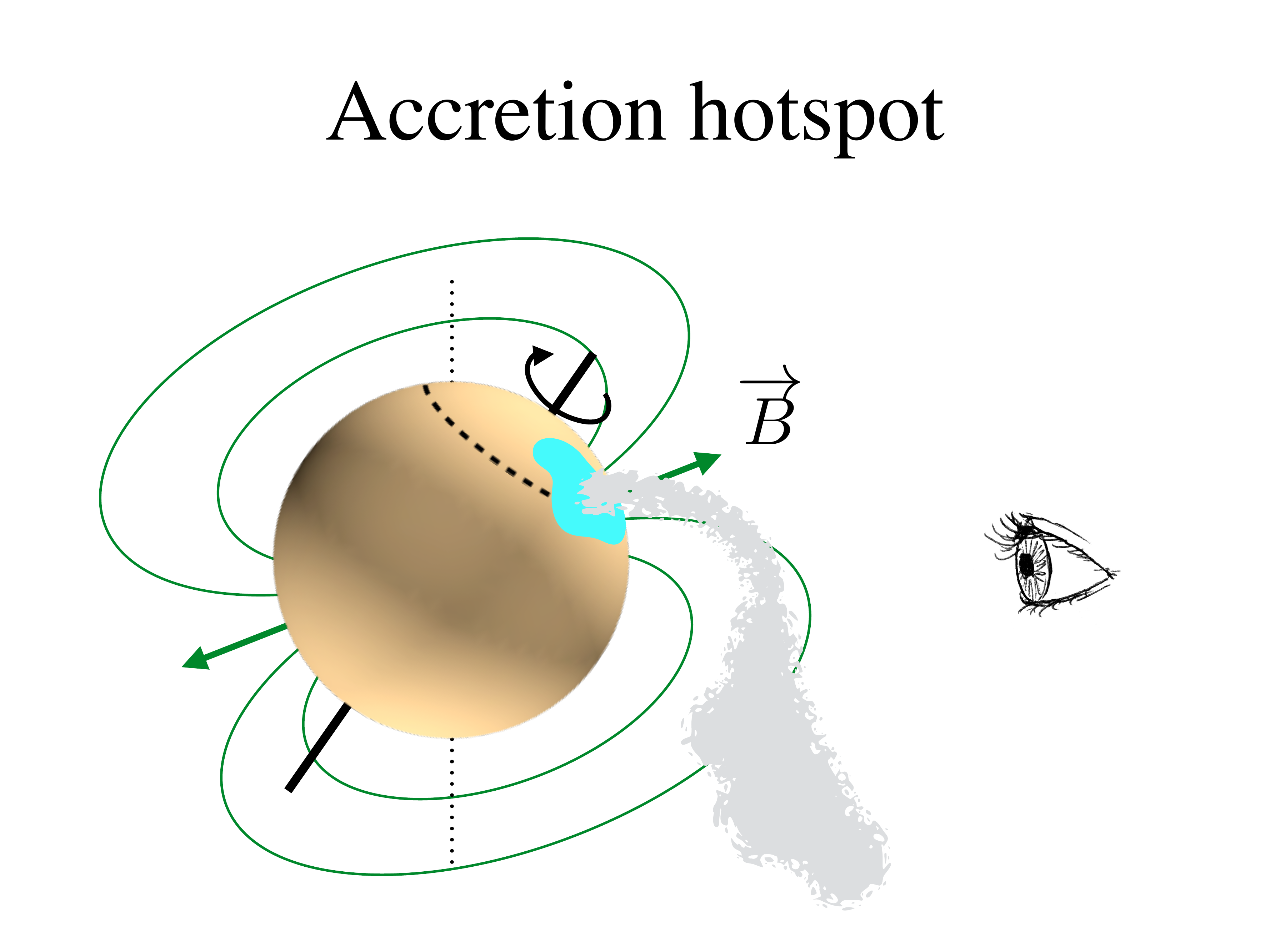}
\endminipage\hfill
\caption{Illustrations of the five hypotheses discussed in Section~\ref{sec:discussion}.}
\label{fig:models} 
\end{figure*}

In summary, our observations have significantly reduced the
credibility of the planet hypothesis.  However, because the hypothesis
invokes an unusual planet in unusual circumstances, it is difficult to
rule out definitively.  It may be possible to find reasons for the
failure of each of the individual tests. For example, the predicted
occultation times might have been incorrect, because the planet has a
highly eccentric orbit (see \S~\ref{sec:spitzer}).  Or perhaps the
values of the key parameters $(R_p/R_\star)^2$ and $R_\star/a$ are
smaller than the values postulated by \cite{barnes13}, which would
reduce the predicted occultation depth. The planet's atmosphere might
have deep absorption features near 1.7~$\mu$m and 4.5~$\mu$m that
rendered the planetary occultations undetectable.  The planet's orbit
might have been nearly perpendicular to the stellar equator at the
time of our attempt to detect the Rossiter-McLaughlin effect. It
remains possible that a comprehensive search of parameter space of the
model proposed by \cite{barnes13} --- including the effects of gravity
darkening and orbital precession --- could reveal a configuration that
possesses these properties and is also compatible with the lack of
detectable change in $v\sin i_\star$, as well as the unexpectedly
bland morphologies, strongly chromatic depths, lack of occultation
signals, and timing irregularities that are seen in the new light
curves. We leave such a computationally intensive search for future
work. It is also important to try and develop alternative hypotheses
for the fading events of PTFO\,8-8695. Below we describe four
alternatives, along with their apparent strengths and
weaknesses. Figure~\ref{fig:models} illustrates these hypotheses.

\subsection{Dust-emitting rocky planet}
\label{sec:kic1255}

The seemingly erratic variations in the depth and duration of the
fading events, along with the slight asymmetries in the phase-folded
light curves (Figure~\ref{fig:phasefolded}), bring to mind the case of
KIC~12557548 \citep{rappaport12}. This object was identified in data
from the {\it Kepler} spacecraft. It exhibits transit-like flux dips
with a very short period (15.7~hours) and duration (1.5~hours), an
erratically varying depth ranging from 0--1.3\%, and an egress of
longer duration than ingress. \citet{rappaport12} interpreted the dips
as transits by a dusty tail being emitted by a small rocky planet. Two
other similar cases have since been identified
\citep{rappaport14,sanchisojeda15}. In at least one case the transit
depth has been shown to be smaller at infrared wavelengths, as
expected for extinction by small dust grains.

PTFO\,8-8695 shares much of the phenomenology that has just been
described. The main difference is that the other systems are not
rapidly-rotating young stars. They appear to be slowly-rotating
main-sequence stars, and are not even close to being synchronized with
the transit period. Furthermore it is not clear whether this
hypothesis could be reconciled with the apparent change in period that
was seen in the most recent season of observations.

It is tempting to invoke tidal dissipation as a mechanism for
gradually shrinking the orbit and shortening the period, but this is
implausible for a low-mass rocky body. Conceivably, orbital decay
could be a consequence of the dust emission. If the dust acquires
additional specific angular momentum from radiation pressure while
leaving the system, it would be driven into a higher orbit. There it
would pull back on the planet and potentially extract angular momentum
from the planetary orbit. However, the magnitude of this effective
drag is difficult to estimate from first principles, particularly
because the dust may represent only a modest fraction of the total
mass loss, and the gas need not behave the same way as dust as it
leaves the system. In any case the lifetime of $\sim$10$^4$ years
implied by the observations (see \S~\ref{sec:omc}) is uncomfortably
short. It would require a special coincidence to observe such a
short-lived phase of evolution.

\subsection{Starspots}
\label{sec:starspots}

The synchronization of the stellar rotation period and the period of
the fading events, along with the changing depth and duration of the
fading events, raises the suspicion that the periodic dips are caused
by starspots being carried around by rotation. The star is expected to
be heavily spotted, given its youth. Moreover, the photometric
variations produced by stellar activity are expected to be weaker in
the infrared than in the optical, consistent with our
observations. Gradual changes in the spot pattern could be invoked to
explain the changes in depth, duration, and timing of the fading
events.

VE+12 have already pointed out the main weakness of this hypothesis.
Flux variations caused by starspots have a natural timescale of half
the rotation period, the interval over which a patch of the stellar
surface is typically visible to the observer. In contrast, the fading
events of PTFO\,8-8695 last only $\approx$15\% of the rotation period.
One can arrange for short-duration dips by locating the starspots near
one of the rotation poles, and tipping the star such that the spots
are only on the visible hemisphere over a narrow range of
longitudes. Indeed, T~Tauri stars are well known for displaying stable
spot patterns near the magnetic poles
\citep[e.g.][]{granzer00}. However, in such a circumstance the spot is
highly foreshortened and limb-darkened, and it is difficult to produce
variations of several percent. It is also difficult to produce the
apparently sharp features that have been occasionally seen in the
light curves, such as the apparent point of ``fourth contact'' in our
Magellan $H$-band time series (see the bottom panel of
Figure~\ref{fig:multiband}). It might be possible to save this
hypothesis by invoking multiple spots and in a complex and stable
pattern, such that the summation of their photometric variations is
coincidentally narrow in time. This model, though, would be rather
contrived.

\subsection{Eclipses by a circumstellar disk or dust}
\label{sec:accretion}

As we have just seen, the basic difficulty of any model in which the
variations are due to features on the stellar photosphere is the
relatively short duration of the fading events. Eclipses by orbiting
bodies avoid this problem because the time spent in front of the star
scales as $R_\star/a$, and the orbital distance $a$ can be adjusted to
match the observations. This is the basis of the planet hypothesis,
for which \citet{barnes13} found $R_\star/a=0.5$--0.6. However the
eclipsing body need not be a planet.  Perhaps it is a feature in the
innermost, corotating portion of the stellar magnetosphere.

At an age of a few million years, low-mass T Tauri stars are often
still actively accreting material from their circumstellar disks. PTFO\,8-8695 lacks any evidence for an accretion disk, yet, as
noted in \S~\ref{sec:emission}, the strength and breadth of the H$\alpha$
line profile places the star in between the traditional categories of ``classical''
and ``weak-lined'' T~Tauri stars and suggest that PTFO\,8-8695 may be weakly accreting. So we will consider the implications of a small amount of dust that would not be detected as an IR excess, but would still be able to produce the flux dips. The accretion process is thought to
proceed as follows \citep[see, e.g.,][for a review]{bouvier07}.
First, matter spirals inward through a thin disk, until it reaches an
orbital distance of a few stellar radii, at which point the disk is
disrupted by the star's magnetosphere.  Then, within the
magnetosphere, the material is magnetically funneled into narrow tubes
or columns, extending from the inner edge of the accretion disk onto
the star's magnetic poles. The material falls freely along these
columns and crashes onto the star, producing shock waves and a
luminous hotspot.  In many models, the stellar rotation rate becomes
synchronized with the Keplerian orbital velocity at the inner radius
of the magnetosphere (the corotation radius), a phenomenon known as
disk locking.

This picture contains several elements that could naturally lead to
flux variations with a period equal to the stellar rotation
period. For example if the accretion disk is warped or has other
non-axisymmetric variations in thickness near the innermost,
corotating portion of the disk, then the star may be periodically
eclipsed by these irregularities.  This is thought to be the basic
explanation for the quasiperiodic eclipses of AA Tau \citep{bouvier99,
  bouvier03}. Alternatively, periodic eclipses could be produced by
stable patterns or concentrations of dust in the accretion flow. This
may explain the observed variability of the ``short-duration dippers''
recently identified by \citet{cody14} and \citet{stauffer15}. Those
authors studied a number of T~Tauri stars in the young cluster
NGC~2264 that exhibit short-duration fading events. They found that
the flux dips are quasi-periodic and exhibit changes in depth and
shape from epoch to epoch over a period of several years. The dips
appear shallower in the infrared than in the optical, and the light
curves have rounded minima rather than being
flat-bottomed \citep{stauffer15}. The flux dip periods are usually
equal to the rotation periods of the stars, and some dips have been
observed to persist for years.  All these properties have been
attributed to extinction by infalling material from the innermost
portion of the accretion disk \citep{mcginnis15}. And, all these
properties are consistent with our observations of PTFO\,8-8695.

In some respects, though, PTFO\,8-8695 is different from the rest of
the dippers.  Its rotation period of 0.45~days is shorter than the
3--10 day periods of most of the stars observed by \citet{cody14} and
\citet{stauffer15}, or the 8~day period of AA~Tau. The duration of the
fading events is also relatively shorter at 15\% of the period,
compared to the more typical value of $\sim$30\%. It is also unclear
whether disk warps or dust concentrations could produce the apparently
sharp features and flat bottoms observed in a few of our light curves.
Furthermore, the dippers all have SEDs with a detectable infrared
excess, but the currently available data for PTFO\,8-8695 show no
evidence for any infrared excess (Figure~\ref{fig:sed}).

Furthermore, it is questionable whether dust can exist in solid form
so close to the star, with an orbital distance less than 2~$R_\star$.
Assuming that stellar radiation is the dominant
mechanism of heating, the dust sublimation radius $R_s$ is given by
\citep{monnier02}
\begin{equation}
R_s= \frac{1}{2} \sqrt{Q_R} \left( \frac{T_{\star}}{T_s} \right)^2 R_{\star}
\end{equation}
where $T_s$ is the dust sublimation temperature ($\approx$1500~K), and
$Q_R = Q_{\rm abs} (T_{\star}) / Q_{\rm abs} (T_s)$ is the ratio of
the dust absorption efficiencies for radiation at the color
temperatures of the incident and reemitted fields, respectively.
Adopting the customary value of $T_s = 1500$~K, and assuming
$Q_R$ to be of order unity (as expected in this case for
silicate grains), this formula gives $R_s = 2.7~R_\star$. Therefore
the hypothesized dust with $R_\star/a$=0.5--0.6, corresponding
to $a =$1.7--2~$R_\star$, would likely be vaporized.

\subsection{Occultations of an accretion hotspot}
\label{sec:hotspot}

Hotspots are another aspect of the magnetospheric accretion model that
has previously been invoked to explain some of the quasiperiodic
variations of T~Tauri stars \citep{herbst94}. We advance here a
related hypothesis for the case of PTFO\,8-8695: perhaps the fading
events represent occultations of one of the hotspots that is produced
by ongoing low-level accretion. In this scenario, the accreting
material is funnelled onto a hotspot near one of the star's magnetic
poles, which is displaced from the star's rotation pole.  Furthermore,
the star's rotation pole is tipped toward the observer such that the
hotspot is on the visible hemisphere for $\approx$85\% of each
rotation period. When the hotspot is hidden from view, we observe a
fading event.

This is similar to the starspot hypothesis (\S~\ref{sec:starspots}) in
that the photometric modulations are the result of the rotation of a
photospheric feature of contrasting intensity. However, replacing the
dark starspot with a luminous hotspot could potentially solve some of
the problems faced by the starspot model. Hotspots have been inferred
to have effective temperatures ranging up to $10^4$~K, and thereby
present much higher contrast than starspots with the surrounding
photosphere. Furthermore, hotspots are probably not confined to a
vertically thin surface layer of the photosphere, and as such they are
not subject to the effects of limb-darkening or foreshortening.
Together these factors may make it easier to produce abrupt
modulations of a few percent in the total light even for a small
feature near the stellar limb.

If this model is correct, then the mass accretion rate can be estimated
from the loss of light during fading events, which is observed to be
of order 5\%. This requires the accretion luminosity to be
\be
L_{\rm acc} = \frac{GM_\star \dot{M}}{R} \sim 0.05~L_{\rm bol}.
\ee
Using the nominal values $M_\star=0.4~M_\odot$, $R=1.4~R_\odot$, and
$L_{\rm bol}=0.25~L_\odot$, we find $\dot{M} \sim
10^{-9}$~$M_\odot$~yr$^{-1}$,
at the low end of the
range of mass accretion rates that has been inferred for accreting
T~Tauri stars ($10^{-9}$--$10^{-7}$~$M_\odot$~yr$^{-1}$). This seems
reasonable: a relatively low accretion rate is
also consistent with the relatively weak H$\alpha$ equivalent width of
8.7~\AA{} and the absence of a detectable infrared excess.

One concern with this model is that in the systems where hotspots have
been previously invoked to explain photometric variability, the
pattern of variability is not as consistent or long-lasting as is seen
in PTFO\,8-8695 \citep{herbst94}. Oftentimes the hotspot variability
shows no periodicity or, at best, short-lived periodicity, sometimes
with period changes of 20\% or more. However, those previous results
pertain to classical T~Tauri stars with much higher inferred accretion
rates; perhaps we are seeing the different and more stable behavior of
a hotspot in a more weakly accreting system.

\subsection{Summary}
\label{sec:summary}

We have discussed five hypotheses for the fading events of
PTFO\,8-8695. The precessing giant-planet model has failed several key
tests, the most serious of which are probably the nondetection of the
planetary occultations, and the apparent shift in the phase of the
fading events in the most recent season of observations. The planet
hypothesis also struggles to explain the observed coincidence of the
rotation and orbital periods. This same problem afflicts the
hypothesis of the dust-emitting rocky planet.

The other three models share the virtue of a natural explanation for
the equality of the rotation period and the period of the fading
events. However, the starspot model has difficulty reproducing the
observed duration and occasionally sharp ingress/egress of the fading
events. The other two models invoke the presence of an accretion disk,
for which the evidence is ambiguous or negative: the H$\alpha$ line
profile does extend to higher velocities than can be explained by
stellar rotation, but the equivalent width is relatively low and there
is no detectable infrared excess.

The occulted-hotspot model seems quantitatively promising, as it is
consistent with a low accretion rate of
$\sim$10$^{-9}$~$M_\odot$~yr$^{-1}$. There is no deterministic theory
for the expected photometric variations due to magnetically-funnelled
accretion, making it difficult to achieve a firm confirmation of this
hypothesis.  Nevertheless all our observations seem at least
consistent with this picture. At present, we consider this hypothesis
to be the best explanation for PTFO\,8-8695.  To come to a firmer
conclusion will probably require more photometric and spectroscopic
observations, seeking changes in the timing and behavior of the fading
events, variations in the H$\alpha$ line profile, and more sensitive
searches for any infrared (or ultraviolet) excess or other indicators
of low-level accretion.

At the outset of this project, and throughout most of this paper, we
have been mainly concerned with the status of the planetary hypothesis
for this intriguing planetary candidate. In fact this object may turn
out to be useful for understanding magnetospheric accretion, due to a
fortuitous geometry, thereby joining the ranks of such systems as
AA~Tau \citep{bouvier99} and KH~15D \citep{hamilton12}.


\acknowledgments We thank the referee, Jason W.\ Barnes, for his
comments on the manuscript. We are grateful to Geoff Marcy for helping
to arrange the Keck observations, to Aur\'elie Fumel for helping with
the TRAPPIST observations, and to John Johnson and Jon Swift for
attempting to perform photometric observations for this project.  We
thank Greg Herczeg, Hans Moritz G\"unther and Nikku Madhusudhan for helpful discussions, and
Julian van Eyken for providing the PTF data in a convenient
format. Work by JNW was supported by the NASA Origins program (grant
NNX11AG85G). MG and EJ are Research Associates at the Belgian Fund for
Scientific Research (Fond National de la Recherche Scientifique,
F.R.S-FNRS); LD received the support of the F.R.I.A. fund of the FNRS.
TRAPPIST is a project funded by the F.R.S-FNRS under grant FRFC
2.5.594.09.F, with the participation of the Swiss National Science
Foundation (SNF). BTM is supported by the National Science Foundation
Graduate Research Fellowship under Grant No. DGE1144469. The authors
wish to extend special thanks to those of Hawai'ian ancestry on whose
sacred mountain of Mauna Kea we are privileged to be guests. Without
their generous hospitality, the Keck observations presented herein
would not have been possible.

{\it Facilities:} \facility{FLWO:1.2m}, \facility{Magellan:Baade},
\facility{Euler1.2m}, \facility{Keck:I (HIRES)}, \facility{Spitzer}

\bibliography{ms}

\begin{thebibliography}{}
\expandafter\ifx\csname natexlab\endcsname\relax\def\natexlab#1{#1}\fi

\bibitem[{{Albrecht} {et~al.}(2012){Albrecht}, {Winn}, {Johnson}, {Howard},
  {Marcy}, {Butler}, {Arriagada}, {Crane}, {Shectman}, {Thompson}, {Hirano},
  {Bakos}, \& {Hartman}}]{albrecht12}
{Albrecht}, S., {Winn}, J.~N., {Johnson}, J.~A., {et~al.} 2012,
  \href{http://dx.doi.org/10.1088/0004-637X/757/1/18}{\apj},
  \href{http://adsabs.harvard.edu/abs/2012ApJ...757...18A}{757},
  \href{http://adsabs.harvard.edu/abs/2012ApJ...757...18A}{18}

\bibitem[{{Barnes}(2009)}]{barnes09}
{Barnes}, J.~W. 2009,
  \href{http://dx.doi.org/10.1088/0004-637X/705/1/683}{\apj},
  \href{http://adsabs.harvard.edu/abs/2009ApJ...705..683B}{705},
  \href{http://adsabs.harvard.edu/abs/2009ApJ...705..683B}{683}

\bibitem[{{Barnes} {et~al.}(2013){Barnes}, {van Eyken}, {Jackson}, {Ciardi}, \&
  {Fortney}}]{barnes13}
{Barnes}, J.~W., {van Eyken}, J.~C., {Jackson}, B.~K., {Ciardi}, D.~R., \&
  {Fortney}, J.~J. 2013,
  \href{http://dx.doi.org/10.1088/0004-637X/774/1/53}{\apj},
  \href{http://adsabs.harvard.edu/abs/2013ApJ...774...53B}{774},
  \href{http://adsabs.harvard.edu/abs/2013ApJ...774...53B}{53}

\bibitem[{{Bouvier} {et~al.}(1999){Bouvier}, {Chelli}, {Allain}, {Carrasco},
  {Costero}, {Cruz-Gonzalez}, {Dougados}, {Fern{\'a}ndez}, {Mart{\'{\i}}n},
  {M{\'e}nard}, {Mennessier}, {Mujica}, {Recillas}, {Salas}, {Schmidt}, \&
  {Wichmann}}]{bouvier99}
{Bouvier}, J., {Chelli}, A., {Allain}, S., {et~al.} 1999, \aap,
  \href{http://adsabs.harvard.edu/abs/1999A%26A...349..619B}{349},
  \href{http://adsabs.harvard.edu/abs/1999A%26A...349..619B}{619}

\bibitem[{{Bouvier} {et~al.}(2003){Bouvier}, {Grankin}, {Alencar}, {Dougados},
  {Fern{\'a}ndez}, {Basri}, {Batalha}, {Guenther}, {Ibrahimov}, {Magakian},
  {Melnikov}, {Petrov}, {Rud}, \& {Zapatero Osorio}}]{bouvier03}
{Bouvier}, J., {Grankin}, K.~N., {Alencar}, S.~H.~P., {et~al.} 2003,
  \href{http://dx.doi.org/10.1051/0004-6361:20030938}{\aap},
  \href{http://adsabs.harvard.edu/abs/2003A%26A...409..169B}{409},
  \href{http://adsabs.harvard.edu/abs/2003A%26A...409..169B}{169}

\bibitem[{{Bouvier} {et~al.}(2007){Bouvier}, {Alencar}, {Boutelier},
  {Dougados}, {Balog}, {Grankin}, {Hodgkin}, {Ibrahimov}, {Kun}, {Magakian}, \&
  {Pinte}}]{bouvier07}
{Bouvier}, J., {Alencar}, S.~H.~P., {Boutelier}, T., {et~al.} 2007,
  \href{http://dx.doi.org/10.1051/0004-6361:20066021}{\aap},
  \href{http://adsabs.harvard.edu/abs/2007A%26A...463.1017B}{463},
  \href{http://adsabs.harvard.edu/abs/2007A%26A...463.1017B}{1017}

\bibitem[{{Brice{\~n}o} {et~al.}(2005){Brice{\~n}o}, {Calvet}, {Hern{\'a}ndez},
  {Vivas}, {Hartmann}, {Downes}, \& {Berlind}}]{briceno05}
{Brice{\~n}o}, C., {Calvet}, N., {Hern{\'a}ndez}, J., {et~al.} 2005,
  \href{http://dx.doi.org/10.1086/426911}{\aj},
  \href{http://adsabs.harvard.edu/abs/2005AJ....129..907B}{129},
  \href{http://adsabs.harvard.edu/abs/2005AJ....129..907B}{907}

\bibitem[{{Castelli} \& {Kurucz}(2004)}]{castelli04}
{Castelli}, F., \& {Kurucz}, R.~L. 2004, ArXiv Astrophysics e-prints,
  \href{http://adsabs.harvard.edu/abs/2004astro.ph..5087C}{astro-ph/0405087}

\bibitem[{{Ciardi} {et~al.}(2015){Ciardi}, {van Eyken}, {Barnes}, {Beichman},
  {Carey}, {Crockett}, {Eastman}, {Johns-Krull}, {Howell}, {Kane}, {Mclane},
  {Plavchan}, {Prato}, {Stauffer}, {van Belle}, \& {von Braun}}]{ciardi15}
{Ciardi}, D.~R., {van Eyken}, J.~C., {Barnes}, J.~W., {et~al.} 2015, ArXiv
  e-prints,
  arXiv:\href{http://adsabs.harvard.edu/abs/2015arXiv150608719C}{1506.08719}

\bibitem[{{Claret} {et~al.}(2012){Claret}, {Hauschildt}, \&
  {Witte}}]{claret2012}
{Claret}, A., {Hauschildt}, P.~H., \& {Witte}, S. 2012,
  \href{http://dx.doi.org/10.1051/0004-6361/201219849}{\aap},
  \href{http://adsabs.harvard.edu/abs/2012A%26A...546A..14C}{546},
  \href{http://adsabs.harvard.edu/abs/2012A%26A...546A..14C}{A14}

\bibitem[{{Clemens} {et~al.}(2007){Clemens}, {Sarcia}, {Grabau}, {Tollestrup},
  {Buie}, {Dunham}, \& {Taylor}}]{clemens07}
{Clemens}, D.~P., {Sarcia}, D., {Grabau}, A., {et~al.} 2007,
  \href{http://dx.doi.org/10.1086/524775}{\pasp},
  \href{http://adsabs.harvard.edu/abs/2007PASP..119.1385C}{119},
  \href{http://adsabs.harvard.edu/abs/2007PASP..119.1385C}{1385}

\bibitem[{{Cody} {et~al.}(2014){Cody}, {Stauffer}, {Baglin}, {Micela},
  {Rebull}, {Flaccomio}, {Morales-Calder{\'o}n}, {Aigrain}, {Bouvier},
  {Hillenbrand}, {Gutermuth}, {Song}, {Turner}, {Alencar}, {Zwintz},
  {Plavchan}, {Carpenter}, {Findeisen}, {Carey}, {Terebey}, {Hartmann},
  {Calvet}, {Teixeira}, {Vrba}, {Wolk}, {Covey}, {Poppenhaeger}, {G{\"u}nther},
  {Forbrich}, {Whitney}, {Affer}, {Herbst}, {Hora}, {Barrado}, {Holtzman},
  {Marchis}, {Wood}, {Medeiros Guimar{\~a}es}, {Lillo Box}, {Gillen},
  {McQuillan}, {Espaillat}, {Allen}, {D'Alessio}, \& {Favata}}]{cody14}
{Cody}, A.~M., {Stauffer}, J., {Baglin}, A., {et~al.} 2014,
  \href{http://dx.doi.org/10.1088/0004-6256/147/4/82}{\aj},
  \href{http://adsabs.harvard.edu/abs/2014AJ....147...82C}{147},
  \href{http://adsabs.harvard.edu/abs/2014AJ....147...82C}{82}

\bibitem[{{Eastman} {et~al.}(2010){Eastman}, {Siverd}, \& {Gaudi}}]{eastman10}
{Eastman}, J., {Siverd}, R., \& {Gaudi}, B.~S. 2010,
  \href{http://dx.doi.org/10.1086/655938}{\pasp},
  \href{http://adsabs.harvard.edu/abs/2010PASP..122..935E}{122},
  \href{http://adsabs.harvard.edu/abs/2010PASP..122..935E}{935}

\bibitem[{{Gillon} {et~al.}(2011){Gillon}, {Jehin}, {Magain}, {Chantry},
  {Hutsem{\'e}kers}, {Manfroid}, {Queloz}, \& {Udry}}]{gillon11}
{Gillon}, M., {Jehin}, E., {Magain}, P., {et~al.} 2011, in European Physical
  Journal Web of Conferences, Vol.~11, European Physical Journal Web of
  Conferences, 6002,
  arXiv:\href{http://adsabs.harvard.edu/abs/2011EPJWC..1106002G}{1101.5807}

\bibitem[{{Gillon} {et~al.}(2013){Gillon}, {Anderson}, {Collier-Cameron},
  {Doyle}, {Fumel}, {Hellier}, {Jehin}, {Lendl}, {Maxted}, {Montalb{\'a}n},
  {Pepe}, {Pollacco}, {Queloz}, {S{\'e}gransan}, {Smith}, {Smalley},
  {Southworth}, {Triaud}, {Udry}, \& {West}}]{gillon13}
{Gillon}, M., {Anderson}, D.~R., {Collier-Cameron}, A., {et~al.} 2013,
  \href{http://dx.doi.org/10.1051/0004-6361/201220561}{\aap},
  \href{http://adsabs.harvard.edu/abs/2013A%26A...552A..82G}{552},
  \href{http://adsabs.harvard.edu/abs/2013A%26A...552A..82G}{A82}

\bibitem[{{Granzer} {et~al.}(2000){Granzer}, {Sch{\"u}ssler}, {Caligari}, \&
  {Strassmeier}}]{granzer00}
{Granzer}, T., {Sch{\"u}ssler}, M., {Caligari}, P., \& {Strassmeier}, K.~G.
  2000, \aap, \href{http://adsabs.harvard.edu/abs/2000A%26A...355.1087G}{355},
  \href{http://adsabs.harvard.edu/abs/2000A%26A...355.1087G}{1087}

\bibitem[{{Hamilton} {et~al.}(2012){Hamilton}, {Johns-Krull}, {Mundt},
  {Herbst}, \& {Winn}}]{hamilton12}
{Hamilton}, C.~M., {Johns-Krull}, C.~M., {Mundt}, R., {Herbst}, W., \& {Winn},
  J.~N. 2012, \href{http://dx.doi.org/10.1088/0004-637X/751/2/147}{\apj},
  \href{http://adsabs.harvard.edu/abs/2012ApJ...751..147H}{751},
  \href{http://adsabs.harvard.edu/abs/2012ApJ...751..147H}{147}

\bibitem[{{Herbst} {et~al.}(1994){Herbst}, {Herbst}, {Grossman}, \&
  {Weinstein}}]{herbst94}
{Herbst}, W., {Herbst}, D.~K., {Grossman}, E.~J., \& {Weinstein}, D. 1994,
  \href{http://dx.doi.org/10.1086/117204}{\aj},
  \href{http://adsabs.harvard.edu/abs/1994AJ....108.1906H}{108},
  \href{http://adsabs.harvard.edu/abs/1994AJ....108.1906H}{1906}

\bibitem[{{Husser} {et~al.}(2013){Husser}, {Wende-von Berg}, {Dreizler},
  {Homeier}, {Reiners}, {Barman}, \& {Hauschildt}}]{husser2013}
{Husser}, T.-O., {Wende-von Berg}, S., {Dreizler}, S., {et~al.} 2013,
  \href{http://dx.doi.org/10.1051/0004-6361/201219058}{\aap},
  \href{http://adsabs.harvard.edu/abs/2013A%26A...553A...6H}{553},
  \href{http://adsabs.harvard.edu/abs/2013A%26A...553A...6H}{A6}

\bibitem[{{Jehin} {et~al.}(2011){Jehin}, {Gillon}, {Queloz}, {Magain},
  {Manfroid}, {Chantry}, {Lendl}, {Hutsem{\'e}kers}, \& {Udry}}]{jehin11}
{Jehin}, E., {Gillon}, M., {Queloz}, D., {et~al.} 2011, The Messenger,
  \href{http://adsabs.harvard.edu/abs/2011Msngr.145....2J}{145},
  \href{http://adsabs.harvard.edu/abs/2011Msngr.145....2J}{2}

\bibitem[{{Kamiaka} {et~al.}(2015){Kamiaka}, {Masuda}, {Xue}, {Suto},
  {Nishioka}, {Murakami}, {Inayama}, {Saitoh}, {Tanaka}, \&
  {Yonehara}}]{kamiaka15}
{Kamiaka}, S., {Masuda}, K., {Xue}, Y., {et~al.} 2015, ArXiv e-prints,
  arXiv:\href{http://adsabs.harvard.edu/abs/2015arXiv150604829K}{1506.04829}

\bibitem[{{Knutson} {et~al.}(2008){Knutson}, {Charbonneau}, {Allen}, {Burrows},
  \& {Megeath}}]{knutson+08}
{Knutson}, H.~A., {Charbonneau}, D., {Allen}, L.~E., {Burrows}, A., \&
  {Megeath}, S.~T. 2008, \href{http://dx.doi.org/10.1086/523894}{\apj},
  \href{http://adsabs.harvard.edu/abs/2008ApJ...673..526K}{673},
  \href{http://adsabs.harvard.edu/abs/2008ApJ...673..526K}{526}

\bibitem[{{Koen}(2015)}]{koen15}
{Koen}, C. 2015, \href{http://dx.doi.org/10.1093/mnras/stv906}{\mnras},
  \href{http://adsabs.harvard.edu/abs/2015MNRAS.450.3991K}{450},
  \href{http://adsabs.harvard.edu/abs/2015MNRAS.450.3991K}{3991}

\bibitem[{{Lendl} {et~al.}(2012){Lendl}, {Anderson}, {Collier-Cameron},
  {Doyle}, {Gillon}, {Hellier}, {Jehin}, {Lister}, {Maxted}, {Pepe},
  {Pollacco}, {Queloz}, {Smalley}, {S{\'e}gransan}, {Smith}, {Triaud}, {Udry},
  {West}, \& {Wheatley}}]{lendl12}
{Lendl}, M., {Anderson}, D.~R., {Collier-Cameron}, A., {et~al.} 2012,
  \href{http://dx.doi.org/10.1051/0004-6361/201219585}{\aap},
  \href{http://adsabs.harvard.edu/abs/2012A%26A...544A..72L}{544},
  \href{http://adsabs.harvard.edu/abs/2012A%26A...544A..72L}{A72}

\bibitem[{{Mandel} \& {Agol}(2002)}]{mandelagol02}
{Mandel}, K., \& {Agol}, E. 2002,
  \href{http://dx.doi.org/10.1086/345520}{\apjl},
  \href{http://adsabs.harvard.edu/abs/2002ApJ...580L.171M}{580},
  \href{http://adsabs.harvard.edu/abs/2002ApJ...580L.171M}{L171}

\bibitem[{{McGinnis} {et~al.}(2015){McGinnis}, {Alencar}, {Guimar{\~a}es},
  {Sousa}, {Stauffer}, {Bouvier}, {Rebull}, {Fonseca}, {Venuti}, {Hillenbrand},
  {Cody}, {Teixeira}, {Aigrain}, {Favata}, {F{\H u}r{\'e}sz}, {Vrba},
  {Flaccomio}, {Turner}, {Gameiro}, {Dougados}, {Herbst},
  {Morales-Calder{\'o}n}, \& {Micela}}]{mcginnis15}
{McGinnis}, P.~T., {Alencar}, S.~H.~P., {Guimar{\~a}es}, M.~M., {et~al.} 2015,
  \href{http://dx.doi.org/10.1051/0004-6361/201425475}{\aap},
  \href{http://adsabs.harvard.edu/abs/2015A%26A...577A..11M}{577},
  \href{http://adsabs.harvard.edu/abs/2015A%26A...577A..11M}{A11}

\bibitem[{{McLaughlin}(1924)}]{mclaughlin24}
{McLaughlin}, D.~B. 1924, \href{http://dx.doi.org/10.1086/142826}{\apj},
  \href{http://adsabs.harvard.edu/abs/1924ApJ....60...22M}{60},
  \href{http://adsabs.harvard.edu/abs/1924ApJ....60...22M}{22}

\bibitem[{{Monnier} \& {Millan-Gabet}(2002)}]{monnier02}
{Monnier}, J.~D., \& {Millan-Gabet}, R. 2002,
  \href{http://dx.doi.org/10.1086/342917}{\apj},
  \href{http://adsabs.harvard.edu/abs/2002ApJ...579..694M}{579},
  \href{http://adsabs.harvard.edu/abs/2002ApJ...579..694M}{694}

\bibitem[{{Rappaport} {et~al.}(2014){Rappaport}, {Barclay}, {DeVore}, {Rowe},
  {Sanchis-Ojeda}, \& {Still}}]{rappaport14}
{Rappaport}, S., {Barclay}, T., {DeVore}, J., {et~al.} 2014,
  \href{http://dx.doi.org/10.1088/0004-637X/784/1/40}{\apj},
  \href{http://adsabs.harvard.edu/abs/2014ApJ...784...40R}{784},
  \href{http://adsabs.harvard.edu/abs/2014ApJ...784...40R}{40}

\bibitem[{{Rappaport} {et~al.}(2013){Rappaport}, {Sanchis-Ojeda}, {Rogers},
  {Levine}, \& {Winn}}]{rappaport+13}
{Rappaport}, S., {Sanchis-Ojeda}, R., {Rogers}, L.~A., {Levine}, A., \& {Winn},
  J.~N. 2013, \href{http://dx.doi.org/10.1088/2041-8205/773/1/L15}{\apjl},
  \href{http://adsabs.harvard.edu/abs/2013ApJ...773L..15R}{773},
  \href{http://adsabs.harvard.edu/abs/2013ApJ...773L..15R}{L15}

\bibitem[{{Rappaport} {et~al.}(2012){Rappaport}, {Levine}, {Chiang}, {El
  Mellah}, {Jenkins}, {Kalomeni}, {Kite}, {Kotson}, {Nelson},
  {Rousseau-Nepton}, \& {Tran}}]{rappaport12}
{Rappaport}, S., {Levine}, A., {Chiang}, E., {et~al.} 2012,
  \href{http://dx.doi.org/10.1088/0004-637X/752/1/1}{\apj},
  \href{http://adsabs.harvard.edu/abs/2012ApJ...752....1R}{752},
  \href{http://adsabs.harvard.edu/abs/2012ApJ...752....1R}{1}

\bibitem[{{Rossiter}(1924)}]{rossiter24}
{Rossiter}, R.~A. 1924, \href{http://dx.doi.org/10.1086/142825}{\apj},
  \href{http://adsabs.harvard.edu/abs/1924ApJ....60...15R}{60},
  \href{http://adsabs.harvard.edu/abs/1924ApJ....60...15R}{15}

\bibitem[{{Ruan} {et~al.}(2000){Ruan}, {Zhang}, {Zhu}, \& {Zhu}}]{ruan00}
{Ruan}, S.~C., {Zhang}, F., {Zhu}, Q., \& {Zhu}, G.~P. 2000, in SPIE Conference
  Series, Vol. 4087, SPIE Conference Series, ed. A.~R. {Lessard} \& G.~A.
  {Lampropoulos}, 817R

\bibitem[{{Sanchis-Ojeda} {et~al.}(2015){Sanchis-Ojeda}, {Rappaport},
  {Pall{\'e}}, {Delrez}, {DeVore}, {Gandolfi}, {Fukui}, {Ribas}, {Stassun},
  {Albrecht}, {Dai}, {Gaidos}, {Gillon}, {Hirano}, {Holman}, {Howard},
  {Isaacson}, {Jehin}, {Kuzuhara}, {Mann}, {Marcy}, {Miles-P{\'a}ez},
  {Monta{\~n}{\'e}s-Rodr{\'{\i}}guez}, {Murgas}, {Narita}, {Nowak}, {Onitsuka},
  {Paegert}, {Van Eylen}, {Winn}, \& {Yu}}]{sanchisojeda15}
{Sanchis-Ojeda}, R., {Rappaport}, S., {Pall{\'e}}, E., {et~al.} 2015, ArXiv
  e-prints,
  arXiv:\href{http://adsabs.harvard.edu/abs/2015arXiv150404379S}{1504.04379}

\bibitem[{{Stauffer} {et~al.}(2015){Stauffer}, {Cody}, {McGinnis}, {Rebull},
  {Hillenbrand}, {Turner}, {Carpenter}, {Plavchan}, {Carey}, {Terebey},
  {Morales-Calder{\'o}n}, {Alencar}, {Bouvier}, {Venuti}, {Hartmann}, {Calvet},
  {Micela}, {Flaccomio}, {Song}, {Gutermuth}, {Barrado}, {Vrba}, {Covey},
  {Padgett}, {Herbst}, {Gillen}, {Lyra}, {Medeiros Guimaraes}, {Bouy}, \&
  {Favata}}]{stauffer15}
{Stauffer}, J., {Cody}, A.~M., {McGinnis}, P., {et~al.} 2015,
  \href{http://dx.doi.org/10.1088/0004-6256/149/4/130}{\aj},
  \href{http://adsabs.harvard.edu/abs/2015AJ....149..130S}{149},
  \href{http://adsabs.harvard.edu/abs/2015AJ....149..130S}{130}

\bibitem[{{Stevenson} {et~al.}(2012){Stevenson}, {Harrington}, {Fortney},
  {Loredo}, {Hardy}, {Nymeyer}, {Bowman}, {Cubillos}, {Bowman}, \&
  {Hardin}}]{stevenson+12}
{Stevenson}, K.~B., {Harrington}, J., {Fortney}, J.~J., {et~al.} 2012,
  \href{http://dx.doi.org/10.1088/0004-637X/754/2/136}{\apj},
  \href{http://adsabs.harvard.edu/abs/2012ApJ...754..136S}{754},
  \href{http://adsabs.harvard.edu/abs/2012ApJ...754..136S}{136}

\bibitem[{{Szab{\'o}} {et~al.}(2012){Szab{\'o}}, {P{\'a}l}, {Derekas}, {Simon},
  {Szalai}, \& {Kiss}}]{szabo12}
{Szab{\'o}}, G.~M., {P{\'a}l}, A., {Derekas}, A., {et~al.} 2012,
  \href{http://dx.doi.org/10.1111/j.1745-3933.2012.01219.x}{\mnras},
  \href{http://adsabs.harvard.edu/abs/2012MNRAS.421L.122S}{421},
  \href{http://adsabs.harvard.edu/abs/2012MNRAS.421L.122S}{L122}

\bibitem[{{Szab{\'o}} {et~al.}(2011){Szab{\'o}}, {Szab{\'o}}, {Benk{\H o}},
  {Lehmann}, {Mez{\H o}}, {Simon}, {K{\H o}v{\'a}ri}, {Hodos{\'a}n},
  {Reg{\'a}ly}, \& {Kiss}}]{szabo11}
{Szab{\'o}}, G.~M., {Szab{\'o}}, R., {Benk{\H o}}, J.~M., {et~al.} 2011,
  \href{http://dx.doi.org/10.1088/2041-8205/736/1/L4}{\apjl},
  \href{http://adsabs.harvard.edu/abs/2011ApJ...736L...4S}{736},
  \href{http://adsabs.harvard.edu/abs/2011ApJ...736L...4S}{L4}

\bibitem[{{Triaud} {et~al.}(2010){Triaud}, {Collier Cameron}, {Queloz},
  {Anderson}, {Gillon}, {Hebb}, {Hellier}, {Loeillet}, {Maxted}, {Mayor},
  {Pepe}, {Pollacco}, {S{\'e}gransan}, {Smalley}, {Udry}, {West}, \&
  {Wheatley}}]{triaud+10}
{Triaud}, A.~H.~M.~J., {Collier Cameron}, A., {Queloz}, D., {et~al.} 2010,
  \href{http://dx.doi.org/10.1051/0004-6361/201014525}{\aap},
  \href{http://adsabs.harvard.edu/abs/2010A%26A...524A..25T}{524},
  \href{http://adsabs.harvard.edu/abs/2010A%26A...524A..25T}{A25}

\bibitem[{{van Eyken} {et~al.}(2012){van Eyken}, {Ciardi}, {von Braun}, {Kane},
  {Plavchan}, {Bender}, {Brown}, {Crepp}, {Fulton}, {Howard}, {Howell},
  {Mahadevan}, {Marcy}, {Shporer}, {Szkody}, {Akeson}, {Beichman}, {Boden},
  {Gelino}, {Hoard}, {Ram{\'{\i}}rez}, {Rebull}, {Stauffer}, {Bloom}, {Cenko},
  {Kasliwal}, {Kulkarni}, {Law}, {Nugent}, {Ofek}, {Poznanski}, {Quimby},
  {Walters}, {Grillmair}, {Laher}, {Levitan}, {Sesar}, \&
  {Surace}}]{vaneyken12}
{van Eyken}, J.~C., {Ciardi}, D.~R., {von Braun}, K., {et~al.} 2012,
  \href{http://dx.doi.org/10.1088/0004-637X/755/1/42}{\apj},
  \href{http://adsabs.harvard.edu/abs/2012ApJ...755...42V}{755},
  \href{http://adsabs.harvard.edu/abs/2012ApJ...755...42V}{42}

\bibitem[{{Vogt} {et~al.}(1994){Vogt}, {Allen}, {Bigelow}, {Bresee}, {Brown},
  {Cantrall}, {Conrad}, {Couture}, {Delaney}, {Epps}, {Hilyard}, {Hilyard},
  {Horn}, {Jern}, {Kanto}, {Keane}, {Kibrick}, {Lewis}, {Osborne},
  {Pardeilhan}, {Pfister}, {Ricketts}, {Robinson}, {Stover}, {Tucker}, {Ward},
  \& {Wei}}]{vogt1994}
{Vogt}, S.~S., {Allen}, S.~L., {Bigelow}, B.~C., {et~al.} 1994, in SPIE
  Conference Series, ed. {D.~L.~Crawford \& E.~R.~Craine}, Vol. 2198, 362

\bibitem[{{Winn}(2010)}]{winn_review}
{Winn}, J.~N. 2010, ArXiv e-prints,
  arXiv:\href{http://adsabs.harvard.edu/abs/2010arXiv1001.2010W}{1001.2010}

\end{thebibliography}

\end{document}